\documentclass[aps, prd, twocolumn, superscriptaddress, longbibliography]{revtex4-2}
\usepackage{xspace}
\usepackage{pgfplotstable}
\usepackage{array} 
\usepackage{xcolor}
\usepackage{makecell}
\usepackage{amsmath}
\usepackage{subfigure}
\usepackage{float}
\usepackage{orcidlink}

\usepackage[normalem]{ulem}

\newcolumntype{C}{>{\centering\arraybackslash}m{2.5em}}

\def\Tobs{T_{\textrm{\mbox{\tiny{obs}}}}}
\def\Tcoh{T_{\textrm{\mbox{\tiny{coh}}}}}

\def\fdot{\dot f}
\def\fddot{\ddot f}
\def\paramset{($f$,$\fdot$,$\fddot$)}

\newif\ifshowfigs
\showfigstrue

\newcommand{\F}{2\mathcal{F}}
\newcommand{\FM}{2\mathcal{\hat{F}}}
\newcommand{\FMthresh}{2\hat{\mathcal{F}}_{\mathrm{thresh}}}

\newcommand{\Nseg}{N_{\rm seg}}
\newcommand{\mcoh}{m_{\rm coh}}
\newcommand{\msemicoh}{m_{\rm semi}}

\newcommand{\SNR}{\textrm{SNR}}
\newcommand{\Fstat}{$\mathcal{F}$-statistic\xspace}
\newcommand{\depth}{\mathcal{D}}
\newcommand{\dd}{\mathcal{D}^{95\%}}

\newcommand{\psdwt}{{\bar S_h}}

\newcommand{\asdwtoff}{\sqrt{{\bar S_h(f)}}}

\newcommand\T{\rule{0pt}{2.6ex}}       
\newcommand\B{\rule[-1.2ex]{0pt}{0pt}} 

\def\weave{{\footnotesize\textsc{Weave}}\xspace}

\def\fmin{20\xspace}
\def\fmax{475\xspace}

\def\Re{\textrm{Re}}

\def\fgw{f}
\def\frot{f_{\rm rot}}

\def\hul{h_0^{95\%}}

\def\sci#1#2{#1\times10^{#2}}

\newcommand{\apjl}{Astrophys. J. Lett.}

\newcommand{\mnras}{Mon. Not. R. Astron. Soc.}

\newcommand{\aj}{Astron. J.}

\begin{document}

\title{Search for continuous gravitational waves from neutron stars in five globular clusters in the first part of the fourth LIGO-Virgo-KAGRA observing run}

\author{Damon~H.~T.~Cheung\,\orcidlink{0000-0003-3905-0665}} 
\email{damoncht@umich.edu}
\affiliation{Department of Physics, University of Michigan, 450 Church St, Ann Arbor, Michigan 48109, USA}

\author{Keith~Riles\,\orcidlink{0000-0002-6418-5812}} 
\affiliation{Department of Physics, University of Michigan, 450 Church St, Ann Arbor, Michigan 48109, USA}

\author{Rafel~Amengual\,\orcidlink{0000-0001-8691-3166}}
\affiliation{Departament de F\'isica, Universitat de les Illes Balears, IAC3--IEEC, Carretera de Valldemossa, km 7.5, 07122 Palma, Spain}

\author{Preet~Baxi} 
\affiliation{Department of Physics, University of Michigan, 450 Church St, Ann Arbor, Michigan 48109, USA}

\author{Alicia~Calafat\,\orcidlink{0009-0008-7515-6305}}
\affiliation{Departament de F\'isica, Universitat de les Illes Balears, IAC3--IEEC, Carretera de Valldemossa, km 7.5, 07122 Palma, Spain}

\author{Anamaria~Effler\,\orcidlink{0000-0001-8242-3944}}
\affiliation{LIGO Livingston Observatory, Livingston, Louisiana 70754, USA}

\author{Tabata~Aira~Ferreira\,\orcidlink{0000-0002-1166-2005}}
\affiliation{Instituto Nacional de Pesquisas Espaciais, 12227-010 S\~{a}o Jos\'{e} dos Campos, S\~{a}o Paulo, Brazil}

\author{Evan~Goetz\,\orcidlink{0000-0003-2666-721X}}
\affiliation{Department of Physics and Astronomy, University of British Columbia, Vancouver, British Columbia V6T 1Z1, Canada}

\author{David~Keitel\,\orcidlink{0000-0002-2824-626X}}
\affiliation{Departament de F\'isica, Universitat de les Illes Balears, IAC3--IEEC, Carretera de Valldemossa, km 7.5, 07122 Palma, Spain}

\author{Tom~Kimpson\,\orcidlink{0000-0002-6542-6032}}
\affiliation{School of Physics, University of Melbourne, Parkville, Victoria 3010, Australia}
\affiliation{Australian Research Council Centre of Excellence for Gravitational Wave Discovery (OzGrav), Hawthorn, Victoria 3122, Australia}

\author{Alan~M.~Knee\,\orcidlink{0000-0003-0703-947X}}
\affiliation{Department of Physics, University of Michigan, 450 Church St, Ann Arbor, Michigan 48109, USA}

\author{Joan-Ren\'{e}~M\'{e}rou\,\orcidlink{0000-0002-5776-6643}}
\affiliation{Departament de F\'isica, Universitat de les Illes Balears, IAC3--IEEC, Carretera de Valldemossa, km 7.5, 07122 Palma, Spain}

\author{Quynh~Lan~Nguyen\,\orcidlink{0000-0002-1828-3702}}
\affiliation{Phenikaa University, Nguyen Trac Street, Duong Noi, Hanoi, Vietnam}

\author{Joseph~O'Leary\,\orcidlink{0000-0002-6547-2039}}
\affiliation{School of Physics, University of Melbourne, Parkville, Victoria 3010, Australia}
\affiliation{Australian Research Council Centre of Excellence for Gravitational Wave Discovery (OzGrav), Hawthorn, Victoria 3122, Australia}

\author{Ornella~J.~Piccinni\,\orcidlink{0000-0001-5478-3950}}
\affiliation{Departament de F\'isica, Universitat de les Illes Balears, IAC3--IEEC, Carretera de Valldemossa, km 7.5, 07122 Palma, Spain}

\author{Alicia~M.~Sintes\,\orcidlink{0000-0001-9050-7515}}
\affiliation{Departament de F\'isica, Universitat de les Illes Balears, IAC3--IEEC, Carretera de Valldemossa, km 7.5, 07122 Palma, Spain}

\author{Rebekah~Slocum\,\orcidlink{0009-0005-8232-6833}}
\affiliation{Louisiana State University, Baton Rouge, Louisiana 70803, USA}

\author{Karl~Wette\,\orcidlink{0000-0002-6418-5812}} 
\affiliation{Centre for Gravitational Astrophysics, Australian National University, Canberra, Australian Capital Territory 2601, Australia}
\affiliation{Australian Research Council Centre of Excellence for Gravitational Wave Discovery (OzGrav), Hawthorn, Victoria 3122, Australia}
\date{\today}

\begin{abstract}
We present the results of directed searches for continuous gravitational waves from unknown neutron stars in five Milky Way globular clusters.
We carry out these searches in the LIGO data from the first eight months of the fourth LIGO-Virgo-KAGRA observing run using the \weave semicoherent program, which sums matched-filter detection-statistic values over many time segments spanning the observation period.
No gravitational wave signal is detected in the search band of \fmin--\fmax~Hz.
Injections of simulated continuous wave signals in the data indicate that we achieve the most sensitive results to date across most of the explored parameter space volume, obtaining median 95\% confidence level upper limits as low as $\sim 4.2 \times 10^{-26}$ near 282~Hz for NGC~6397. 
We also derive upper limits on neutron star ellipticity and $r$-mode amplitudes, reaching $\lesssim 10^{-5}$ and $\lesssim 10^{-3}$, respectively, at frequencies above 200 Hz. 

\end{abstract}

\maketitle

\section{Introduction} \label{sec:introduction}
Continuous gravitational waves (CWs) are faint, long-duration, quasimonochromatic signals that remain undetected despite extensive searches \citep[for recent reviews of searches, see][]{bib:TenorioKeitelSintesreview,bib:PiccinniReview,bib:RilesReview,bib:WetteReview}. 
Rapidly rotating, nonaxisymmetric neutron stars (NSs) in the Milky Way are promising sources of CWs \citep[for reviews of emission mechanisms, see][]{Lasky_review,GandG_review}. 
Young NSs are particularly attractive targets because they have had less time to anneal structural deformations. 
Furthermore, many known young pulsars exhibit high spin-down magnitudes, which are compatible with a detectable contribution from gravitational-wave (GW) energy loss~\citep{ATNF_2005}.

On the other hand, millisecond pulsars (MSPs) with high spin frequencies, thought to form via ``recycling'' of a NS's rotation by accretion from a binary companion, have much lower spin-down magnitudes, consistent with low asymmetry~\citep{Lorimer_2008}. 
Yet MSPs are attractive targets precisely because their high spins means that a low asymmetry can still produce detectable radiation.
In the following, we use ``MSP'' as a shorthand for a fast-spinning ($f_{\rm rot}\gtrsim100$ Hz) NS, whether or not the star emits pulsations and whether or not emitted pulses are beamed toward the Earth.

With these considerations in mind, we target NSs in globular clusters (GCs). 
The high stellar density in GC cores increases the likelihood of encounters with debris disks~\citep{Wang_2006}
or planets~\citep{Bailes_2011,Wolszczan1992,Spiewak2018,Behrens2020,Ni_u_2022}, triggering bombardment episodes. 
This scenario could lead to an old, annealed isolated NS acquiring a new nonaxisymmetry and hence the high spin-down rate characteristic of young stars. Another possibility is that successive close encounters lead to MSP formation in a newly formed accreting binary system, followed by binary disruption, leaving an isolated MSP~\citep{Dunn2025}.

In this work, we analyze the first eight months (O4a) of LIGO~\citep{LIGO, DetectorPaper, o4a_data, Capote_2025} data from the fourth LIGO-Virgo-KAGRA observing run to search for CW signals from unknown NSs in the central regions of the Terzan~10, NGC104, NGC~6397, NGC~6544, and NGC~6540 GCs.
Using a template-based semicoherent search based on the \(\mathcal{F}\)-statistic~\citep{JKS}, we probe a wide range of spin-down parameters, but find no evidence of an astrophysical signal. Hence, upper limits on strain amplitude and astrophysical constraints on fiducial ellipticity $\epsilon_0$ and $r$-mode amplitude $\alpha_0$ are set.

Our analysis achieves improved sensitivity compared to previous searches for NGC~6544~\citep{LVC2017GC} and the recent results for NGC~6397, NGC~6544, and NGC~6540~\citep{Dunn2025}.
We report the most stringent strain upper limits to date for these targets while covering a broader spin-down range. 
Furthermore, we present the first directed searches for Terzan~10 and NGC~104.
Our upper limits beat the age-based limit on GW strain amplitude across all frequencies $> 100$~Hz for sources younger than 50~kyr.
This limit is the indirect upper bound obtained by assuming that all rotational energy lost since birth is emitted as GWs.

The remainder of this article is organized as follows. Section~\ref{sec:data} details the data set used. Section~\ref{sec:targets} outlines the selection criteria for the GCs in this search. 
Section~\ref{sec:model} describes the signal model used. 
Section~\ref{sec:method} briefly describes the \Fstat, the semicoherent search method, and the \weave infrastructure. Section~\ref{sec:search} outlines the search strategy, including follow-up analysis using \weave. Section~\ref{sec:results} presents the results of the search, including the astrophysical constraints. 
Finally, Sec.~\ref{sec:conclusions} concludes with a discussion of the results and prospects for future searches.

\section{Dataset} \label{sec:data} 
\begin{figure*}[htb]
  \centering
  \includegraphics[width=1\textwidth]{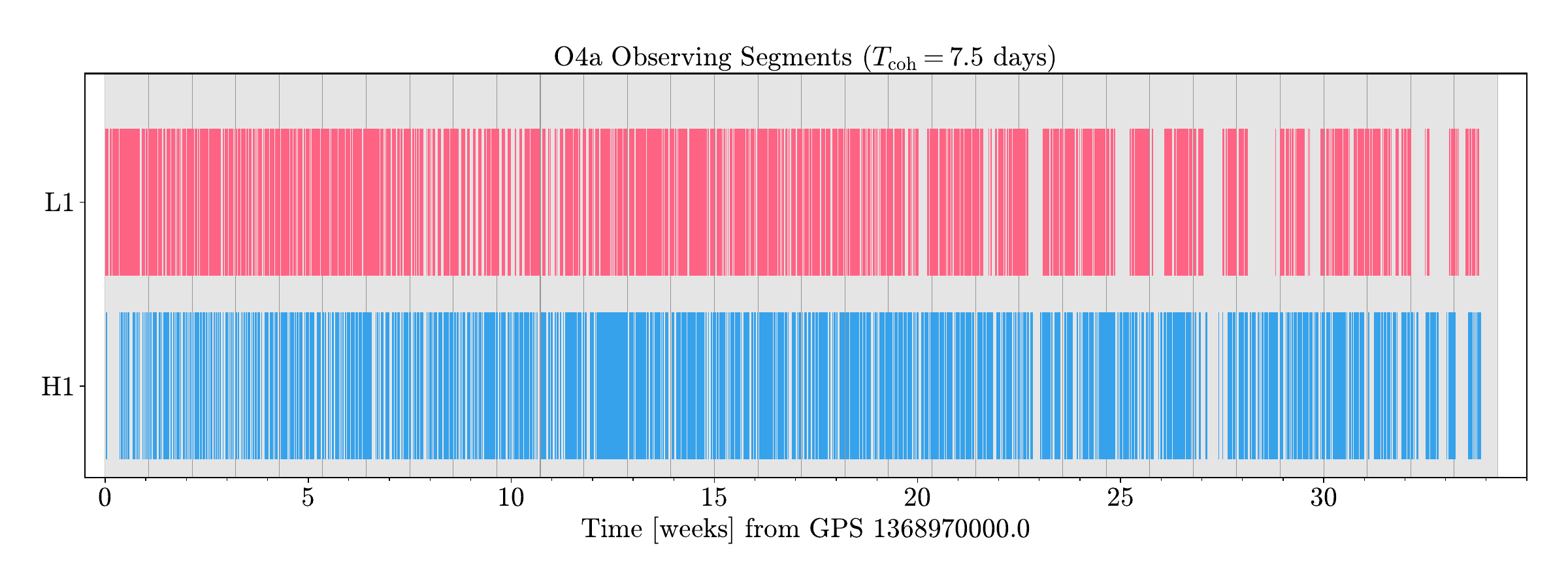}
  \caption{
  Data collected from the H1 and L1 detectors during the first eight months of the fourth observing run (O4a), spanning from May 24, 2023 (15:00:00 UTC) to January 16, 2024 (16:00:00 UTC). 
  The shaded segments represent the 7.5-day coherence segments used in the initial stage of the search (detailed in Sec.~\ref{sec:method}).
  }
  \label{fig:data segment}
\end{figure*}

We analyze data from the LIGO Livingston (L1) and LIGO Hanford (H1) detectors~\citep{LIGO2015} collected during the first eight months of the fourth Advanced LIGO and Virgo observing run (O4a).
It began May 24, 2023 (15:00:00 UTC) and ended January 16, 2024 (16:00:00 UTC)~\citep{o4a_data}.
Data were used only when the detectors were in science observing mode~\citep{science_mode_O4a}, corresponding to duty factors of 69.0\% for L1 and 67.5\% for H1. 
Figure~\ref{fig:data segment} illustrates the timeline of the data collected from both detectors over this period.
The Virgo detector~\citep{Virgo2014} was not included in this analysis as it joined the O4 run later, on April 10, 2024, while the KAGRA detector~\citep{kagra_2021} is scheduled to join by the end of the observing run. 
For a detailed description of the upgrades to the Advanced LIGO, Advanced Virgo, and KAGRA detectors in preparation for O4, we refer the reader to Appendix A of \citet{Abac_2024}.

Detector sensitivity improved significantly across all frequencies in this observing run compared to O3, particularly above 400\,Hz~\citep{Capote_2025, Jia_2024, Ganapathy_2023, Wade_2025}. 
The LIGO detectors are calibrated using photon radiation pressure actuation, in which an amplitude-modulated laser beam is directed onto the end test masses to induce a known change in arm length~\citep{photoncalibration,Karki_2016, Viets_2018}. 
The maximum systematic error in strain amplitude and phase calibration is estimated to be within 10\% and 10 degrees (68\% confidence interval), respectively, for both detectors over the entire frequency band analyzed in this search~\citep{O4cal}.

Prior to the search, the dataset underwent cleaning procedures to assess data quality and mitigate the effects of instrumental artifacts~\citep{soni_2025}. 
As in previous Advanced LIGO observing runs~\citep{linespaper}, instrumental ``lines'' (sharp peaks in fine-resolution, run-averaged H1 and L1 spectra) are marked, and where possible, their instrumental or environmental sources are identified~\citep{known_lines_O4a, science_mode_O4a}. 
The resulting database of artifacts was used to veto spurious signal candidates arising from the search; however, no frequency bands were vetoed \textit{a priori}. 
Consistent with the O3 run, the number of instrumental lines identified in H1 data for O4a is significantly larger than in L1.

Another class of artifacts observed in both detectors consists of frequent, loud ``glitches'' (short, high-amplitude instrumental transients), with the majority of their spectral power concentrated below $\sim$500\,Hz~\citep{O3aAllSky}. 
To mitigate their impact on CW searches, a glitch-gating algorithm was applied~\citep{o4gating} to excise these transients. 
For this analysis, we use the G02 version of gated 30-minute short Fourier transforms (SFTs)~\citep{sft_spec_2025}.

\section{Targets} \label{sec:targets}

In the frequency band where current ground-based detectors are most sensitive, the canonical sources of CWs are rapidly rotating, nonaxisymmetric NSs in the Milky Way. 
GCs are promising environments for such sources because of their dense stellar cores and high rates of dynamical interactions. 
These environments frequently give birth to low-mass x-ray binaries (LMXBs)~\citep{Katz_1975, Clark_1975, Pooley_2003}, which are progenitors of MSPs via accretion~\citep{msp_1982}. 
Additionally, close encounters in dense cluster cores can destabilize debris disks or planetary orbits around NSs, triggering bombardment episodes that funnel material into magnetic field-aligned ``mountains'' that emit CWs.
In this paper, we implicitly consider both scenarios; we seek ``young'' (postdisruption) stars at low frequencies and moderately fast-spinning MSPs that exist as isolated NSs following the disruption of a recycled binary system by a secondary encounter~\citep{Dunn2025}.

Because of limited computational resources, we cannot search for CW signals from all known GCs. 
Therefore, we employ specific figures of merit to rank clusters based on their likelihood of hosting detectable CW sources.
First, we consider the age-based upper limit on CW strain amplitude for a compact object. 
This limit is derived by assuming that the star's current rotation frequency is significantly lower than its birth frequency and that its spin-down history has been dominated by GW energy loss~\citep{cwcasamethod},
\begin{equation}
  h_{\rm age}  = 2.3\times10^{-24}\Bigg(\frac{1\>{\rm kpc}}{d}\Bigg)\Bigg(\dfrac{1000\;\mathrm{yr}}{\tau}\Bigg)^{1/2}\Bigg(\dfrac{I_{zz}}{I_0}\Bigg)^{1/2};
\label{eq:age-based limit}
\end{equation}
where $\tau$ is the source age, $d$ is the distance, and $I_{zz}$ denotes the principal moment of inertia about the rotation axis, with a fiducial value $I_0 = 10^{38}\text{ kg}\cdot\text{m}^{2}$.
To apply this relation to a cluster population, we use an effective age that scales inversely with the stellar encounter rate, representing the mean time since the last bombardment. 
This yields a figure of merit proportional to the number of potentially GW-emitting ``young'' NSs~\citep{LVC2017GC},
\begin{equation} \label{eq:encounter merit}
    \Gamma^{1/2}  d^{-1} = \rho_c ^{3/4} r_c d^{-1} ;
\end{equation}
where $\Gamma$ represents the stellar encounter rate, $\rho_c$ is the core density, and $r_c$ is the core radius.

We also consider the figure of merit introduced by~\citet{Dunn2025}, which accounts for the probability of binary disruption by replacing the age with the inverse of the binary encounter rate,
\begin{equation} \label{eq:binary encounter merit}
    \gamma^{1/2}  d^{-1} = \rho_c ^{1/4} r_c^{-1/2} d^{-1};
\end{equation}
where $\gamma$ denotes the binary encounter rate.
This metric is inversely proportional to the square root of a binary's lifetime before disruption by a secondary encounter. 
Consequently, a higher value implies a larger expected population of isolated MSPs. 
We ranked the GCs from the Harris catalog~\citep{Harris1996} using both Eq.~\eqref{eq:encounter merit} and Eq.~\eqref{eq:binary encounter merit}.

The top three GCs based on Eq.~\eqref{eq:encounter merit} are Terzan~10, NGC~6544, and NGC~104. 
Based on Eq.~\eqref{eq:binary encounter merit}, the top three candidates are NGC~6397, NGC~6544, and NGC~6540. 
Since NGC~6544 appears in the top tier for both metrics, we select a total of five GCs as targets for this search. 
Detailed properties of these targets, including sky locations, core radii, central luminosity densities, tidal radii, and distances, are provided in Table~\ref{tab:target param}.
Here, the core radius $r_c$ characterizes the size of the dense central region, defined as the distance from the center at which the apparent surface brightness falls to half its central value. 
The tidal radius $r_t$ defines the theoretical outer boundary of the cluster, beyond which stars are gravitationally stripped by the tidal field of the Milky Way.

\begin{table*}[htb]
\begin{center}
  \begin{tabular}{lccccccc}\hline
    \T\B                            & $\alpha$ (right ascension) & $\delta$ (declination) & $r_c$ [arcmin]  & $\log_{10} (\rho_c/L_\odot \mathrm{pc}^{-3}) $ & $r_t$ (arcmin) & $d$ [kpc] \\
    \hline\hline 
    \T\B Terzan~10 & $18^\mathrm{h}03^\mathrm{m}36.4^\mathrm{s}$ & $-26^\circ04^\prime21^{\prime\prime}$ & $0.90^\prime$ & 4.64 & $1.91^\prime$ & 5.8 \\
    \T\B NGC~104 (47 Tuc)& $00^\mathrm{h}24^\mathrm{m}05.67^\mathrm{s}$ & $-72^\circ04^\prime52.6^{\prime\prime}$ & $0.36^\prime$ & 4.88 & $2.85^\prime$ & 4.5 \\
    \hline\hline
    \T\B NGC~6397 & $17^\mathrm{h}40^\mathrm{m}42.09^\mathrm{s}$ & $-53^\circ40^\prime27.6^{\prime\prime}$ & $0.05^\prime$ & 5.76 & $0.61^\prime$ & 2.3 \\
    \T\B NGC~6544 & $18^\mathrm{h}07^\mathrm{m}20.58^\mathrm{s}$ & $-24^\circ59^\prime50.4^{\prime\prime}$ & $0.05^\prime$ & 6.06 & $0.26^\prime$ & 3.0 \\
    \T\B NGC~6540 & $18^\mathrm{h}06^\mathrm{m}08.60^\mathrm{s}$ & $-27^\circ45^\prime55.0^{\prime\prime}$ & $0.03^\prime$ & 5.85 & $0.37^\prime$ & 5.3 \\
\hline 
\end{tabular}
  \caption{Sky locations (right ascension $\alpha$ and declination $\delta$), core radius $r_c$, central luminosity density $\rho_c$, tidal radius $r_t$, and distance $d$ for each targeted globular cluster from the Harris catalog~\citep{Harris1996}.}
\label{tab:target param}
\end{center}
\end{table*}

\begin{figure*}[htb]
  \centering
  \includegraphics[width=\textwidth]{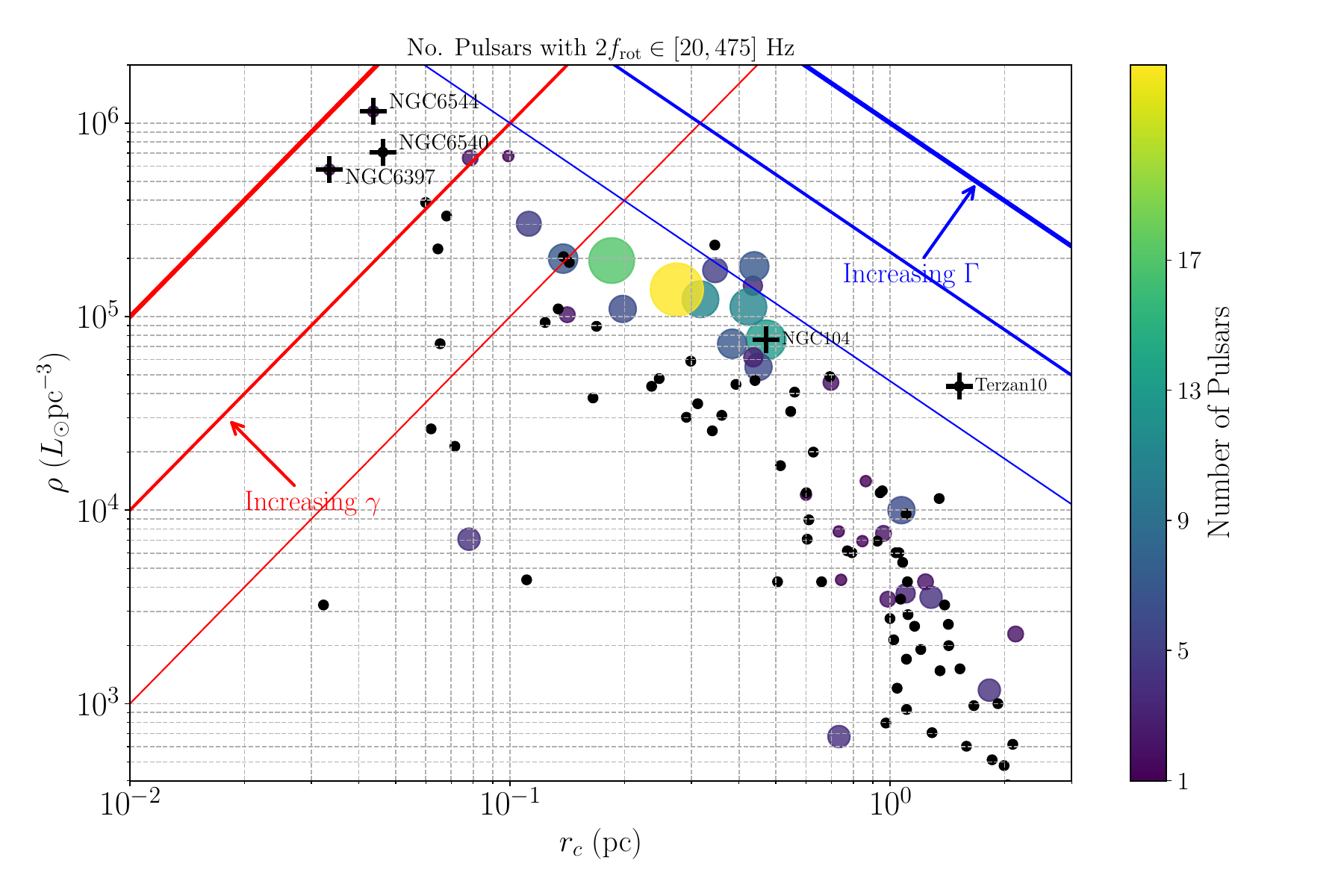}
  \caption{Central density as a function of core radius for Milky Way globular clusters from~\citet{Harris1996}. 
  The color of each point indicates the number of pulsars observed in that cluster.
  Blue lines indicate lines of constant stellar encounter rate $\Gamma$ [Eq.~\eqref{eq:encounter merit}], and red lines indicate constant binary encounter rate $\gamma$ [Eq.~\eqref{eq:binary encounter merit}]. The targets selected for this search are marked with black crosses. The figures of merit [Eqs.~\eqref{eq:encounter merit}--\eqref{eq:binary encounter merit}] used to prioritize target selection also consider the distance $d$ to each cluster, not shown in this diagram.}
  \label{fig:interaction rate}
\end{figure*}

Figure~\ref{fig:interaction rate} displays the central density versus core radius for Milky Way GCs, with data points colored according to the number of known pulsars (with $2f_\mathrm{rot} \in [20, 475]\,\mathrm{Hz}$) discovered in each cluster. 
The blue contours represent lines of constant stellar encounter rate $\Gamma$, while the red contours indicate lines of constant binary encounter rate $\gamma$. 
Clusters with large known pulsar populations generally lie in the upper-right region where $\Gamma$ is high, consistent with the success of $\Gamma$ in predicting pulsar abundance~\citep{Verbunt_2013}. Conversely, clusters with comparably high binary encounter rates do not currently show rich pulsar populations. 
This discrepancy may stem from observational biases or the simplifying assumptions inherent in the binary encounter rate model. 
Furthermore, the recent radio detection of a heavily obscured millisecond pulsar in NGC~6397 suggests that core-collapsed GCs may host a hidden population of extremely faint binary pulsars that evade traditional electromagnetic surveys~\citep{Zhang_2022}. This makes these dense cluster environments particularly compelling targets for GW searches, as these signals are negligibly affected by absorption or scattering.
Therefore, we target top candidates from both figures of merit to ensure comprehensive coverage in our search.

\section{Signal model} \label{sec:model}
We consider a rapidly rotating, nonaxisymmetric neutron star with a time-varying quadrupole moment. This source emits circularly polarized gravitational radiation along the rotation axis, linearly polarized radiation in directions perpendicular to the rotation axis, and elliptically polarized radiation in the general case.
The strain signal $h(t)$ measured by the detector is given by
\begin{align} \label{eq:signalmodel}
  h(t) = h_0 \bigg[ & F_+(t, \alpha, \delta, \psi)\frac{1+\cos^2\iota}{2}\cos\Phi(t) \nonumber \\
  & + F_\times(t, \alpha, \delta, \psi)\cos\iota\sin \Phi(t) \bigg];
\end{align}
where $h_0$ is the intrinsic strain amplitude and $\Phi(t)$ is the signal phase. 
The functions $F_+$ and $F_\times$ characterize the detector's response to ``$+$'' and ``$\times$'' polarizations, respectively,~\citep{S4allsky} determined by the source sky location (right ascension $\alpha$, declination $\delta$) and the polarization angle $\psi$. 
The inclination angle $\iota$ describes the orientation of the star's rotation axis relative to the line of sight.
The linear polarization case ($\iota=\pi/2$) is the most unfavorable, with the GW flux impinging on the detectors possessing eight times less incident strain power than for circularly polarized waves ($\iota = 0,\>\pi$) with the same intrinsic strain amplitude $h_0$.

The phase evolution of the signal is described in the Solar System barycenter (SSB) reference frame. 
We approximate the phase using a Taylor expansion in time,
\begin{equation} \label{eq:phase_evolution}
\Phi(t) = \phi_0 + 2\pi \sum_{k=0}^{s} \frac{f^{(k)}(t_0)}{(k+1)!} (t-t_0)^{k+1};
\end{equation}
where $f^{(k)}$ denotes the $k$th time derivative of the frequency at the reference time $t_0$ (taken as the midpoint of O4a; GPS 1379338000). 
In this work, we search up to the second order ($s=2$) of frequency derivatives; frequency $f$, spin-down $\dot{f}$, and the second derivative $\ddot{f}$.
When expressed as a function of the local time of ground-based detectors, Eq.~\eqref{eq:phase_evolution} acquires sky-position-dependent Doppler shift terms due to the relative motion between the source and the detector, as well as the relativistic Einstein and Shapiro time delays.

The relationship between the gravitational-wave frequency $f$ and the stellar rotation frequency $\frot$ depends on the specific emission mechanism. For example, mass quadrupoles emit at $f=2\frot$~\citep{ushomirsky2000deformations, Cutler, Lasky2013}, while $r$-modes emit at $f \approx 4\frot /3$~\citep{Andersson, OwenEtal, Idrisy_2015, caride2019search, Gittins_2023}.

\subsection{Parameter space} \label{sec:parameter space}

We search a GW frequency band from \fmin\ to \fmax\,Hz.
The lower bound is dictated by the rapid deterioration of detector sensitivity due to seismic noise below \fmin\,Hz.
The upper bound is chosen to avoid the forests of ``violin modes'' near $500$\,Hz (and near integer multiples thereof).
These frequencies correspond to vibration resonances of the 16 suspension fibers supporting the four primary mirrors in each interferometer.
While less severe resonances from suspension fibers supporting the beam splitter mirrors contaminate narrow bands between 300 and 400\,Hz, the region between \fmax\,Hz and the lowest violin mode harmonic (just below 500\,Hz) is particularly problematic due to nonlinear couplings that create frequency ``shoulders'' and sidebands around the violin modes.
Consequently, we set a ceiling of 475\,Hz. 
This range is astrophysically and computationally motivated; it targets the expected frequencies of young isolated neutron stars and the lower-frequency population of MSPs, while significantly reducing the computational cost associated with expanding the parameter space volume.

\begin{figure}[ht]
  \centering
  \includegraphics[width=0.45\textwidth]{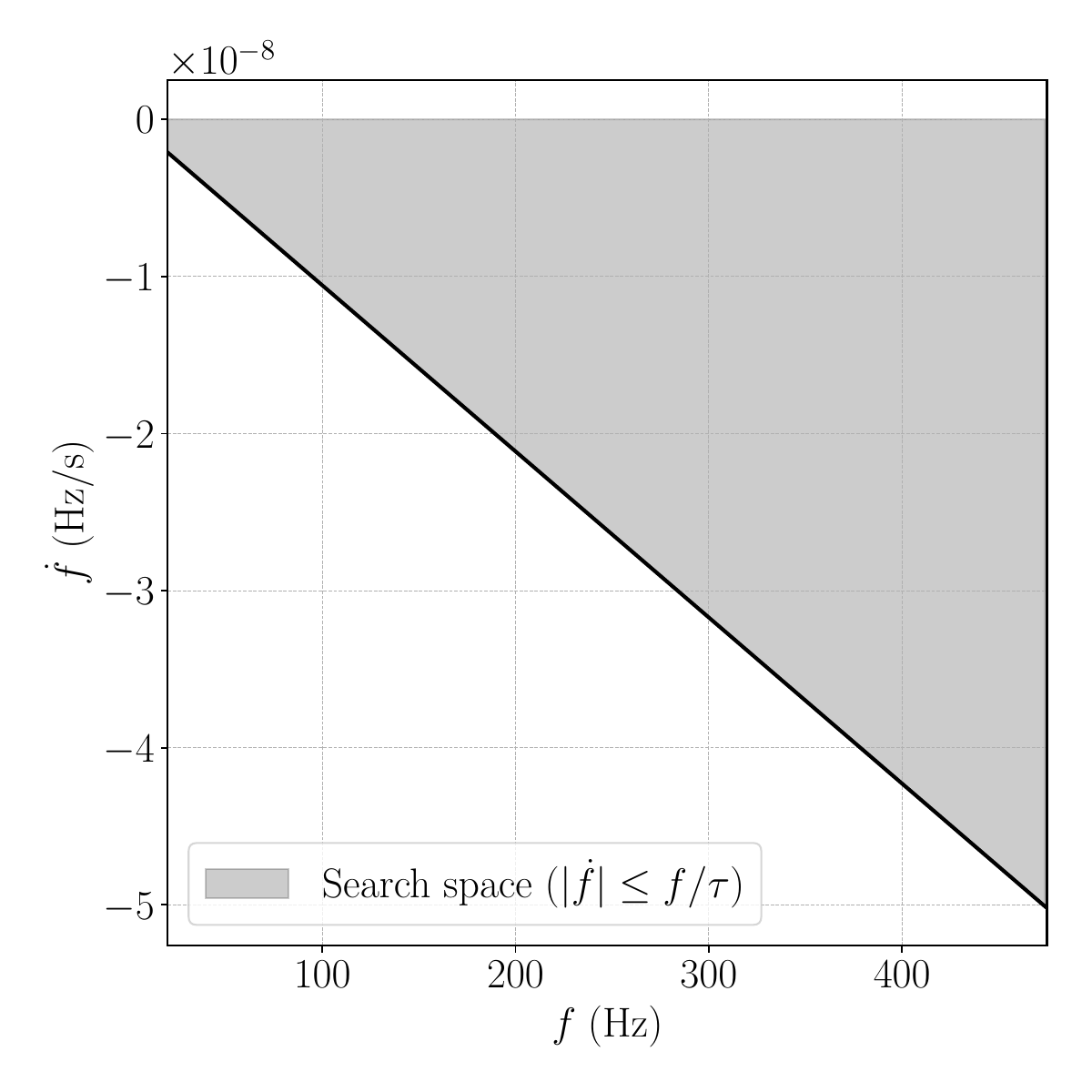}
  \caption{Parameter space for frequency $f$ and spin-down $\dot{f}$ covered in these searches (shaded region). The black solid line indicates the maximum spin-down defined by the minimum age $\tau$.
  }
  \label{fig:spin down limits}
\end{figure}

The search range for frequency derivatives is governed by the assumed minimum age $\tau$ of the source.
Following previous directed CW searches~\citep{abadie2010first, aasi2015searches, Abbott_2019, O3aSNR, O3aSNRweave, fullo3_cass, O4aSNR}, we define the bounds by assuming a power-law spin-down model, $\dot{f} \propto -f^n$, where the braking index $n$ ranges between 2 and 7. This covers multiple spin-down mechanisms, including magnetic dipole emission ($n=3$), GW quadrupole emission (gravitar scenario~\citealt{Palomba_2005}, $n=5$), and $r$-mode emission ($n=7$).
Adopting the conservative approach from~\citep{O3aSNRweave, fullo3_cass, O4aSNR}, we extend the upper limit on $\dot{f}$ to zero,
\begin{align}
\label{eq:weave_ranges}
  -\frac{f}{\tau} \leq \dot{f} \leq 0, \quad
  \frac{2\dot{f}^2}{f} \leq \ddot{f} \leq \frac{7\dot{f}^2}{f}.
\end{align}
Although Galactic and cluster acceleration, as well as the Shklovskii effect, can induce an apparent spin-up, the maximum observed spin-up in known GC pulsars from the ATNF catalog~\citep{ATNF_2005} is $\sim 7 \times 10^{-14}$~Hz/s.
The maximum observed spin-up is smaller than a single $\dot{f}$ grid step ($\sim 8 \times 10^{-14}$~Hz/s) in our template bank (see Sec.~\ref{sec:weave} for details on the template bank construction).
Therefore, capping the search limit for $\dot{f}$ at zero is sufficient to account for possible apparent spin-up. 
Figure~\ref{fig:spin down limits} illustrates the parameter space we covered for frequency and spin-down, while Table~\ref{tab:search range} lists the search ranges for $\dot{f}$ and $\ddot{f}$ at \fmin, 200, and \fmax~Hz.

\begin{table}[ht]
\begin{center}
  \begin{tabular}{lc}\hline
    \hline
    \T\B $\fdot$ range (Hz/s) @\fmin\,Hz & $[-\sci{2.1}{-9},\;0]$ \\
    \T\B $\fddot$ range (Hz/s$^2$) @\fmin\,Hz & $[0,\;\sci{1.6}{-18}]$ \\
    \hline
    \T\B $\fdot$ range (Hz/s) @200\,Hz & $[-\sci{2.1}{-8},\;0]$ \\
    \T\B $\fddot$ range (Hz/s$^2$) @200\,Hz & $[0,\;\sci{1.5}{-17}]$ \\
    \hline
    \T\B $\fdot$ range (Hz/s) @\fmax\,Hz & $[-\sci{5.0}{-8},\;0]$ \\
    \T\B $\fddot$ range (Hz/s$^2$) @\fmax\,Hz & $[0,\;\sci{3.7}{-17}]$ \\
\hline
\end{tabular}
  \caption{The search ranges for spin-down $\fdot$ and second order derivative $\fddot$ at \fmin, 200, and \fmax\,Hz.}
\label{tab:search range}
\end{center}
\end{table}

\section{Method} \label{sec:method}

This search relies upon a semicoherent averaging of the \Fstat~\citep{JKS,Cutler_2005} computed over segments spanning the observation period using the \weave\ infrastructure~\citep{WetteEtal}.
Here, we provide a brief summary of the detection statistic and search method used; we refer the interested reader to the cited references for full details.

\subsection{Detection statistic} \label{sec:fstat}

The search for CW signals buried in detector noise is formulated as a maximum likelihood detection problem.
We assume the detector data $x(t)$ consists of a signal $h(t)$ additive to zero-mean Gaussian noise $n(t)$,
\begin{equation}
    x(t) = n(t) + h(t).
\end{equation}
The log-likelihood ratio $\ln\Lambda$ deciding between the signal hypothesis and the noise hypothesis is given by
\begin{align}
    \ln\Lambda &= (x|h) - \frac{1}{2} (h|h).
\end{align}
Here, $(a|b)$ denotes the standard noise-weighted inner product (scalar product), defined as
\begin{equation}
    (a|b) = 4 \Re \int_{0}^{\infty} \frac{\tilde{a}(f) \tilde{b}^*(f)}{S_n(f)} \, df;
\end{equation}
where $\tilde{a}(f)$ is the Fourier transform of $a(t)$, where ``$^*$'' denotes complex conjugation, and where $S_n(f)$ is the one-sided power spectral density of the detector noise.

Maximizing this likelihood over the full parameter space is computationally prohibitive for long observation periods. However, the signal $h(t)$ depends linearly on four amplitude parameters $\mathcal{A}^\mu$ (functions of $h_0, \cos \iota, \psi, \phi_0$). Following \citet{JKS}, we decompose the signal into a linear combination of four basis functions $h_\mu(t)$,
\begin{align}
    h(t) = \sum_{\mu=1}^4 \mathcal{A}^\mu h_\mu(t);
\end{align}
where the basis functions $h_\mu(t)$ depend only on the phase evolution parameters $\vec{\lambda}=(\alpha, \delta, f, \dot{f}, \dots)$ and detector geometry.
Substituting this decomposition into the log-likelihood ratio allows us to rewrite it in vector notation,
\begin{align}
  \ln\Lambda &= \vec{\mathcal{A}}\cdot \vec{x} - \frac{1}{2} \vec{\mathcal{A}}\cdot \mathcal{M} \cdot \vec{\mathcal{A}}.
\end{align}
Here, $\vec{x}$ represents the projections of the data onto the basis functions, and $\mathcal{M}$ is the antenna pattern matrix,
\begin{align}
    x_\mu &= (x|h_\mu); \\
    \mathcal{M}_{\mu\nu} &= (h_\mu|h_\nu).
\end{align}
To analytically marginalize over the unknown amplitude parameters, we find the estimators $\vec{\mathcal{A}}_\mathrm{ML}$ that maximize the likelihood by solving $\partial (\ln \Lambda) / \partial \mathcal{A}^\mu = 0$. This yields $\mathcal{A}^\mu_\mathrm{ML} = (\mathcal{M}^{-1})^{\mu\nu} x_\nu$.
Substituting these estimators back into the likelihood equation yields the $\mathcal{F}$-statistic,
\begin{align} \label{eqn:fstat}
    2\mathcal{F} = 2\ln\Lambda(\vec{\mathcal{A}}_\mathrm{ML}) = \vec{x} \cdot \mathcal{M}^{-1}\cdot \vec{x}.
\end{align}
Using $2\mathcal{F}$ as the detection statistic significantly reduces the computational burden by eliminating the need to search explicitly over $h_0$, $\iota$, $\psi$, and $\phi_0$.

In Gaussian noise with no signal present, $2\mathcal{F}$ follows a central $\chi^2$ distribution with four degrees of freedom.
In the presence of a signal, it follows a noncentral $\chi^2$ distribution with a noncentrality parameter $\rho^2 \propto h_0^2 \Tcoh / S_n(f)$. This parameter encapsulates the signal strength relative to the noise, scaled by the coherence time $\Tcoh$ and modified by the source orientation and detector response.

\subsection{Semicoherent search}

Despite the efficiency of the analytical maximization, a fully coherent search over the entire observing run $\Tobs$ remains computationally intractable because the density of templates required to cover the parameter space scales with a high power of the observation time.
To address this, we employ a semicoherent approach. We partition the total observation time $\Tobs$ into $\Nseg$ segments of duration $\Tcoh$ and compute $2\mathcal{F}$ coherently for each segment. We then sum these values incoherently across segments while maintaining consistency in the frequency evolution.
Rather than a simple sum at fixed parameter-space points, this process involves mapping templates from a finely spaced semicoherent grid to the nearest templates in coarser, segment-specific grids (as detailed in Sec.~\ref{sec:weave}).

The semicoherent detection statistic, denoted $\FM$, is defined as the mean of the $2\mathcal{F}$-statistic values over the $\Nseg$ segments,
\begin{equation}
  \FM = \frac{1}{\Nseg}\sum_{i=1}^{\Nseg}\, 2\mathcal{F}_i.
\end{equation}
In the absence of a signal, the quantity $\Nseg\FM$ follows a central $\chi^2$ distribution with $4\Nseg$ degrees of freedom.
Consequently, $\FM$ has an expected value of 4 and a standard deviation of $\sqrt{8/\Nseg}$.
The presence of a signal induces an offset in the mean proportional to the noncentrality parameter averaged over the segments.

\subsection{The \weave\ infrastructure} \label{sec:weave}

To use the detection statistic to search for CW signals over a range of parameter space, a template bank is required.
The \weave\ software infrastructure provides a systematic approach to covering the parameter space volume in a templated search to ensure acceptable loss of signal-to-noise ratio (\SNR) for true signals lying between template points~\citep{WetteEtal}.
The \weave\ program combines recent developments in template placement to use an optimal parameter-space metric~\citep{WettePrix,WetteMetrics} and optimal template lattices~\citep{WetteLattice}.

In brief, a template grid in the parameter space is created for each time segment---a grid suitable for computing the $\F$ for a coherence time $\Tcoh = \Tobs / \Nseg$.
The actual fractional loss in the squared SNR, $\rho^2$, due to a true signal with parameters $(\vec{\mathcal{A}}, \vec\lambda^s)$ not exactly coinciding with a search template at $\vec\lambda^s + \Delta \vec \lambda$ is quantified by the actual mismatch $\mu$:
The spacing of the grid points for the phase evolution parameters $\Delta \vec \lambda$ is determined by the maximum mismatch parameter,
\begin{align}
    \mu &\equiv 1-\frac{\rho^2(\vec{\mathcal{A}}, \vec\lambda^s + \Delta \vec \lambda)}{\rho^2(\vec{\mathcal{A}},\vec \lambda^s)} \nonumber \\ 
    &= g_{ij}(\vec{\mathcal{A}},\vec \lambda^s)\Delta \lambda_i \Delta \lambda_j +\mathcal{O}(\Delta\lambda^3);
\end{align}
where $g_{ij}(\vec{\mathcal{A}},\vec \lambda^s)$ is the exact parameter-space metric.
A useful approximation for constructing template banks is the \textit{phase metric}~\cite{Brady_1998,Astone_2002,Prix_2006}, which neglects the amplitude modulation of the signal. 
This discards the dependence on $\vec{\mathcal{A}}$, retaining only the dependence on the phase evolution parameters $\vec{\lambda}$,
\begin{align} 
    g_{ij}(\vec \lambda^s) \equiv \left\langle \frac{\partial \Phi}{\partial \lambda_i}\frac{\partial \Phi}{\partial \lambda_j} \right\rangle - \left\langle \frac{\partial \Phi}{\partial \lambda_i} \right\rangle\left\langle \frac{\partial \Phi}{\partial \lambda_j} \right\rangle;
\end{align}
where the angle brackets denote an average over the observation time. The spacing of the grid points $\Delta \vec \lambda$ is determined by specifying a maximum mismatch parameter, $\mcoh$. 
The template bank is constructed by enforcing the condition $g_{ij}(\vec \lambda)\Delta \lambda_i \Delta \lambda_j \le \mcoh$, which serves as a design threshold guaranteeing that the actual mismatch between any potential signal and its nearest template does not exceed $\mcoh$.

Separately, a much finer grid is defined for the full observation period with respect to the reference time $t_0$. This grid uses a semicoherent mismatch parameter $\msemicoh$, analogous to $\mcoh$, but defined as the average of the coherent mismatch values over all segments.
The choice of both $\mcoh$ and $\msemicoh$ is an empirical trade-off between sensitivity and computational cost.
During initialization, the \weave\ package creates a mapping between each point in the fine semicoherent template grid and the nearest corresponding point in each of the coarser segment grids, accounting for frequency evolution.
The semicoherent detection statistic $\FM$ is then constructed for each semicoherent template using this mapping.

\section{Search configuration} \label{sec:search}

In this section, we describe how to set up and conduct the search, as well as how to follow up candidates for signal validation using the \weave program.

\subsection{Initial stage} \label{sec:initial stage}
We adopt the same mismatch parameters, $\mcoh=0.1$ and $\msemicoh=0.2$, as used in \cite{O3aSNR, O4aSNR} for \weave.
The initial search stage uses a coherence time of $\Tcoh = 7.5$ days, targeting the center of the cluster using a single sky-location template. 
While a grid of sky templates would be required to maintain uniform sensitivity across the entire parameter space, we restricted the initial search to a single central template due to computational costs. 
For NGC~6397, NGC~6544, and NGC~6540, we target CWs from NSs within the tidal radius ($r_t$).
However, for Terzan~10 and NGC~104, the tidal radii are too large to be covered effectively by a single sky template.
We therefore restrict the target region for these two clusters to the core radius ($r_c$), where the sky area is smaller, and the sensitivity loss due to the offset between the source and the central search template is reduced. 
The resulting sensitivity across the target region is shown through dedicated software injections, as described in Sec.~\ref{sec:upper limits}.

Search jobs are carried out in 0.1 Hz bands of $f$, with further divisions in $\fdot, \fddot$ to keep each job's computational duration less than 24 hours, for practical reasons.
Table~\ref{tab:configparameters} summarizes the \weave\ mismatch configurations, the coherence time setup, and the parameter space partitions for individual jobs in the initial search stage.

\begin{table}[htb]
\begin{center}
  \begin{tabular}{lc}
    \hline\hline
    \T\B Coherent mismatch $\mcoh$        &  0.1\\

    \T\B Semicoherent mismatch $\msemicoh$  &  0.2\\

    \T\B Coherence time $\Tcoh$    &  7.5 days\\

    \T\B Number of segments $\Nseg$       & 32\\
    
    \hline
    
    \T\B $\fdot$ width     &  $2\times 10^{-9}$ Hz/s\\

    \T\B $\fddot$ width       &  $5\times 10^{-18}$ Hz/s$^2$\\
    
    \hline
\end{tabular}
  \caption{Summary of the \weave\ mismatch parameter configurations, along with the coherence time, number of segments, and parameter space partitions ($\fdot$ and $\fddot$ widths) for individual jobs in the initial search stage.}
  \label{tab:configparameters}
\end{center}
\end{table}

Each individual job returns the \paramset\ values of the 1000 templates (``top-list'') with the largest (``loudest'') $\FM$ values.

To select the candidates from the initial search result, we sum up the number of templates for each 1-Hz band and calculate the nominal threshold,
\begin{equation} \label{eq:weave threshold}
\FMthresh = \frac{1}{N_\mathrm{seg}} \cdot \mathrm{CDF}^{-1}_{\chi^2}\bigg(1 - \frac{1}{N_\mathrm{temp}}, 4N_\mathrm{seg}\bigg);
\end{equation}
where $\mathrm{CDF}^{-1}_{\chi^2}$ is the inverse cumulative distribution function (CDF) of the $\chi^2$ distribution with $4N_\mathrm{seg}$ degrees of freedom, evaluated at $1 - 1/N_\mathrm{temp}$ percentile, and $N_\mathrm{temp}$ is the number of templates in a 1-Hz band.
This threshold is set based on a signal-free $\chi^2$ distribution such that the expected number of candidates is one per 1 Hz band, assuming Gaussian noise (In practice, non-Gaussian artifacts lead to much higher outlier counts).
Templates with $\FM$ values exceeding this threshold are considered candidates.
In some cases, strong instrumental lines can lead to more than 1000
templates from a single job that exceed the threshold.
We refer to those cases as ``saturated'' since potentially interesting templates may be suppressed by the top-list cap.

\subsection{Follow-up stage} \label{sec:followup stage}

For nonsaturated jobs, outliers exceeding the threshold $\FMthresh(f)$ are followed up in a sequential procedure.
The outliers are first clustered by grouping those within $3 \Delta f^{(k)}$, where $\Delta f^{(k)}$ represents the average template spacing for the $k$th spin-down order.
We retain only the loudest outlier in each cluster as the seed for the next stage.
In each subsequent follow-up step, the coherence time $\Tcoh$ is doubled (and hence the number of segments $\Nseg$ is halved).
To better capture the signal, spin-down terms up to order $s$ are included in the phase model $\Phi(t)$, depending on the coherence time configured for each stage (see Table~\ref{tab:follow-up config}).
Because the noncentrality parameter for the $\FM$ detection statistic scales approximately linearly with $\Tcoh$, one expects a nominal doubling of the $(\FM-4)$ value.

To determine the expected increase in the $\FM$ for true signals, we perform 50 software injections per 1 Hz band.
Signals are injected with random frequency parameters and sky positions drawn from the target's central region.
Each injection uses a strain amplitude equal to the upper limit $\hul$ corresponding to 95\% detection efficiency (see Sec.~\ref{sec:upper limits}).
Using these injections, we determine the threshold $\FM$ values required to recover the signals.
The resulting required fractional increases in $\FM$ to recover injected signals above the threshold are listed in Table~\ref{tab:mean2F fractional increase requirement}.

\begin{table}[htb]
\begin{center}
\begin{tabular}{lcc}
\hline
\T\B  & $\Tcoh$  & Spin-down order $s$ \\
\hline\hline
\T\B Initial search   & $7.5\;\mathrm{days}$ & 2 \\
\hline
\T\B First follow-up   & 15 & 2 \\
\hline
\T\B Second follow-up    & 30 & 2 \\
\hline
\T\B Third follow-up  & 60 & 3 \\
\hline
\T\B Fourth follow-up   & 120 & 4 \\
\hline
\end{tabular}
\caption{Coherence time $\Tcoh$ and spin-down order $s$ configuration for the initial search stage and subsequent follow-up stages.}
\label{tab:follow-up config}
\end{center}
\end{table}

\begin{table}[tbh]
\begin{center}
\begin{tabular}{ccccc}\hline
  & First follow-up & Second & Third & Fourth \\
  & (\%) & (\%) & (\%) & (\%) \\
\hline\hline
\T\B Terzan~10 & 46 & 43 & 62 & 55 \\
\T\B NGC~104 & 35 & 41 & 60 & 62 \\
\hline\hline
\T\B NGC~6397 & 35 & 42 & 58 & 57 \\
\T\B NGC~6544 & 35 & 42 & 60 & 65 \\
\T\B NGC~6540 & 34 & 41 & 58 & 62 \\
\hline 
\end{tabular}
  \caption{Required fractional increases in $\FM$ across four follow-up stages for the five targeted globular clusters.}
  \label{tab:mean2F fractional increase requirement}
\end{center}
\end{table}

For targets other than Terzan~10, we use a single sky template (fixed at the cluster center) for both the initial search and all subsequent follow-up stages. 
As the coherence time increases in the follow-up stages, the template grid becomes finer, and signals originating farther from the center experience a greater fractional loss in $\FM$ due to mismatch. 
We account for this by adjusting our detection criteria rather than adding more sky points. 
The required fractional increase in $\FM$ is empirically determined via software injections; signals are injected with random source positions within the radius we aim to cover but are recovered using only the single central sky search template. 
Consequently, the injection analysis yields lower, less stringent thresholds for targets with larger radii to maintain the targeted signal recovery efficiency. 
For example, as shown in Table~\ref{tab:mean2F fractional increase requirement}, we consider an NGC~6397 outlier with a $\FM$ fractional increase above 57\% in the fourth stage to be a surviving candidate. This threshold is lower than the 62\% required for NGC~6540 and NGC~104, and the 65\% for NGC~6544, consistent with their respective radii (where NGC~6397 $>$ NGC~6540 $\sim$ NGC~104 $>$ NGC~6544).

For Terzan~10, the initial search is performed using the single central sky template, just as with the other targets. 
This single template is capable of detecting candidates from the wider region, albeit with reduced efficiency for off-center sources.
Because Terzan~10 has the largest angular radius we aim to cover, continuing to use a single sky template for follow-up would result in such severe signal loss that a true signal's $\FM$ increase would be indistinguishable from a noise-induced outlier. 
Therefore, for every Terzan~10 outlier above the initial $\FMthresh(f)$ threshold, we generate nine follow-up jobs for the first stage instead of one. 
These copies cover the same frequency search range but use distinct sky positions: one fixed at the cluster center and eight surrounding points positioned roughly at the midpoint between the center and the edge of the target area. This nine-point grid ensures that at least one template is close enough to an off-center source to effectively recover it. 
We calculate the required detection ratio using the loudest $\FM$ from these nine jobs relative to the initial seed. The search sky position is then fixed to the position of this loudest job for all subsequent follow-up stages. This strategy allows us to significantly reduce the offset between the source position and the search direction at an early stage, before the coherence time becomes too long.

At later follow-up stages, we retain only the top 10 loudest survivors for each seed from previous stage.
Thus, while all outliers above the $\FMthresh(f)$ threshold are evaluated in the initial search stage, only the top 10 loudest survivors are followed up in successive stages until the coherence time $\Tcoh$ reaches 120 days.
We have confirmed that these threshold criteria achieve over 99\% recovery efficiency for the injected signals, with source right ascension and declination randomly drawn within the tidal/core radius region of the GCs.
This corresponds to a strict false-dismissal rate of $\lesssim 1\%$ introduced by the multistage follow-up pipeline.

The loudest outlier in each saturated sub-range is followed up using the same procedure.
This ensures that an unusually strong signal causing the saturation is not discarded due to the top-list cap.

\section{Results} \label{sec:results}

\subsection{Candidates} \label{sec:candidates}
We find no significant evidence for CW signals from any of the five targeted GCs.
All the candidates were discarded after the follow-up procedures and the final veto procedure described in this section.
We searched the GCs Terzan~10, NGC~6544, NGC~104, NGC~6397 and NGC~6540 using O4a data.
The initial search yielded approximately $3.4$, $5.2$, $5.3$, $3.2$, and $3.7\times 10^{5}$ outliers above the $\FM$ threshold, respectively, in bands not excluded by severe instrumental artifacts.
These outliers were then subjected to the clustering and hierarchical follow-up procedure described in Sec.~\ref{sec:followup stage}.
The number of clustered outliers at each stage is presented in Table~\ref{tab:no of outlier clusters}. 
The candidates surviving the follow-up stage 4 ($\Tcoh=120$ days) have $\FM \gtrsim 43$. For Terzan~10, which had no survivors in stage 4, candidates passing the stage 3 ($\Tcoh=60$ days) required $\FM \gtrsim 31$.
Parameters of the surviving outliers after the last follow-up stage are listed in Appendix~\ref{appendix: candidate}. 

\begin{table}[tbh]
\begin{center}
  \begin{tabular}{cccccc}\hline
    & Initial search stage & \multicolumn{4}{c}{Follow-up stages} \\
  & Stage 0 & First & Second & Third & Fourth \\
\hline\hline
Terzan~10 & $1.8\times10^{5}$ & $3.0\times10^{3}$ & 362 & 28  & 0 \\
NGC~104 & $ 2.3\times10^{5}$ & $1.2\times10^{4}$ & 2,922 & 231 & 18 \\
\hline\hline
NGC~6397 & $ 2.6\times10^{5}$ & 9,864 & 2,644 & 1,492 & 72 \\
NGC~6544 & $ 1.7\times10^{5}$ & 5,697 & 1,301 & 119 & 10 \\
NGC~6540 & $ 1.9\times10^{5}$ & 6,970 & 1,509 & 1,106 & 87 \\
\hline 
\end{tabular}
  \caption{Number of outlier clusters surviving each follow-up stage for the five targeted globular clusters.}
  \label{tab:no of outlier clusters}
\end{center}
\end{table}

\begin{figure}[h!]
  \centering
  \includegraphics[width=0.5\textwidth]{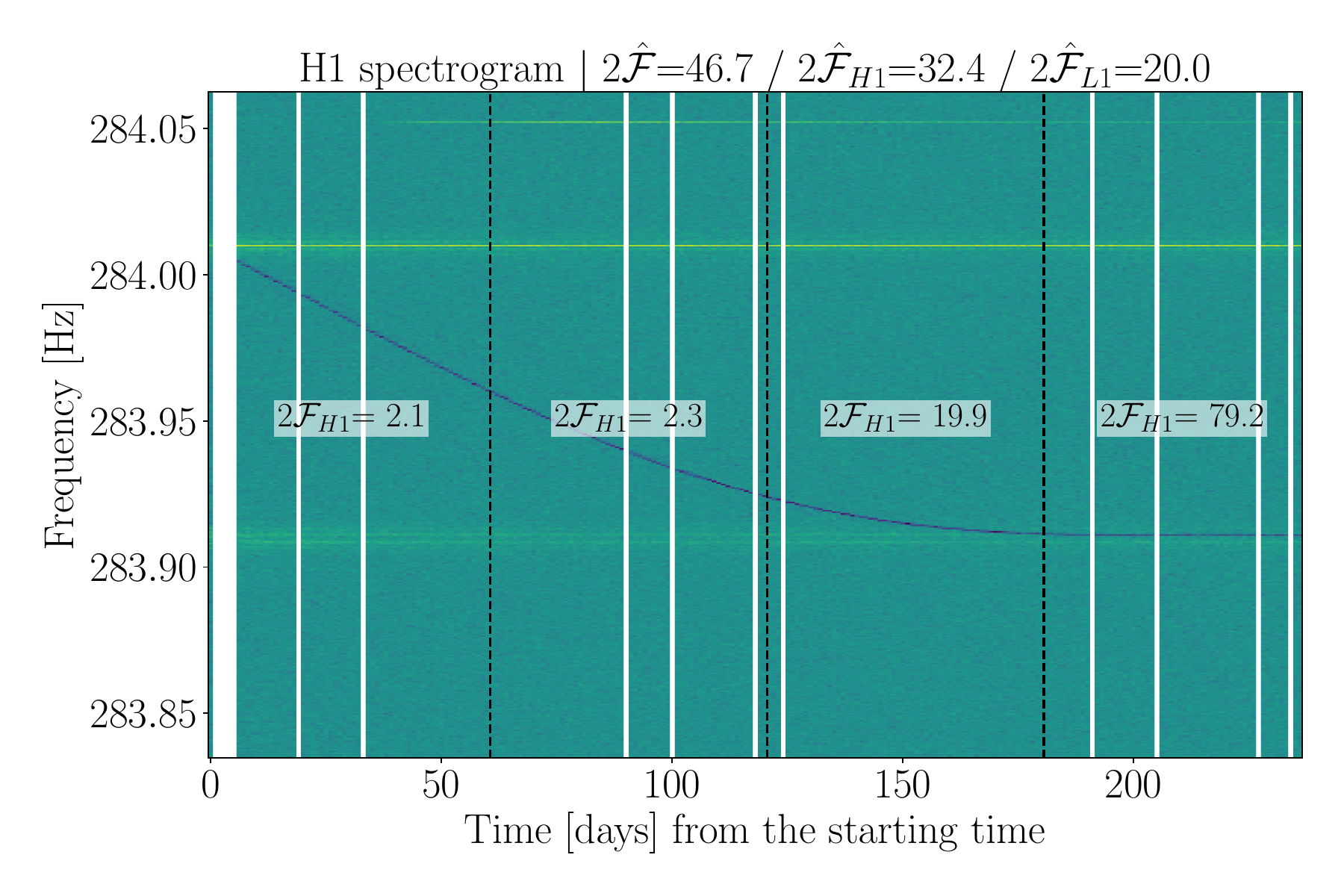}
  \caption{
  Spectrogram of O4a H1 data near 284~Hz, generated by taking the daily sum of the cumulative averaged power across all 1800-second SFTs within each day. The dark track represents the frequency evolution of a simulated signal injected with the same parameters as the loudest NGC~6397 outlier found within the 283--284~Hz band at the final $\Tcoh=120$ days follow-up stage. Bright horizontal bands represent instrumental line artifacts. 
  The artifact at 283.91 Hz intersects the simulated signal track during the final 60 days. The dashed vertical black lines denote the boundaries of four consecutive 60-day time segments. The text markers displayed within each segment indicate the single-detector statistic $2\mathcal{F}_\mathrm{H1}$ evaluated over that specific 60-day period.
  }
  \label{fig:spectrogram284Hz}
\end{figure}

To assess the credibility of these remaining outliers at $\Tcoh=120$ days as coming from an astrophysical source or from instrumental contamination, we construct spectrograms using O4a data overlaid with strong simulated signals with the same frequency parameters as the loudest outlier in each cluster.
The spectrogram is plotted using the cumulative averaged power, which is then summed daily across all 1800-second SFTs within each day of the O4a observation period.

For each surviving candidate, we visually inspect the spectrogram alongside a consistency check of the $\mathcal{F}$-statistic values. 
Specifically, we partition the observation period into four 60-day segments and compute the segment-wise coherent single-detector statistics $\F_\mathrm{H1}$ and $\F_\mathrm{L1}$.
A true astrophysical signal should accumulate significance consistently across the observation period and between detectors.
Based on the segmentwise $\F_\mathrm{H1}$ and $\F_\mathrm{L1}$, we can determine whether an artificially high $\FM$ value is caused by an intersecting instrumental line, which manifests as highly asymmetric single-detector statistics or sudden spikes in localized time segments.
We veto candidates for which there is a gross discrepancy in the $\F_\mathrm{H1}$ or $\F_\mathrm{L1}$ value of a single segment with respect to what is expected from other segments for the same or the other detector.

Figure~\ref{fig:spectrogram284Hz} shows an example narrowband H1 spectrogram used to veto the loudest outlier within the 283--284~Hz band for NGC~6397. 
In the figure, prominent instrumental line artifacts (bright horizontal bands) are visible.
The artifact near 283.91 Hz overlaps with the simulated signal (dark curve) during approximately the last 60-day segment of the O4a observation period---a period when the signal template's Doppler modulation is nearly stationary.
The simulated signal is based on the loudest real outlier's parameters, which has $\FM=46.7$ at $\Tcoh=120$ days, but is inconsistent between detectors ($2\hat{\mathcal{F}}_\mathrm{H1}=32.4$ versus $2\hat{\mathcal{F}}_\mathrm{L1}=20.0$).
Furthermore, the $\F_\mathrm{H1}$ values evaluated for the four 60-day segments, displayed between the vertical dashed lines in the spectrogram, reveal a significant bias toward the final segment, confirming that the $\FM$ is predominantly driven by the instrumental line artifact in H1 data intersecting the simulated signal track during that specific period. 
Hence, we do not consider it to be a credible astrophysical signal.

After applying this combined spectrogram and segment-consistency veto procedure to all outliers surviving the final stage of follow-up, no candidates remained for any of the five targets.
We conclude that there is no significant evidence in this analysis for a CW signal from the compact objects in the central region of the targeted GCs.

\subsection{Upper limits} \label{sec:upper limits}
Given the absence of a detection, we determine 95\% efficiency upper limits on strain amplitude $\hul$ for each 1 Hz band, excluding saturated 0.1 Hz sub-bands.

To obtain $\hul$, we conduct software injections of signals with varying amplitudes, drawing frequencies uniformly within each 1-Hz band (excluding saturated bands). 
The parameters $\cos \iota$, $\psi$, $\fdot$, and $\fddot$ are also drawn uniformly, while the sky locations $(\alpha, \delta)$ are drawn isotropically from the center region of the GC within the tidal/core radius.
An injection is deemed detectable if its $\FM \geq \FMthresh(f)$ with the initial search setup (assuming all outliers above this threshold have been followed up and excluded). 
This procedure is repeated for different signal amplitudes $ h_0 $, and the detection fraction is fitted using a sigmoid function,
\begin{equation} \label{eq:sigmoid}
  p(h_0) = \frac{1}{1 + \exp\left(\frac{a - h_0}{b}\right)};
\end{equation}
where $ p(h_0) $ represents the fraction of detected injections for a given $ h_0 $, and $ a $ and $ b $ are the parameters of the sigmoid function. 
The number of injections is chosen to ensure that the statistical uncertainty in the estimated $ h_0^{\rm 95\%} $ is below 5\%.

The quoted $h_0^{95\%}$ values represent the strain amplitude at which a signal has a 95\% probability of passing the initial stage threshold. As detailed in Sec.~\ref{sec:followup stage}, the subsequent multistage follow-up introduces a false-dismissal rate of $\lesssim 1\%$. Consequently, the true end-to-end detection efficiency for a signal at the quoted $h_0^{95\%}$ amplitude is approximately 94\%. Achieving a strictly 95\% end-to-end recovery through the entire pipeline would require an amplitude only negligibly higher than the quoted limits.

\begin{figure*}[htb!] 
	\centering
	\subfigure[][Terzan~10, NGC~6544, and NGC~104]
	{
		\scalebox{0.4}{\includegraphics{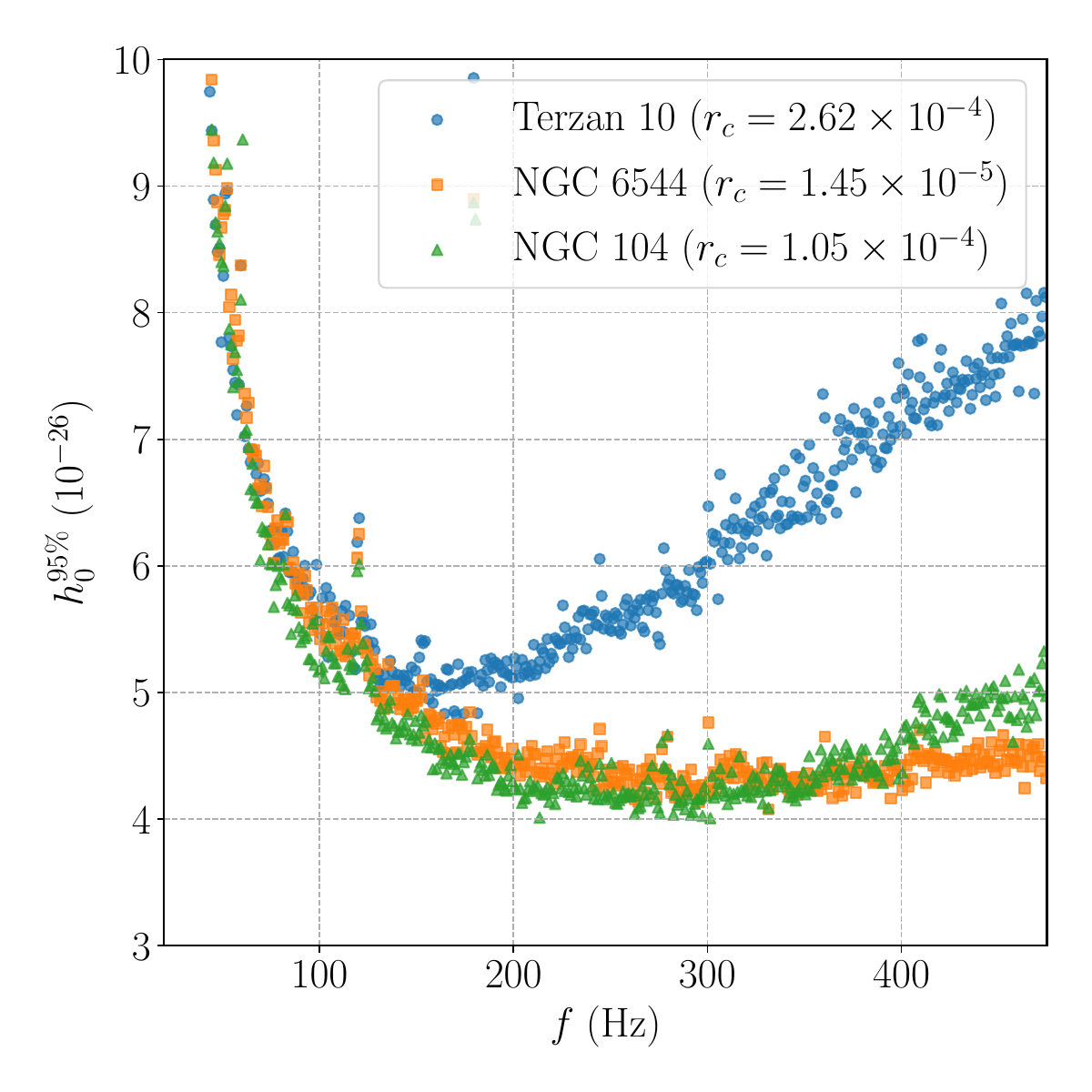}}
        \label{fig:upper limits rc}
	}
	\subfigure[][NGC~6397, NGC~6544, and NGC~6540]
	{
		\scalebox{0.4}{\includegraphics{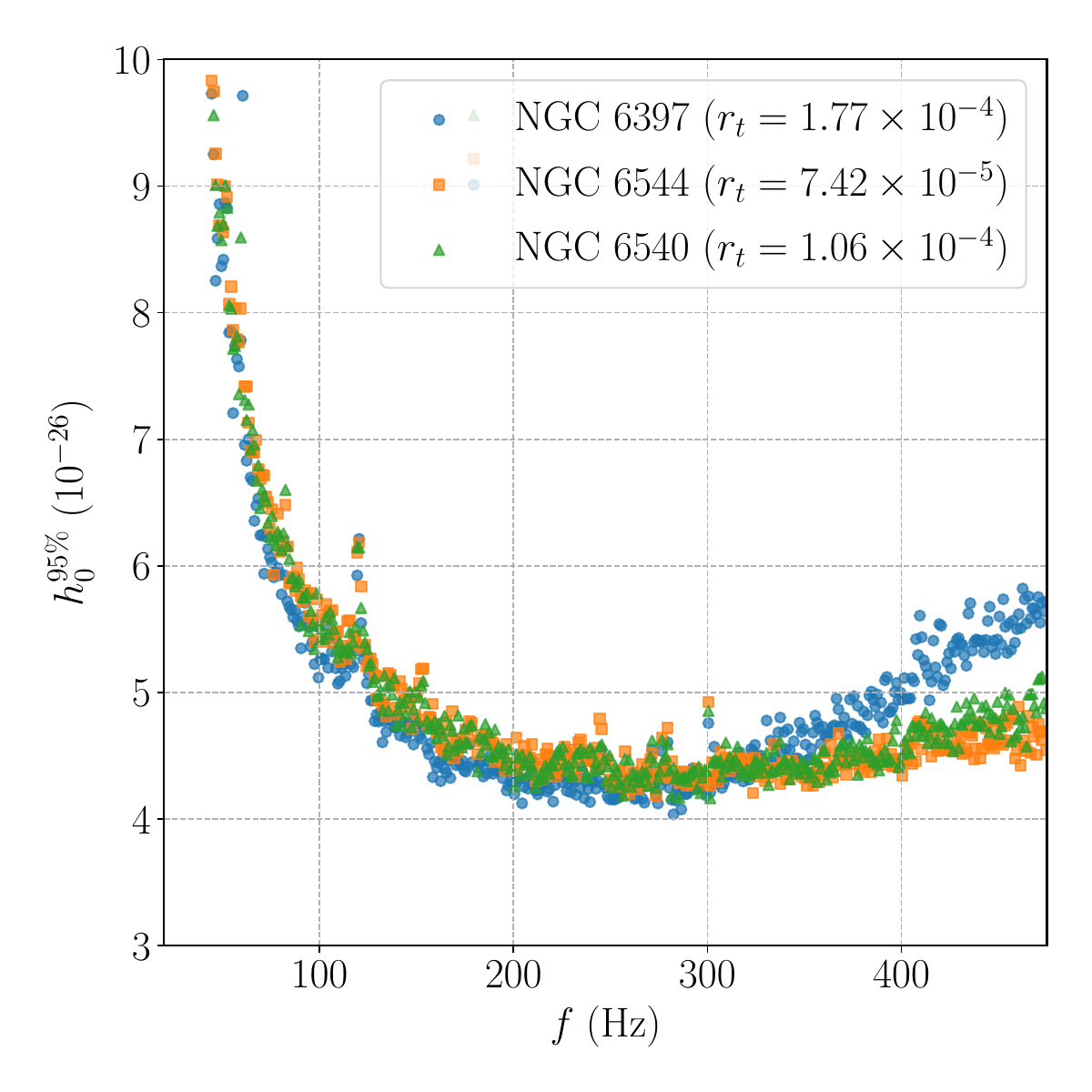}}
        \label{fig:upper limits rt}
	}
    \caption{Gravitational wave strain amplitude upper limits (95\% efficiency) in each 1-Hz band. 
	(a) Results for Terzan~10 (blue circle), NGC~6544 (orange square), and NGC~104 (green triangle) assuming signals originate from a central region with radius equal to the core radius $r_c$. 
	(b) Results for NGC~6397 (blue circle), NGC~6544 (orange square), and NGC~6540 (green triangle) assuming signals originate from within the tidal radius $r_t$.
    The degradation of the Terzan 10 upper limits becomes more severe at high frequencies due to the size of the cluster and the use of a single sky position in the initial stage; see text for details.
}
    \label{fig:upper limits}
\end{figure*}

\begin{figure}[htb]
  \centering
  \includegraphics[width=0.5\textwidth]{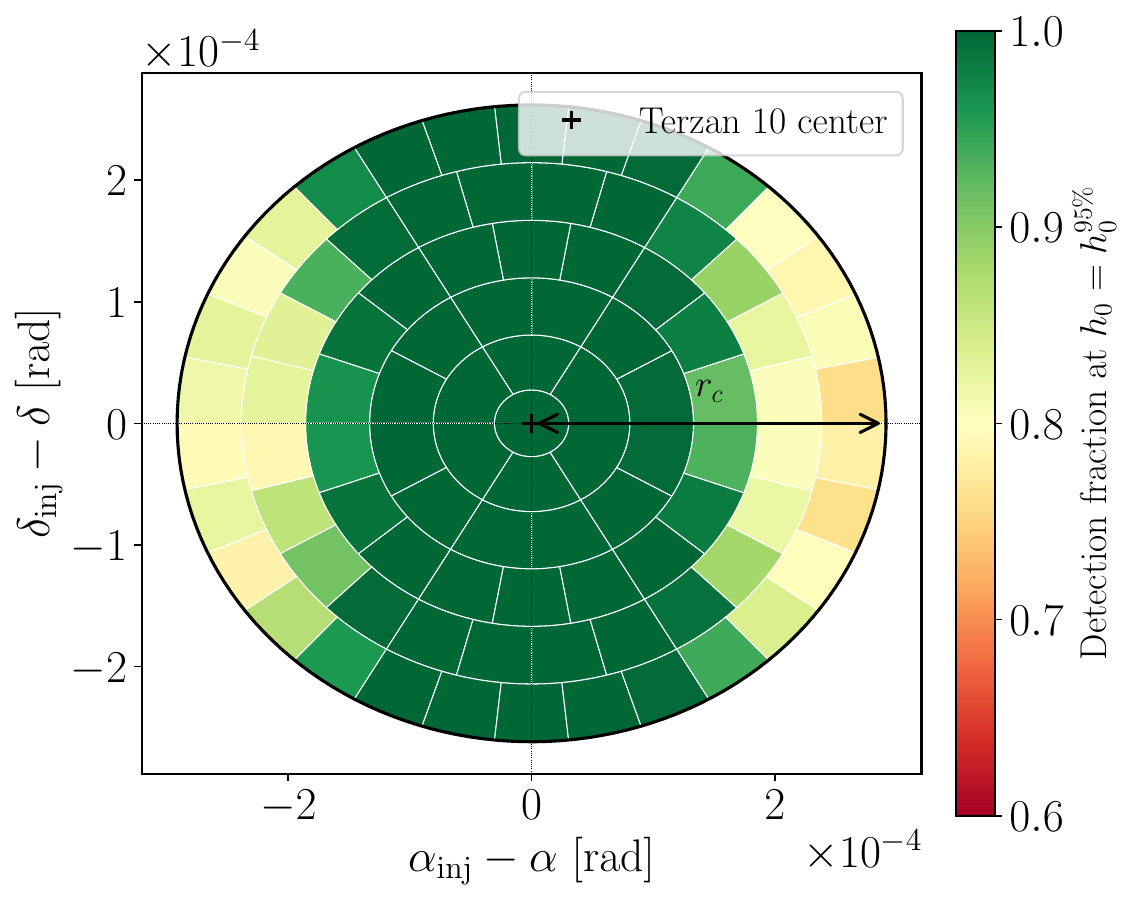}
  \caption{Measured detection efficiency for Terzan~10 at 474~Hz using the initial search configuration with a single sky template fixed at the cluster center $(\alpha, \delta)$. 
  The sky is divided into spatial bins, each containing 500 signal injections with randomly sampled sky coordinates $(\alpha_{\mathrm{inj}}, \delta_{\mathrm{inj}})$ and strain amplitude $h_0=h_0^{95\%}$.
  The color indicates the fraction recovered above the initial-stage detection threshold. 
  Recovery is approximately 99--100\% near the cluster center and decreases gradually toward the edge of the core-radius region. The reduction is anisotropic, remaining close to 99\% along the declination direction while falling to approximately 80\% along right ascension at the largest offsets.}
  \label{fig:Terzan10 sky coverage}
\end{figure}

Figure~\ref{fig:upper limits} presents the 95\% confidence level upper limits on the GW strain amplitude $h_0$ as a function of frequency. 
Panel~(a) displays the results for Terzan~10, NGC~6544, and NGC~104, assuming signals originate from the central region defined by the core radius, $r_c$. 
Panel~(b) illustrates the upper limits for NGC~6397, NGC~6544, and NGC~6540, assuming a source distribution extending out to the tidal radius, $r_t$. 
These $h_0$ values represent population-averaged upper limits over the target's spatial distribution, as the simulated signals were drawn uniformly across the core or tidal regions.
Note that the upper limits derived for Terzan~10 degrade more significantly at high frequencies. 
This degradation arises because the signal population is distributed over a larger sky area, while we use a single sky template fixed at the cluster center for the initial search stage. 
The components of the metric associated with sky position scale with the square of the frequency ($g_{\alpha\alpha}, g_{\delta\delta} \propto f^2$). Consequently, the mismatch due to spatial offsets is more severe at higher frequencies, causing the effective coverage of a single template to shrink. 
To demonstrate the actual sky coverage, we performed software injections with $h_0=h_0^{95\%}$ in the 474--475~Hz band.
Signals were injected with sky positions $(\alpha_{\mathrm{inj}}, \delta_{\mathrm{inj}})$ randomly sampled across the entire core-radius region and searched using only the single central sky template. 
Figure~\ref{fig:Terzan10 sky coverage} shows the measured recovery fraction as a function of sky-position offset from the cluster center.
The overall recovery efficiency across the full core-radius region is approximately 95\%, exhibiting an anisotropic dependence on sky location.
Specifically, the recovery remains at 99--100\% near the cluster center and along the declination direction out to the core radius, but decreases to approximately 80\% along right ascension at the largest offsets.

\begin{figure}[htp] 
  \centering
  \includegraphics[width=0.5\textwidth]{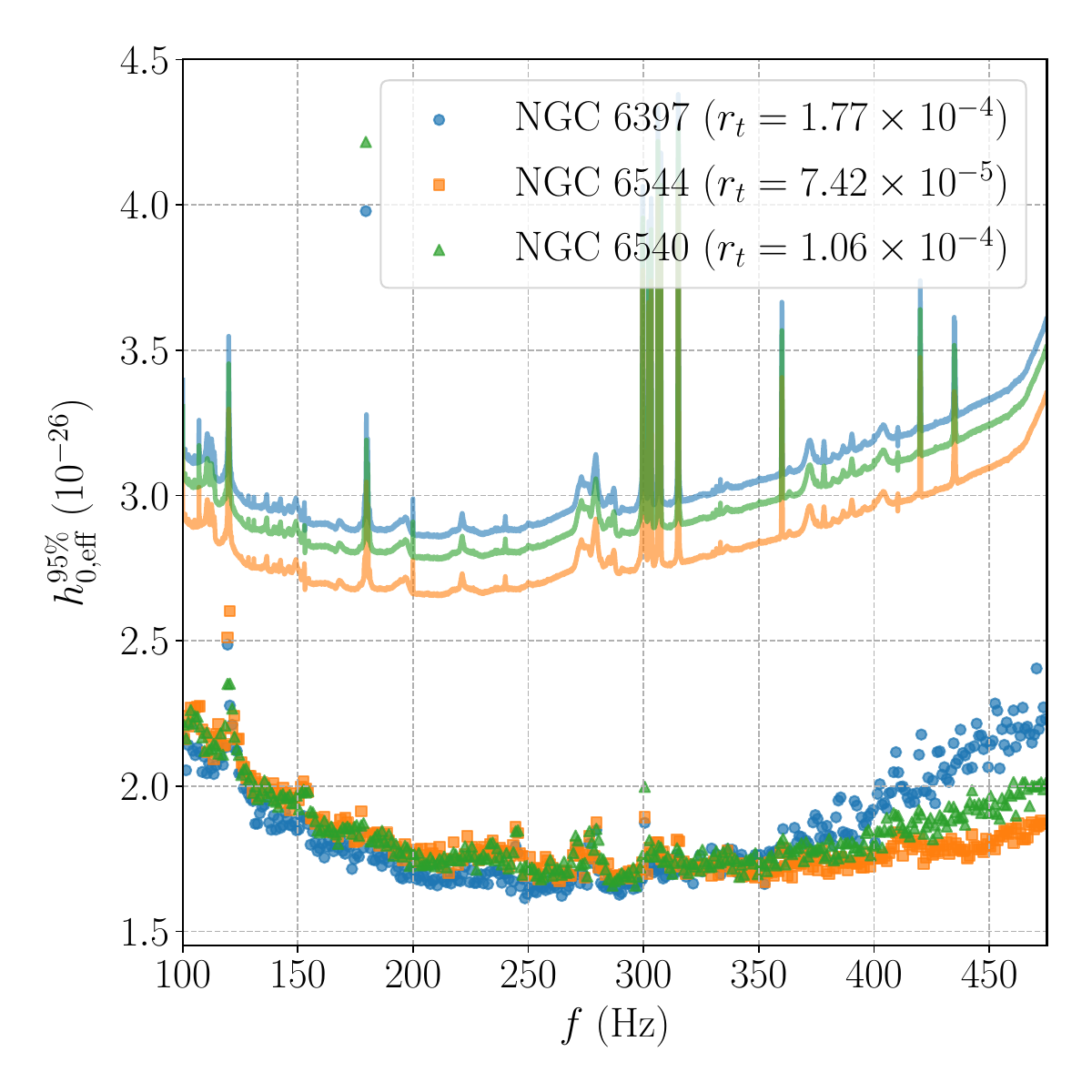}
  \caption{Effective gravitational wave strain amplitude upper limits (95\% efficiency) in each 1 Hz band for NGC~6397 (blue circle), NGC~6544 (orange square), and NGC~6540 (green triangle). The solid lines represent the corresponding sensitivity from~\citep{Dunn2025}, with the spin-down range restricted to $|\dot{f}| \leq 5 \times 10^{-10}$ Hz/s.
  }
  \label{fig:eff upper limits}
\end{figure}

We also compare our upper limits to the sensitivity from~\cite{Dunn2025}, as shown in Fig.~\ref{fig:eff upper limits}. 
To facilitate a direct comparison of the absolute improvements on strain amplitude constraints with the previous search on O3 data by~\citet{Dunn2025}, we calculate the effective upper limits to match the reporting convention published in~\cite{Dunn2025}.
The effective strain amplitude $h_{0,\mathrm{eff}}$ is defined as
\begin{equation} \label{eq:effective h95} 
    \bigl( h_{0,\mathrm{eff}} \bigr)^2 = h_0^2  \dfrac{\left[( 1 + \cos^2 \iota )/2\right]^2 + \cos^2 \iota }{2}.
\end{equation}
Similar to the procedure described above for $\hul$, we determine the 95\% effective upper limits ($h_{0,\mathrm{eff}}^{95\%}$) directly through software injections. 
However, we vary $h_{0,\mathrm{eff}}$ values rather than $h_0$. For each simulated signal at a given $h_{0,\mathrm{eff}}$, we draw the inclination angle uniformly in $\cos \iota$ and invert Eq.~\eqref{eq:effective h95} to calculate the corresponding intrinsic strain $h_0$ required for the injection. 
All other injection parameters are drawn as previously described. We then fit the detection fraction and estimate the 95\% effective upper limits, ensuring the statistical uncertainty in the estimated $h_{0,\mathrm{eff}}^{95\%}$ is below 5\%. 
Our results improve upon those from~\citet{Dunn2025} across 100--475~Hz. 
Notably, we achieve stricter effective upper limits of $\sim 1.7$--$2.3 \times 10^{-26}$, improving upon their limits by $\sim 30$--$40$\%. 
Furthermore, our search explores a spin-down range 1--2 orders of magnitude broader than the parameter space considered in their work.

To quantify the sensitivity of a search pipeline, independent of the data quality, a figure of merit known as the sensitivity depth $\depth$~\citep{sensitivitydepth} is commonly used,
\begin{equation}
  \depth(f) \equiv \frac{\asdwtoff}{ h_0^{\rm 95\%}};
  \label{eq:depth}
\end{equation}
\noindent where $\asdwtoff$ is an estimate of the effective strain amplitude spectral noise density. 
For nonstationary detector noise, we use an inverse-noise weighted estimate for each frequency bin $j$ from the two interferometers,
\begin{equation} \label{eq:weightedasd}
  \psdwt(f_j) = \frac{\sum_i w_{ij} S_h(f_i)}{\sum_i w_{ij}};
\end{equation}
where $i$ ranges over Fourier transforms of 30-minute segments of the H1 and L1 data, and $w_{ij}$ is a weight equal to the average inverse power spectral density for 50 neighboring frequency bins $j'\ne j$
in the same Fourier transform $i$,
\begin{equation}
  w_{ij} \equiv \frac{1}{50} \sum_{j'}\frac{1}{S_h(f_{j'})}
\end{equation}
\noindent for $|j'-j|\le25$ and $j'\ne j$. 
This weighting deemphasizes noisy segments of data, similarly to the weighting used to define the \Fstat. 

Table~\ref{tab:depth} shows the resulting sensitivity depths at 200 Hz for the five targets, as well as the averaged sensitivity depth where frequencies below 50 Hz have been excluded from consideration because the combination of highly disturbed bands and substantial mismatch between H1 and L1 strain noise levels would lead to artificially inflated sensitivity depth values.

\begin{table}[h!]
\centering
\begin{tabular}{lcc}
\hline
\T\B                        & $\dd$ at 200~Hz [$\mathrm{Hz}^{-1/2}$] &  $\bar{\mathcal{D}}^{95\%}$ [$\mathrm{Hz}^{-1/2}$]   \\
\hline\hline
\T\B Terzan~10    \strut    &  70.5  &  69.8 \\
\hline
\T\B NGC~104      \strut    &  87.2  &  89.3 \\
\hline\hline
\T\B NGC~6397     \strut    &  88.6  &  86.5 \\
\hline
\T\B NGC~6544     \strut    &  84.3  &  87.1 \\
\hline
\T\B NGC~6540     \strut    &  87.3  &  86.7 \\
\hline 
\end{tabular}
  \caption{Sensitivity depths $\dd$ for the targeted globular clusters at 200~Hz and the averaged depth $\bar{\mathcal{D}}^{95\%}$ over 50--\fmax~Hz.}
\label{tab:depth}
\end{table}

\subsection{Astrophysical constraints} 
\label{sec:astrolimits}

The GW strain upper limits derived in the previous section can be converted into constraints on the physical properties of the NSs. We consider two emission scenarios; a nonaxisymmetric deformation of the star (ellipticity) and unstable $r$-mode oscillations.

First, for a triaxial neutron star, the strain amplitude is directly related to the fiducial ellipticity, $\epsilon_0$~\citep{Jaranowski1998}. Assuming a canonical moment of inertia with respect to the rotation axis ($I_{zz} = 10^{38}$ kg m$^2$), our 95\% confidence strain limits ($\hul$) constrain $\epsilon_0$ as a function of frequency via
\begin{equation}
  \epsilon_0^{95\%}  = 9.46\times10^{-5}\bigg({\hul\over10^{-24}}\bigg)\bigg({d\over1\>{\rm kpc}}\bigg)
\bigg({100\>{\rm Hz}\over f}\bigg)^2.
\label{eq:ellipticity}
\end{equation}

Alternatively, we consider emission via unstable $r$-modes~\citep{Andersson,Bildsten,FriedmanMorsink,OwenEtal,Kojima}. In this model, the gravitational wave frequency is $f \approx (4/3)\frot$, and the signal strength is governed by the dimensionless amplitude $\alpha_0$~\citep{Owen2010},

\begin{equation}
  h_0  = 3.6\times10^{-23}\bigg({\alpha_0\over0.001}\bigg)\bigg({\fgw\over1\>{\rm kHz}}\bigg)^3
\bigg({1\>{\rm kpc}\over d}\bigg).
\label{eq:hrmodes}
\end{equation}

By inverting this relationship and substituting our upper limits $\hul$ for $h_0$, we obtain the 95\% confidence limits on the $r$-mode amplitude,

\begin{equation}
  \alpha_0^{95\%}  \simeq 0.028\bigg({\hul\over10^{-24}}\bigg)\bigg({d\over1\>{\rm kpc}}\bigg)
\bigg({100\>{\rm Hz}\over f}\bigg)^3.
\label{eq:rmodes}
\end{equation}

\begin{figure*}[htb!] 
	\centering
	\subfigure[]
	{
		\scalebox{0.4}{\includegraphics{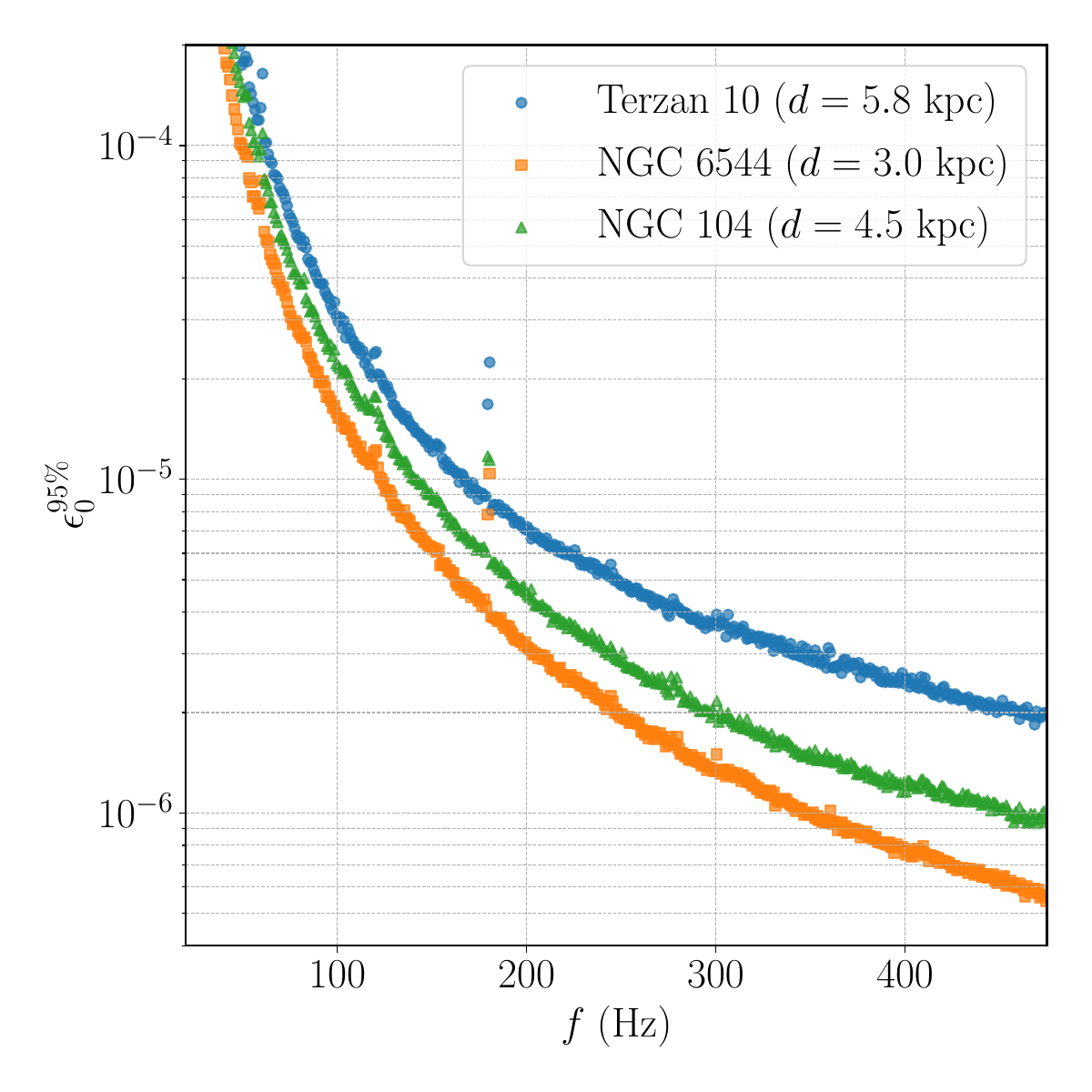}}
	}
	\subfigure[]
	{
		\scalebox{0.4}{\includegraphics{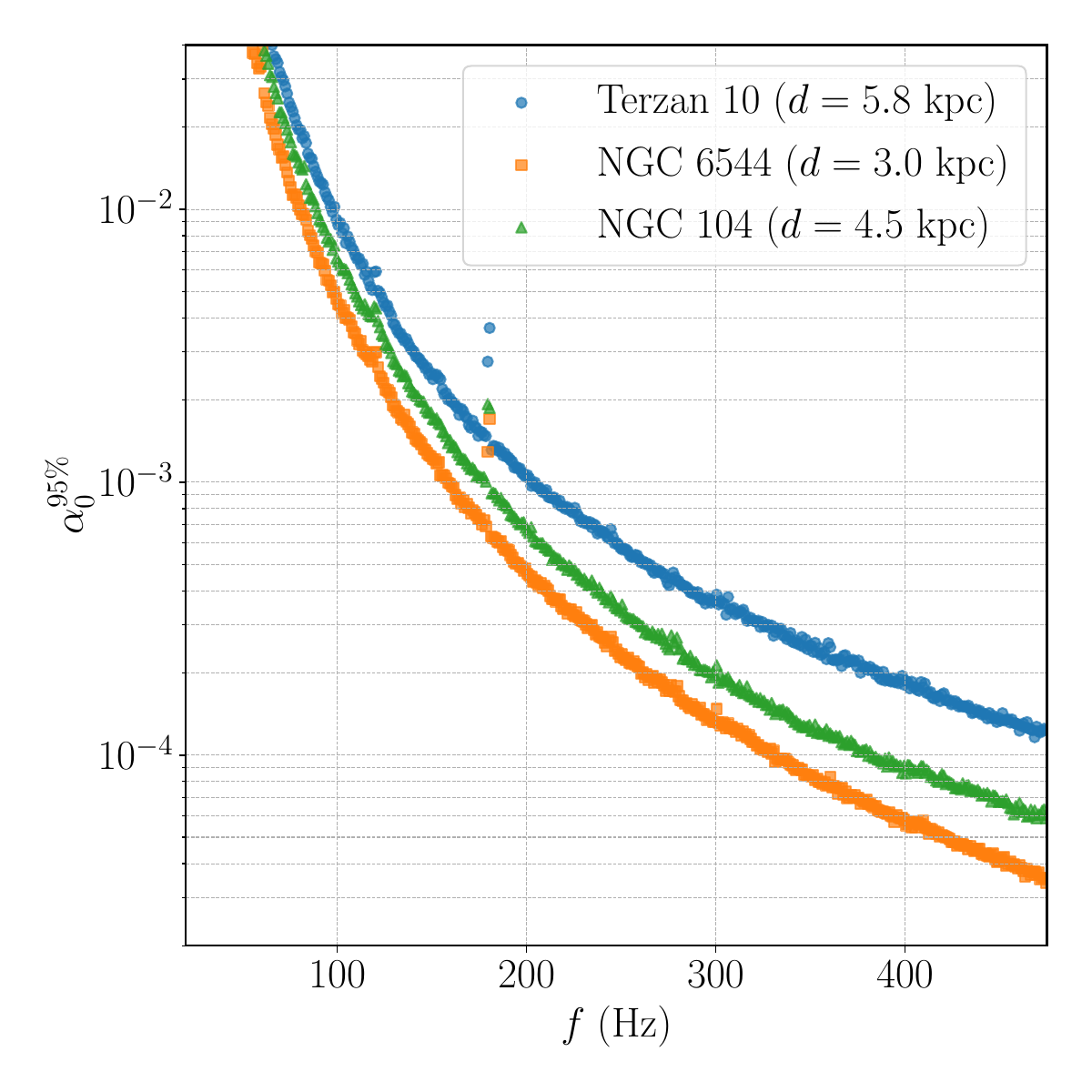}}
	}
	\caption{Equatorial ellipticity and $r$-mode amplitude $\alpha_0$ upper limits (95\% confidence level) in each 1 Hz band for Terzan~10 (blue circle), NGC~6544 (orange square), and NGC~104 (green triangle). These constraints are derived from the strain amplitude upper limits in Fig.~\ref{fig:upper limits rc} and distances from Table~\ref{tab:target param}.}
    \label{fig:astro rc}
\end{figure*}

\begin{figure*}[htb!]  
	\centering
	\subfigure[]
	{
		\scalebox{0.4}{\includegraphics{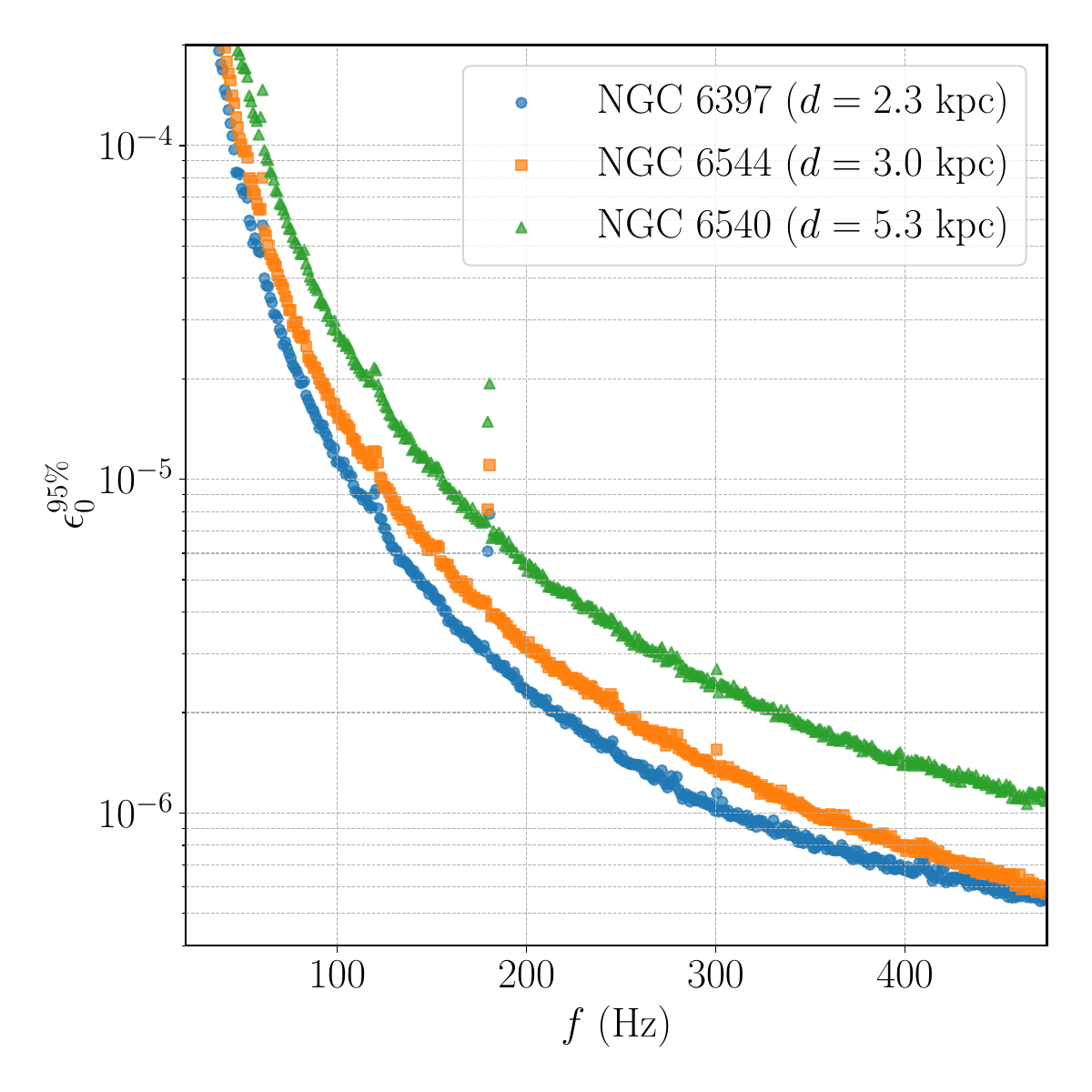}}
	}
	\subfigure[]
	{
		\scalebox{0.4}{\includegraphics{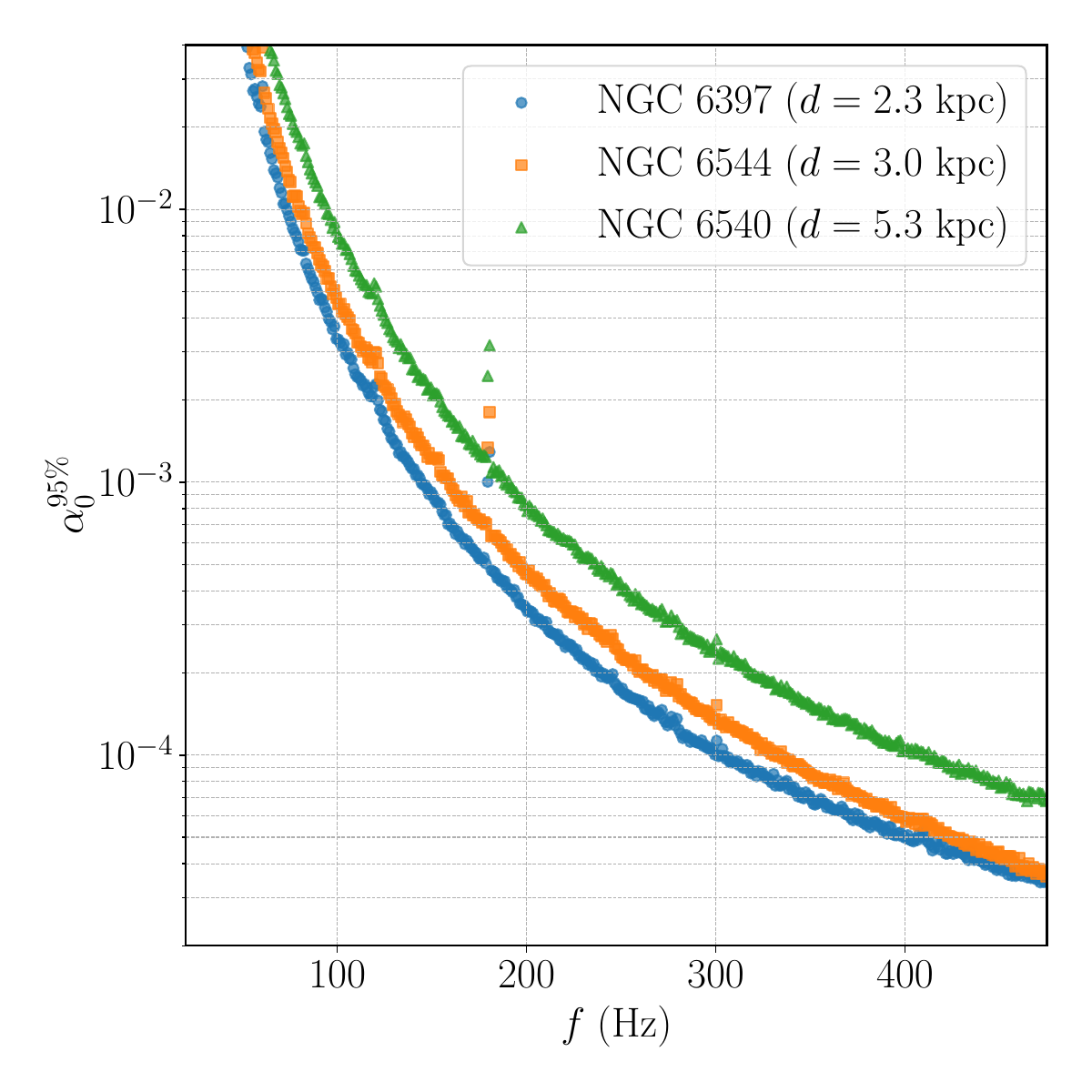}}
	}
	\caption{Equatorial ellipticity and $r$-mode amplitude $\alpha_0$ upper limits (95\% confidence level) in each 1 Hz band for NGC~6397 (blue circle), NGC~6544 (orange square), and NGC~6540 (green triangle). These constraints are derived from the strain amplitude upper limits in Fig.~\ref{fig:upper limits rt} and distances from Table~\ref{tab:target param}.}
    \label{fig:astro rt}
\end{figure*}

The resulting astrophysical constraints are presented in Figs.~\ref{fig:astro rc} and \ref{fig:astro rt}, calculated using the distance estimates listed in Table~\ref{tab:target param}. Figure~\ref{fig:astro rc} displays the limits for the targets searched over the core radius, while Fig.~\ref{fig:astro rt} shows the corresponding results for the targets searched over the tidal radius.
These constraints enter a physically interesting regime. 
Early theoretical models predicted a maximum ellipticity of up to $\sim 10^{-4}$ for exotic compact objects, such as solid strange quark stars~\citep{Owen_2005}.
Our derived $\epsilon_0^{95\%}$ limits, which reach $\lesssim 10^{-5}$ for $f>$ 200~Hz, are probing past this exotica regime and beginning to constrain the theoretically predicted maxima for ``normal'' NS matter.
Depending on the assumed equation of state and breaking strain, these theoretical limits lie in the range of $\sim 10^{-6}$--$10^{-5}$~\citep{Haskell_2007, Johnson-McDaniel2013eps}.
Similarly, our limits on $\alpha_0^{95\%}$ approach $\sim 10^{-3}$ for $f>$ 200~Hz, which corresponds to an estimated theoretical maximum amplitude for newborn neutron stars~\citep{Bondarescu2009rmode}. 
We note, however, that for recycled NSs in GCs, nonlinear saturation during accretion is predicted to limit $r$-mode amplitudes to much smaller values on the order of $\sim 10^{-5}$~\citep{Bondarescu_2007}.

\section{Conclusions} 
\label{sec:conclusions}
We have performed the deepest search for CWs from compact stars in the GCs NGC~6397, NGC~6544, and NGC~6540, covering the region within their tidal radii.
Additionally, we present the first directed search for Terzan~10 and NGC~104, targeting the central regions within their core radii.
These searches resulted in no detections.

Our analysis achieved 95\% confidence level upper limits as low as $\sim 5.1/4.2/4.2/4.4/4.4 \times 10^{-26}$ for Terzan~10/NGC~104/NGC~6397/NGC~6544/NGC~6540 at frequencies near 170/301/282/273/301 Hz, respectively.

We have achieved better sensitivities (and ellipticity/$r$-mode constraints) for NGC~6544 compared to~\citet{LVC2017GC}.
Furthermore, our results for NGC~6397, NGC~6544, and NGC~6540 improve upon those from~\citet{Dunn2025} across the 100--475~Hz band by 30--40\%. 
We also explore a spin-down range 1--2 orders of magnitude broader than the parameter space considered in their work, allowing us to surpass the age-based limit for 300 year-old sources across the entire frequency band.

Our results have begun to constrain the NS ellipticity and $r$-mode amplitude within a physically interesting regime, reaching values below the theoretical predicted maximums. 
As the LIGO, Virgo, and KAGRA detectors continue to improve their strain sensitivities in upcoming observing runs, these searches will dig deeper into the astrophysically plausible parameter space, bringing us progressively closer to the first detection of CWs from GCs.

\section{Acknowledgments}
We gratefully acknowledge useful discussions and long collaboration with current and former colleagues in the LIGO-Virgo-KAGRA continuous waves working group.
We also thank members of the LVK detector characterization group and the spectral line investigations team for valuable identification and mitigation of instrumental artifacts. In addition, we thank Cristiano Palomba and the anonymous referees for their helpful comments and careful reading of the manuscript.
This work was supported in part by National Science Foundation Awards No. PHY-2110181 and No. PHY-2408883, and through computational resources and services provided by Advanced Research Computing at the University of Michigan, Ann Arbor.

A.C., R.J., D.K., J-R.M.M., O.J.P., and A.S. were supported by the Universitat de les Illes Balears (UIB) with funds from the Programa de Foment de la Recerca i la Innovaci\'o de la UIB 2024-2026 (supported by the yearly plan of the Tourist Stay Tax ITS2023-086); the Spanish Agencia Estatal de Investigaci\'on Grants No. PID2022-138626NB-I00, No. RED2024-153978-E, No. RED2024-153735-E, funded by MICIU/AEI/10.13039/501100011033 and the ERDF/EU; and the Comunitat Aut\`onoma de les Illes Balears through the Conselleria d'Educaci\'o i Universitats with funds from the ERDF (SINCO2022/18146 - Plataforma HiTech-IAC3-BIO). 
J-R.M.M. is supported by the Spanish Ministerio de Ciencia, Innovaci\'on y Universidades (FPU22/01187).
O.J.P. is supported by the Spanish Ministerio de Ciencia, Innovaci\'on y Universidades (Ramon y Cajal, RYC2023-044489-I) funded by MCIN/AEI/10.13039/501100011033 and the FSE+ and cofinanced by the Universitat de les Illes Balears (UIB)
T.A.F. acknowledges support from the S\~ao Paulo Research Foundation (FAPESP) under Grant No. 2025/08599-6.
Q.L.N acknowledges support from NAFOSTED Grant No. 103.01-2025.147.

This material is based upon work supported by NSF's LIGO Laboratory
which is a major facility fully funded by the National Science Foundation.
LIGO was constructed and is operated by the California Institute of Technology and Massachusetts
Institute of Technology with funding from the U.S. National Science Foundation under Grant No. PHY-0757058.
The authors also gratefully acknowledge the support of the Science
and Technology Facilities Council (STFC) of the United Kingdom,
the Max-Planck-Society (MPS), and the
State of Niedersachsen/Germany for support of the construction of Advanced LIGO.
The authors are grateful for the computational resources provided by the
LIGO Laboratory and supported by
National Science Foundation Grants No. PHY-0757058 and No. PHY-0823459.

\section{Data availability}
The data that support the findings of this article are publicly available.
The complete catalog of the loudest surviving outliers mentioned in Appendix~\ref{appendix: candidate} is available from Ref.~\citep{cheung_2026_zenodo}.

\onecolumngrid
\appendix

\appendix \label{sec: appendix}

\section{Candidates} \label{appendix: candidate}
We list the loudest surviving outlier in each 1 Hz band for all targets in Tables~\ref{tab:terzan10 outliers}--\ref{tab:ngc6540 outliers}, as the full number of surviving candidates is large; the complete catalog is available on Zenodo~\citep{cheung_2026_zenodo}.
For Terzan 10, all outliers were vetoed after the fourth follow-up stage. 
For all other targets, candidates who survived the fourth follow-up stage were ruled out by spectrogram inspection.

\begin{table}[H]
    \centering
    \caption{Frequency parameters for the surviving Terzan~10 outliers from follow-up stage 3 ($\Tcoh = 60$ days).} 
    \label{tab:terzan10 outliers}

    \begin{tabular}{|c|c|c|c|}
        \hline
        $f$ (Hz) &
        $\dot{f}$ (Hz/s) &
        $\ddot{f}$ (Hz/s$^2$) &
        $\dddot{f}$ (Hz/s$^3$) \\
        \hline

        $30.00487872$ &
        $-6.0846\times10^{-10}$ &
        $1.2329\times10^{-19}$ &
        $2.2965\times10^{-27}$ \\

        $146.56147045$ &
        $-2.9891\times10^{-9}$ &
        $5.8961\times10^{-19}$ &
        $-2.0032\times10^{-27}$ \\

        $329.71690102$ &
        $-6.5599\times10^{-9}$ &
        $6.3227\times10^{-19}$ &
        $-2.6605\times10^{-27}$ \\

        \hline
    \end{tabular}
\end{table}

\begin{table}[H]
    \centering
    \caption{Frequency parameters for the surviving NGC~104 outliers from follow-up stage 4 ($\Tcoh = 120$ days).} 
    \label{tab:ngc104 outliers}

    \begin{tabular}{|c|c|c|c|c|}
        \hline
        $f$ (Hz) &
        $\dot{f}$ (Hz/s) &
        $\ddot{f}$ (Hz/s$^2$) &
        $\dddot{f}$ (Hz/s$^3$) &
        $\ddddot{f}$ (Hz/s$^4$) \\
        \hline
        $40.06316482$ & $-3.1890\times10^{-10}$ & $2.4479\times10^{-19}$ & $-1.7670\times10^{-27}$ & $-1.7552\times10^{-33}$ \\
        $70.03539391$ & $-5.4433\times10^{-10}$ & $-1.9565\times10^{-20}$ & $2.1001\times10^{-27}$ & $1.5305\times10^{-33}$ \\
        $270.52421253$ & $-5.8923\times10^{-9}$ & $1.2256\times10^{-18}$ & $2.6355\times10^{-27}$ & $-2.1168\times10^{-33}$ \\
        $283.92794347$ & $-1.2500\times10^{-9}$ & $-3.1553\times10^{-20}$ & $-1.5304\times10^{-27}$ & $-5.0880\times10^{-34}$ \\
        $315.00274203$ & $-2.8860\times10^{-8}$ & $1.4468\times10^{-17}$ & $6.3296\times10^{-27}$ & $-1.9352\times10^{-35}$ \\
        $379.00551396$ & $-2.7534\times10^{-9}$ & $1.3408\times10^{-19}$ & $-5.8579\times10^{-29}$ & $1.4795\times10^{-33}$ \\
        $445.61972296$ & $-1.8802\times10^{-9}$ & $5.0345\times10^{-20}$ & $-9.8901\times10^{-28}$ & $-9.9413\times10^{-35}$ \\
        \hline
    \end{tabular}
\end{table}

\begin{table}[H]
    \centering
    \caption{Frequency parameters for the surviving NGC~6397 outliers from follow-up stage 4 ($\Tcoh = 120$ days).} 
    \label{tab:ngc6397 outliers}

    \begin{tabular}{|c|c|c|c|c|}
        \hline
        $f$ (Hz) &
        $\dot{f}$ (Hz/s) &
        $\ddot{f}$ (Hz/s$^2$) &
        $\dddot{f}$ (Hz/s$^3$) &
        $\ddddot{f}$ (Hz/s$^4$) \\
        \hline
        $283.94864270$ & $-4.9032\times10^{-9}$ & $1.5520\times10^{-19}$ & $3.0986\times10^{-27}$ & $2.0157\times10^{-33}$ \\
        $379.01949516$ & $-6.2017\times10^{-9}$ & $2.3713\times10^{-19}$ & $-3.4899\times10^{-28}$ & $5.0707\times10^{-34}$ \\
        $388.54366280$ & $-6.5297\times10^{-9}$ & $8.7395\times10^{-20}$ & $1.1702\times10^{-27}$ & $2.6488\times10^{-34}$ \\
        $426.68342796$ & $-7.0999\times10^{-9}$ & $9.5778\times10^{-19}$ & $3.0373\times10^{-27}$ & $5.2953\times10^{-34}$ \\
        $430.01914868$ & $-7.3224\times10^{-9}$ & $6.7886\times10^{-19}$ & $-2.6481\times10^{-27}$ & $-9.8514\times10^{-34}$ \\
        $436.16602540$ & $-7.3649\times10^{-9}$ & $8.7411\times10^{-19}$ & $4.9891\times10^{-28}$ & $-2.1974\times10^{-33}$ \\
        $449.60274465$ & $-7.9046\times10^{-9}$ & $5.3481\times10^{-19}$ & $-1.5320\times10^{-27}$ & $1.5738\times10^{-33}$ \\
        $469.68101671$ & $-3.5000\times10^{-8}$ & $8.8900\times10^{-18}$ & $-6.8578\times10^{-27}$ & $3.2041\times10^{-34}$ \\
        $472.98257180$ & $-4.6830\times10^{-8}$ & $9.8280\times10^{-18}$ & $-6.4670\times10^{-27}$ & $-3.1365\times10^{-34}$ \\
        $473.77676000$ & $-8.0640\times10^{-9}$ & $1.0164\times10^{-18}$ & $-3.8945\times10^{-27}$ & $-9.4315\times10^{-34}$ \\
        \hline
    \end{tabular}
\end{table}

\begin{table}[H]
    \centering
    \caption{Frequency parameters for the surviving NGC~6544 outliers from follow-up stage 4 ($\Tcoh = 120$ days).} 
    \label{tab:ngc6544 outliers}

    \begin{tabular}{|c|c|c|c|c|}
        \hline
        $f$ (Hz) &
        $\dot{f}$ (Hz/s) &
        $\ddot{f}$ (Hz/s$^2$) &
        $\dddot{f}$ (Hz/s$^3$) &
        $\ddddot{f}$ (Hz/s$^4$) \\
        \hline
        $30.00488706$ & $-6.0869\times10^{-10}$ & $2.3406\times10^{-19}$ & $2.0249\times10^{-27}$ & $1.2809\times10^{-33}$ \\
        $473.93176396$ & $-2.1467\times10^{-8}$ & $7.7339\times10^{-18}$ & $2.9838\times10^{-27}$ & $-1.0475\times10^{-33}$ \\
        \hline
    \end{tabular}
\end{table}

\begin{table}[H]
    \centering
    \caption{Frequency parameters for the surviving NGC~6540 outliers from follow-up stage 4 ($\Tcoh = 120$ days).} 
    \label{tab:ngc6540 outliers}

    \begin{tabular}{|c|c|c|c|c|}
        \hline
        $f$ (Hz) &
        $\dot{f}$ (Hz/s) &
        $\ddot{f}$ (Hz/s$^2$) &
        $\dddot{f}$ (Hz/s$^3$) &
        $\ddddot{f}$ (Hz/s$^4$) \\
        \hline

        $70.04033430$ &
        $-1.4219\times10^{-9}$ &
        $3.4912\times10^{-19}$ &
        $-1.5183\times10^{-27}$ &
        $4.9406\times10^{-34}$ \\

        $329.71744877$ &
        $-6.5881\times10^{-9}$ &
        $2.3124\times10^{-19}$ &
        $-6.3276\times10^{-28}$ &
        $3.2736\times10^{-34}$ \\

        $463.94293956$ &
        $-2.1718\times10^{-8}$ &
        $3.3800\times10^{-18}$ &
        $-1.7217\times10^{-27}$ &
        $-1.0776\times10^{-33}$ \\

        \hline
    \end{tabular}
\end{table}

\section{Saturated sub-bands}

Some frequency bands were so badly contaminated by instrumental lines that one or more jobs are saturated ($\ge$1000 candidates) in the initial search.
All 0.1 Hz bands with saturation in at least one $\fdot$ sub-range are listed in Tables~\ref{tab:satbands}.
We do not claim strain upper limits to signals in these sub-bands, which sum to 28.4/28.5/28.5/32.7/28.6 Hz for Terzan~10/NGC~6544/NGC~104/NGC~6397/NGC~6540 over the search range of \fmin--\fmax Hz.

\begin{table}[H]
    \centering
    \caption{
    List of 0.1-Hz frequency bands exhibiting saturation (defined as $\ge$1000 outliers above threshold for at least one job) during the first search stage. Values indicate the start frequency of the band. Superscripts identify the specific target where saturation occurred ($a$: Terzan~10; $b$: NGC~6544; $c$: NGC~104; $d$: NGC~6397; $e$: NGC~6540); the absence of a superscript implies the band was saturated for all targets. These frequencies were excluded from the upper limit curves.
    }
    \label{tab:satbands}
    
    \pgfplotstabletypeset[
        col sep=comma,
        header=has colnames,
        every column/.style={
            string type,
            column type={c@{\hspace{20pt}}}, 
            column name={$f_\mathrm{low}$ (Hz)}
        },
        every head row/.style={before row=\hline,after row=\hline\hline},
        every last row/.style={after row=\hline},
    ]{
col0,col1,col2,col3,col4,col5,col6,col7,col8
{$20.0$},{$20.5$},{$20.9$},{$21.4$},{$21.8$},{$21.9^{\rlap{$\scriptstyle a,b,d,e$}}$},{$22.2^{\rlap{$\scriptstyle a,b,c,e$}}$},{$23.5^{\rlap{$\scriptstyle d$}}$},{$24.3$}
{$24.4$},{$24.5$},{$24.9$},{$25.4^{\rlap{$\scriptstyle a,b,d,e$}}$},{$26.5$},{$26.6$},{$27.2$},{$27.3$},{$27.4$}
{$27.5$},{$27.6$},{$27.7$},{$27.8$},{$27.9$},{$28.1$},{$28.2$},{$28.3$},{$28.4$}
{$28.5^{\rlap{$\scriptstyle c$}}$},{$29.9$},{$30.0$},{$33.2$},{$33.3$},{$33.4$},{$33.9$},{$34.8$},{$34.9$}
{$35.2$},{$35.3$},{$35.4$},{$35.6$},{$35.7$},{$35.8$},{$37.8$},{$37.9$},{$39.5$}
{$39.8$},{$39.9$},{$40.0^{\rlap{$\scriptstyle d$}}$},{$40.1^{\rlap{$\scriptstyle c,d$}}$},{$40.8$},{$40.9$},{$41.5^{\rlap{$\scriptstyle c,d$}}$},{$41.8$},{$41.9$}
{$42.1^{\rlap{$\scriptstyle d$}}$},{$42.8$},{$44.4$},{$44.5$},{$44.6$},{$44.7$},{$44.8$},{$44.9^{\rlap{$\scriptstyle a,b,e$}}$},{$45.1$}
{$45.2$},{$45.3$},{$46.0$},{$46.1$},{$47.3$},{$47.4$},{$47.9^{\rlap{$\scriptstyle c,d$}}$},{$48.3^{\rlap{$\scriptstyle c,d$}}$},{$48.4^{\rlap{$\scriptstyle c,d$}}$}
{$48.7$},{$48.8$},{$49.8$},{$49.9$},{$50.0$},{$51.4^{\rlap{$\scriptstyle c,d$}}$},{$52.8$},{$53.6$},{$53.7$}
{$54.8$},{$55.5$},{$55.8$},{$56.8$},{$59.5$},{$59.7$},{$59.8$},{$59.9$},{$60.0$}
{$61.4^{\rlap{$\scriptstyle c,d$}}$},{$62.7^{\rlap{$\scriptstyle a,b,d,e$}}$},{$62.8$},{$64.2^{\rlap{$\scriptstyle c,d$}}$},{$64.3^{\rlap{$\scriptstyle a,b,d,e$}}$},{$64.7$},{$64.8$},{$66.3$},{$66.6$}
{$69.7$},{$69.8$},{$70.0^{\rlap{$\scriptstyle c,d$}}$},{$74.7$},{$74.8^{\rlap{$\scriptstyle a,b,d,e$}}$},{$75.7^{\rlap{$\scriptstyle c$}}$},{$75.8$},{$76.7$},{$76.8$}
{$77.7$},{$79.7$},{$83.3$},{$83.7$},{$83.8^{\rlap{$\scriptstyle a,b,e$}}$},{$84.6$},{$84.7$},{$85.2$},{$85.3^{\rlap{$\scriptstyle c,d$}}$}
{$88.7$},{$88.8^{\rlap{$\scriptstyle d$}}$},{$89.6$},{$89.7$},{$89.8$},{$89.9$},{$90.7$},{$90.8$},{$94.6$}
{$94.7$},{$97.6$},{$97.7$},{$99.6$},{$99.7^{\rlap{$\scriptstyle a,b,e$}}$},{$99.8^{\rlap{$\scriptstyle a,b,e$}}$},{$99.9$},{$100.0$},{$100.1$}
{$102.0$},{$102.1$},{$104.6$},{$104.7^{\rlap{$\scriptstyle a,b,d,e$}}$},{$105.3^{\rlap{$\scriptstyle a,b,d,e$}}$},{$106.4$},{$106.5^{\rlap{$\scriptstyle c,d$}}$},{$107.1$},{$109.6$}
{$111.6$},{$114.6^{\rlap{$\scriptstyle a,c,d,e$}}$},{$118.6$},{$119.6^{\rlap{$\scriptstyle c$}}$},{$119.8$},{$124.6^{\rlap{$\scriptstyle c,d$}}$},{$125.6$},{$129.5^{\rlap{$\scriptstyle c$}}$},{$129.6^{\rlap{$\scriptstyle d$}}$}
{$132.5^{\rlap{$\scriptstyle c,d$}}$},{$132.6^{\rlap{$\scriptstyle a,b,d,e$}}$},{$139.5$},{$144.4^{\rlap{$\scriptstyle c$}}$},{$146.5^{\rlap{$\scriptstyle c,d$}}$},{$153.5^{\rlap{$\scriptstyle c,d$}}$},{$160.5^{\rlap{$\scriptstyle c$}}$},{$164.4$},{$165.8$}
{$165.9^{\rlap{$\scriptstyle c,d,e$}}$},{$177.7$},{$180.7$},{$180.9^{\rlap{$\scriptstyle a,b,d,e$}}$},{$181.0$},{$181.1^{\rlap{$\scriptstyle a,b,d,e$}}$},{$181.4^{\rlap{$\scriptstyle c,d$}}$},{$200.0$},{$209.8^{\rlap{$\scriptstyle d$}}$}
{$213.0$},{$213.1^{\rlap{$\scriptstyle a,b,e$}}$},{$213.2$},{$213.3^{\rlap{$\scriptstyle a,b,d,e$}}$},{$239.7$},{$257.4^{\rlap{$\scriptstyle c$}}$},{$267.3^{\rlap{$\scriptstyle a,b,e$}}$},{$267.4$},{$268.9^{\rlap{$\scriptstyle c$}}$}
{$269.0^{\rlap{$\scriptstyle c,d$}}$},{$269.5$},{$269.6$},{$269.7^{\rlap{$\scriptstyle c,d$}}$},{$270.5$},{$270.6$},{$270.7$},{$270.8$},{$271.6$}
{$272.0$},{$273.8$},{$273.9$},{$274.0$},{$274.1$},{$274.9$},{$275.5$},{$275.6$},{$275.7$}
{$275.9^{\rlap{$\scriptstyle c$}}$},{$277.1$},{$277.2$},{$277.3$},{$277.4$},{$277.5^{\rlap{$\scriptstyle a,b,e$}}$},{$278.2$},{$278.3$},{$278.8$}
{$278.9$},{$279.0$},{$279.1$},{$279.3^{\rlap{$\scriptstyle c$}}$},{$279.4$},{$280.4^{\rlap{$\scriptstyle c$}}$},{$280.5$},{$280.6$},{$280.7$}
{$280.8^{\rlap{$\scriptstyle a,b,d,e$}}$},{$281.6$},{$282.3^{\rlap{$\scriptstyle c,d$}}$},{$282.4^{\rlap{$\scriptstyle d$}}$},{$283.8$},{$283.9^{\rlap{$\scriptstyle c,d$}}$},{$284.0$},{$284.1^{\rlap{$\scriptstyle d$}}$},{$284.9^{\rlap{$\scriptstyle d$}}$}
{$299.2^{\rlap{$\scriptstyle c,d$}}$},{$299.3$},{$299.4$},{$299.5$},{$299.6$},{$299.7$},{$299.8$},{$299.9$},{$300.0$}
{$301.8^{\rlap{$\scriptstyle b,c,d,e$}}$},{$301.9$},{$302.0$},{$302.1$},{$302.2$},{$302.3$},{$302.4$},{$302.9^{\rlap{$\scriptstyle c,d$}}$},{$303.0$}
{$303.1$},{$303.2$},{$303.3$},{$303.4$},{$303.5$},{$303.6^{\rlap{$\scriptstyle c$}}$},{$305.7$},{$305.8$},{$305.9$}
{$306.0$},{$306.1$},{$306.2$},{$306.3$},{$306.4$},{$306.5$},{$307.0$},{$307.1$},{$307.2$}
{$307.3$},{$307.4$},{$307.5$},{$307.6$},{$307.7$},{$307.8$},{$314.5$},{$314.6$},{$314.7$}
{$314.8$},{$314.9$},{$315.0$},{$315.1$},{$315.2$},{$315.3$},{$315.4$},{$329.6^{\rlap{$\scriptstyle c,d$}}$},{$329.7^{\rlap{$\scriptstyle d$}}$}
{$331.6$},{$331.8$},{$359.6$},{$360.4^{\rlap{$\scriptstyle a,b,d,e$}}$},{$360.5^{\rlap{$\scriptstyle d$}}$},{$369.7^{\rlap{$\scriptstyle c$}}$},{$388.7^{\rlap{$\scriptstyle c$}}$},{$389.6^{\rlap{$\scriptstyle c,d$}}$},{$400.0$}
{$400.1$},{$400.2^{\rlap{$\scriptstyle a,b,d,e$}}$},{$400.3^{\rlap{$\scriptstyle a,b,e$}}$},{$407.4^{\rlap{$\scriptstyle c$}}$},{$410.2^{\rlap{$\scriptstyle a,b,d,e$}}$},{$410.3$},{$410.4$},{$410.5^{\rlap{$\scriptstyle a,b,d,e$}}$},{$410.6^{\rlap{$\scriptstyle a,b,e$}}$}
{$416.8^{\rlap{$\scriptstyle c$}}$},{$416.9^{\rlap{$\scriptstyle d$}}$},{$417.1^{\rlap{$\scriptstyle c$}}$},{$417.2^{\rlap{$\scriptstyle d$}}$},{$426.6^{\rlap{$\scriptstyle c,d$}}$},{$435.8^{\rlap{$\scriptstyle c,d$}}$},{$445.6^{\rlap{$\scriptstyle c$}}$},{$449.5^{\rlap{$\scriptstyle c$}}$},{$454.7^{\rlap{$\scriptstyle d$}}$}
{$454.8^{\rlap{$\scriptstyle a,b,d,e$}}$},{$454.9^{\rlap{$\scriptstyle c,d$}}$},{$455.8^{\rlap{$\scriptstyle c$}}$},{$456.2$},{$456.3^{\rlap{$\scriptstyle c,d,e$}}$},{$457.4^{\rlap{$\scriptstyle c$}}$},{$457.9^{\rlap{$\scriptstyle c$}}$},{$458.6^{\rlap{$\scriptstyle b,c,d$}}$},{$458.7^{\rlap{$\scriptstyle d$}}$}
{$459.7$},{$459.8$},{$459.9^{\rlap{$\scriptstyle d$}}$},{$461.3^{\rlap{$\scriptstyle a,b,e$}}$},{$461.4$},{$461.5^{\rlap{$\scriptstyle c,d$}}$},{$462.5^{\rlap{$\scriptstyle d$}}$},{$462.6^{\rlap{$\scriptstyle d$}}$},{$462.8^{\rlap{$\scriptstyle a,b,d,e$}}$}
{$462.9$},{$463.0$},{$463.9^{\rlap{$\scriptstyle a,b,e$}}$},{$464.0^{\rlap{$\scriptstyle c,d$}}$},{$464.1^{\rlap{$\scriptstyle c$}}$},{$464.2^{\rlap{$\scriptstyle c$}}$},{$464.4^{\rlap{$\scriptstyle a,b$}}$},{$464.5^{\rlap{$\scriptstyle a,b,d,e$}}$},{$464.6$}
{$464.7$},{$465.8^{\rlap{$\scriptstyle c$}}$},{$465.9^{\rlap{$\scriptstyle c,d$}}$},{$466.0^{\rlap{$\scriptstyle b,c,d,e$}}$},{$466.1^{\rlap{$\scriptstyle a,b,d,e$}}$},{$466.2^{\rlap{$\scriptstyle a,b,e$}}$},{$466.5^{\rlap{$\scriptstyle d$}}$},{$466.6^{\rlap{$\scriptstyle d$}}$},{$466.8^{\rlap{$\scriptstyle a,b,d,e$}}$}
{$466.9^{\rlap{$\scriptstyle a,b,d,e$}}$},{$467.0$},{$467.1^{\rlap{$\scriptstyle a,d$}}$},{$467.7^{\rlap{$\scriptstyle c$}}$},{$467.8^{\rlap{$\scriptstyle c$}}$},{$467.9^{\rlap{$\scriptstyle c,d$}}$},{$468.0$},{$468.1$},{$468.2$}
{$469.7^{\rlap{$\scriptstyle a,b,d,e$}}$},{$469.8$},{$469.9^{\rlap{$\scriptstyle c,d$}}$},{$470.9^{\rlap{$\scriptstyle c$}}$},{$471.1$},{$471.2$},{$471.3$},{$471.4$},{$472.2^{\rlap{$\scriptstyle a,b,e$}}$}
{$472.3^{\rlap{$\scriptstyle a,b,d,e$}}$},{$472.4$},{$472.9^{\rlap{$\scriptstyle a,b,d,e$}}$},{$473.0^{\rlap{$\scriptstyle a,b,d,e$}}$},{$473.1$},{$473.2$},{$473.8^{\rlap{$\scriptstyle d$}}$},{$473.9$},{$474.0$}
{$474.1^{\rlap{$\scriptstyle a,b,d,e$}}$},{},{},{},{},{},{},{},{}
    }
\end{table}

\twocolumngrid
\bibliography{O4aGC}

\begin{thebibliography}{84}%
\makeatletter
\providecommand \@ifxundefined [1]{%
 \@ifx{#1\undefined}
}%
\providecommand \@ifnum [1]{%
 \ifnum #1\expandafter \@firstoftwo
 \else \expandafter \@secondoftwo
 \fi
}%
\providecommand \@ifx [1]{%
 \ifx #1\expandafter \@firstoftwo
 \else \expandafter \@secondoftwo
 \fi
}%
\providecommand \natexlab [1]{#1}%
\providecommand \enquote  [1]{``#1''}%
\providecommand \bibnamefont  [1]{#1}%
\providecommand \bibfnamefont [1]{#1}%
\providecommand \citenamefont [1]{#1}%
\providecommand \href@noop [0]{\@secondoftwo}%
\providecommand \href [0]{\begingroup \@sanitize@url \@href}%
\providecommand \@href[1]{\@@startlink{#1}\@@href}%
\providecommand \@@href[1]{\endgroup#1\@@endlink}%
\providecommand \@sanitize@url [0]{\catcode `\\12\catcode `\$12\catcode `\&12\catcode `\#12\catcode `\^12\catcode `\_12\catcode `\%12\relax}%
\providecommand \@@startlink[1]{}%
\providecommand \@@endlink[0]{}%
\providecommand \url  [0]{\begingroup\@sanitize@url \@url }%
\providecommand \@url [1]{\endgroup\@href {#1}{\urlprefix }}%
\providecommand \urlprefix  [0]{URL }%
\providecommand \Eprint [0]{\href }%
\providecommand \doibase [0]{https://doi.org/}%
\providecommand \selectlanguage [0]{\@gobble}%
\providecommand \bibinfo  [0]{\@secondoftwo}%
\providecommand \bibfield  [0]{\@secondoftwo}%
\providecommand \translation [1]{[#1]}%
\providecommand \BibitemOpen [0]{}%
\providecommand \bibitemStop [0]{}%
\providecommand \bibitemNoStop [0]{.\EOS\space}%
\providecommand \EOS [0]{\spacefactor3000\relax}%
\providecommand \BibitemShut  [1]{\csname bibitem#1\endcsname}%
\let\auto@bib@innerbib\@empty
\bibitem [{\citenamefont {Tenorio}\ \emph {et~al.}(2021)\citenamefont {Tenorio}, \citenamefont {Keitel},\ and\ \citenamefont {Sintes}}]{bib:TenorioKeitelSintesreview}%
  \BibitemOpen
  \bibfield  {author} {\bibinfo {author} {\bibfnamefont {R.}~\bibnamefont {Tenorio}}, \bibinfo {author} {\bibfnamefont {D.}~\bibnamefont {Keitel}},\ and\ \bibinfo {author} {\bibfnamefont {A.~M.}\ \bibnamefont {Sintes}},\ }\bibfield  {title} {\bibinfo {title} {{Search Methods for Continuous Gravitational-Wave Signals from Unknown Sources in the Advanced-Detector Era}},\ }\bibfield  {journal} {\bibinfo  {journal} {Universe}\ }\textbf {\bibinfo {volume} {7}},\ \href {https://doi.org/10.3390/universe7120474} {10.3390/universe7120474} (\bibinfo {year} {2021})\BibitemShut {NoStop}%
\bibitem [{\citenamefont {Piccinni}(2022)}]{bib:PiccinniReview}%
  \BibitemOpen
  \bibfield  {author} {\bibinfo {author} {\bibfnamefont {O.~J.}\ \bibnamefont {Piccinni}},\ }\bibfield  {title} {\bibinfo {title} {Status and perspectives of continuous gravitational wave searches},\ }\bibfield  {journal} {\bibinfo  {journal} {Galaxies}\ }\textbf {\bibinfo {volume} {10}},\ \href {https://doi.org/10.3390/galaxies10030072} {10.3390/galaxies10030072} (\bibinfo {year} {2022})\BibitemShut {NoStop}%
\bibitem [{\citenamefont {Riles}(2023)}]{bib:RilesReview}%
  \BibitemOpen
  \bibfield  {author} {\bibinfo {author} {\bibfnamefont {K.}~\bibnamefont {Riles}},\ }\bibfield  {title} {\bibinfo {title} {{Searches for continuous-wave gravitational radiation}},\ }\href {https://doi.org/10.1007/s41114-023-00044-3} {\bibfield  {journal} {\bibinfo  {journal} {Living Rev. Rel.}\ }\textbf {\bibinfo {volume} {26}},\ \bibinfo {pages} {3} (\bibinfo {year} {2023})},\ \Eprint {https://arxiv.org/abs/2206.06447} {arXiv:2206.06447 [astro-ph.HE]} \BibitemShut {NoStop}%
\bibitem [{\citenamefont {Wette}(2023)}]{bib:WetteReview}%
  \BibitemOpen
  \bibfield  {author} {\bibinfo {author} {\bibfnamefont {K.}~\bibnamefont {Wette}},\ }\bibfield  {title} {\bibinfo {title} {{Searches for continuous gravitational waves from neutron stars: A twenty-year retrospective}},\ }\href {https://doi.org/10.1016/j.astropartphys.2023.102880} {\bibfield  {journal} {\bibinfo  {journal} {Astropart. Phys.}\ }\textbf {\bibinfo {volume} {153}},\ \bibinfo {pages} {102880} (\bibinfo {year} {2023})},\ \Eprint {https://arxiv.org/abs/2305.07106} {arXiv:2305.07106 [gr-qc]} \BibitemShut {NoStop}%
\bibitem [{\citenamefont {Lasky}(2015)}]{Lasky_review}%
  \BibitemOpen
  \bibfield  {author} {\bibinfo {author} {\bibfnamefont {P.}~\bibnamefont {Lasky}},\ }\bibfield  {title} {\bibinfo {title} {Gravitational waves from neutron stars: A review},\ }\href@noop {} {\bibfield  {journal} {\bibinfo  {journal} {Pub. Astron. Soc. Aust.}\ }\textbf {\bibinfo {volume} {32}},\ \bibinfo {pages} {e034} (\bibinfo {year} {2015})}\BibitemShut {NoStop}%
\bibitem [{\citenamefont {Glampedakis}\ and\ \citenamefont {Gualtieri}(2018)}]{GandG_review}%
  \BibitemOpen
  \bibfield  {author} {\bibinfo {author} {\bibfnamefont {K.}~\bibnamefont {Glampedakis}}\ and\ \bibinfo {author} {\bibfnamefont {L.}~\bibnamefont {Gualtieri}},\ }\bibinfo {title} {Gravitational waves from single neutron stars: an advanced detector era survey},\ in\ \href@noop {} {\emph {\bibinfo {booktitle} {Astrophys. Space Sci. Lib.}}},\ Vol.\ \bibinfo {volume} {457}\ (\bibinfo {year} {2018})\ pp.\ \bibinfo {pages} {673--736}\BibitemShut {NoStop}%
\bibitem [{\citenamefont {{Aharonian}}\ \emph {et~al.}(2005)\citenamefont {{Aharonian}}, \citenamefont {{Akhperjanian}}, \citenamefont {{Aye}}, \citenamefont {{Bazer-Bachi}}, \citenamefont {{Beilicke}}, \citenamefont {{Benbow}}, \citenamefont {{Berge}}, \citenamefont {{Berghaus}}, \citenamefont {{Bernl{\"o}hr}}, \citenamefont {{Boisson}}, \citenamefont {{Bolz}}, \citenamefont {{Borgmeier}}, \citenamefont {{Braun}}, \citenamefont {{Breitling}}, \citenamefont {{Brown}}, \citenamefont {{Gordo}}, \citenamefont {{Chadwick}}, \citenamefont {{Chounet}}, \citenamefont {{Cornils}}, \citenamefont {{Costamante}}, \citenamefont {{Degrange}}, \citenamefont {{Djannati-Ata{\"i}}}, \citenamefont {{Drury}}, \citenamefont {{Dubus}}, \citenamefont {{Ergin}}, \citenamefont {{Espigat}}, \citenamefont {{Feinstein}}, \citenamefont {{Fleury}}, \citenamefont {{Fontaine}}, \citenamefont {{Funk}}, \citenamefont {{Gallant}}, \citenamefont {{Giebels}}, \citenamefont {{Gillessen}}, \citenamefont {{Goret}}, \citenamefont
  {{Hadjichristidis}}, \citenamefont {{Hauser}}, \citenamefont {{Heinzelmann}}, \citenamefont {{Henri}}, \citenamefont {{Hermann}}, \citenamefont {{Hinton}}, \citenamefont {{Hofmann}}, \citenamefont {{Holleran}}, \citenamefont {{Horns}}, \citenamefont {{de Jager}}, \citenamefont {{Jung}}, \citenamefont {{Kh{\'e}lifi}}, \citenamefont {{Komin}}, \citenamefont {{Konopelko}}, \citenamefont {{Latham}}, \citenamefont {{Le Gallou}}, \citenamefont {{Lemi{\`e}re}}, \citenamefont {{Lemoine}}, \citenamefont {{Leroy}}, \citenamefont {{Lohse}}, \citenamefont {{Marcowith}}, \citenamefont {{Masterson}}, \citenamefont {{McComb}}, \citenamefont {{de Naurois}}, \citenamefont {{Nolan}}, \citenamefont {{Noutsos}}, \citenamefont {{Orford}}, \citenamefont {{Osborne}}, \citenamefont {{Ouchrif}}, \citenamefont {{Panter}}, \citenamefont {{Pelletier}}, \citenamefont {{Pita}}, \citenamefont {{P{\"u}hlhofer}}, \citenamefont {{Punch}}, \citenamefont {{Raubenheimer}}, \citenamefont {{Raue}}, \citenamefont {{Raux}}, \citenamefont
  {{Rayner}}, \citenamefont {{Redondo}}, \citenamefont {{Reimer}}, \citenamefont {{Reimer}}, \citenamefont {{Ripken}}, \citenamefont {{Rob}}, \citenamefont {{Rolland}}, \citenamefont {{Rowell}}, \citenamefont {{Sahakian}}, \citenamefont {{Saug{\'e}}}, \citenamefont {{Schlenker}}, \citenamefont {{Schlickeiser}}, \citenamefont {{Schuster}}, \citenamefont {{Schwanke}}, \citenamefont {{Siewert}}, \citenamefont {{Sol}}, \citenamefont {{Steenkamp}}, \citenamefont {{Stegmann}}, \citenamefont {{Tavernet}}, \citenamefont {{Terrier}}, \citenamefont {{Th{\'e}oret}}, \citenamefont {{Tluczykont}}, \citenamefont {{van der Walt}}, \citenamefont {{Vasileiadis}}, \citenamefont {{Venter}}, \citenamefont {{Vincent}}, \citenamefont {{Visser}}, \citenamefont {{V{\"o}lk}},\ and\ \citenamefont {{Wagner}}}]{ATNF_2005}%
  \BibitemOpen
  \bibfield  {author} {\bibinfo {author} {\bibfnamefont {F.}~\bibnamefont {{Aharonian}}}, \bibinfo {author} {\bibfnamefont {A.~G.}\ \bibnamefont {{Akhperjanian}}}, \bibinfo {author} {\bibfnamefont {K.-M.}\ \bibnamefont {{Aye}}}, \bibinfo {author} {\bibfnamefont {A.~R.}\ \bibnamefont {{Bazer-Bachi}}}, \bibinfo {author} {\bibfnamefont {M.}~\bibnamefont {{Beilicke}}}, \bibinfo {author} {\bibfnamefont {W.}~\bibnamefont {{Benbow}}}, \bibinfo {author} {\bibfnamefont {D.}~\bibnamefont {{Berge}}}, \bibinfo {author} {\bibfnamefont {P.}~\bibnamefont {{Berghaus}}}, \bibinfo {author} {\bibfnamefont {K.}~\bibnamefont {{Bernl{\"o}hr}}}, \bibinfo {author} {\bibfnamefont {C.}~\bibnamefont {{Boisson}}}, \bibinfo {author} {\bibfnamefont {O.}~\bibnamefont {{Bolz}}}, \bibinfo {author} {\bibfnamefont {C.}~\bibnamefont {{Borgmeier}}}, \bibinfo {author} {\bibfnamefont {I.}~\bibnamefont {{Braun}}}, \bibinfo {author} {\bibfnamefont {F.}~\bibnamefont {{Breitling}}}, \bibinfo {author} {\bibfnamefont {A.~M.}\ \bibnamefont {{Brown}}},
  \bibinfo {author} {\bibfnamefont {J.~B.}\ \bibnamefont {{Gordo}}}, \bibinfo {author} {\bibfnamefont {P.~M.}\ \bibnamefont {{Chadwick}}}, \bibinfo {author} {\bibfnamefont {L.-M.}\ \bibnamefont {{Chounet}}}, \bibinfo {author} {\bibfnamefont {R.}~\bibnamefont {{Cornils}}}, \bibinfo {author} {\bibfnamefont {L.}~\bibnamefont {{Costamante}}}, \bibinfo {author} {\bibfnamefont {B.}~\bibnamefont {{Degrange}}}, \bibinfo {author} {\bibfnamefont {A.}~\bibnamefont {{Djannati-Ata{\"i}}}}, \bibinfo {author} {\bibfnamefont {L.~O.}\ \bibnamefont {{Drury}}}, \bibinfo {author} {\bibfnamefont {G.}~\bibnamefont {{Dubus}}}, \bibinfo {author} {\bibfnamefont {T.}~\bibnamefont {{Ergin}}}, \bibinfo {author} {\bibfnamefont {P.}~\bibnamefont {{Espigat}}}, \bibinfo {author} {\bibfnamefont {F.}~\bibnamefont {{Feinstein}}}, \bibinfo {author} {\bibfnamefont {P.}~\bibnamefont {{Fleury}}}, \bibinfo {author} {\bibfnamefont {G.}~\bibnamefont {{Fontaine}}}, \bibinfo {author} {\bibfnamefont {S.}~\bibnamefont {{Funk}}}, \bibinfo {author}
  {\bibfnamefont {Y.~A.}\ \bibnamefont {{Gallant}}}, \bibinfo {author} {\bibfnamefont {B.}~\bibnamefont {{Giebels}}}, \bibinfo {author} {\bibfnamefont {S.}~\bibnamefont {{Gillessen}}}, \bibinfo {author} {\bibfnamefont {P.}~\bibnamefont {{Goret}}}, \bibinfo {author} {\bibfnamefont {C.}~\bibnamefont {{Hadjichristidis}}}, \bibinfo {author} {\bibfnamefont {M.}~\bibnamefont {{Hauser}}}, \bibinfo {author} {\bibfnamefont {G.}~\bibnamefont {{Heinzelmann}}}, \bibinfo {author} {\bibfnamefont {G.}~\bibnamefont {{Henri}}}, \bibinfo {author} {\bibfnamefont {G.}~\bibnamefont {{Hermann}}}, \bibinfo {author} {\bibfnamefont {J.~A.}\ \bibnamefont {{Hinton}}}, \bibinfo {author} {\bibfnamefont {W.}~\bibnamefont {{Hofmann}}}, \bibinfo {author} {\bibfnamefont {M.}~\bibnamefont {{Holleran}}}, \bibinfo {author} {\bibfnamefont {D.}~\bibnamefont {{Horns}}}, \bibinfo {author} {\bibfnamefont {O.~C.}\ \bibnamefont {{de Jager}}}, \bibinfo {author} {\bibfnamefont {I.}~\bibnamefont {{Jung}}}, \bibinfo {author} {\bibfnamefont
  {B.}~\bibnamefont {{Kh{\'e}lifi}}}, \bibinfo {author} {\bibfnamefont {N.}~\bibnamefont {{Komin}}}, \bibinfo {author} {\bibfnamefont {A.}~\bibnamefont {{Konopelko}}}, \bibinfo {author} {\bibfnamefont {I.~J.}\ \bibnamefont {{Latham}}}, \bibinfo {author} {\bibfnamefont {R.}~\bibnamefont {{Le Gallou}}}, \bibinfo {author} {\bibfnamefont {A.}~\bibnamefont {{Lemi{\`e}re}}}, \bibinfo {author} {\bibfnamefont {M.}~\bibnamefont {{Lemoine}}}, \bibinfo {author} {\bibfnamefont {N.}~\bibnamefont {{Leroy}}}, \bibinfo {author} {\bibfnamefont {T.}~\bibnamefont {{Lohse}}}, \bibinfo {author} {\bibfnamefont {A.}~\bibnamefont {{Marcowith}}}, \bibinfo {author} {\bibfnamefont {C.}~\bibnamefont {{Masterson}}}, \bibinfo {author} {\bibfnamefont {T.~J.~L.}\ \bibnamefont {{McComb}}}, \bibinfo {author} {\bibfnamefont {M.}~\bibnamefont {{de Naurois}}}, \bibinfo {author} {\bibfnamefont {S.~J.}\ \bibnamefont {{Nolan}}}, \bibinfo {author} {\bibfnamefont {A.}~\bibnamefont {{Noutsos}}}, \bibinfo {author} {\bibfnamefont {K.~J.}\ \bibnamefont
  {{Orford}}}, \bibinfo {author} {\bibfnamefont {J.~L.}\ \bibnamefont {{Osborne}}}, \bibinfo {author} {\bibfnamefont {M.}~\bibnamefont {{Ouchrif}}}, \bibinfo {author} {\bibfnamefont {M.}~\bibnamefont {{Panter}}}, \bibinfo {author} {\bibfnamefont {G.}~\bibnamefont {{Pelletier}}}, \bibinfo {author} {\bibfnamefont {S.}~\bibnamefont {{Pita}}}, \bibinfo {author} {\bibfnamefont {G.}~\bibnamefont {{P{\"u}hlhofer}}}, \bibinfo {author} {\bibfnamefont {M.}~\bibnamefont {{Punch}}}, \bibinfo {author} {\bibfnamefont {B.~C.}\ \bibnamefont {{Raubenheimer}}}, \bibinfo {author} {\bibfnamefont {M.}~\bibnamefont {{Raue}}}, \bibinfo {author} {\bibfnamefont {J.}~\bibnamefont {{Raux}}}, \bibinfo {author} {\bibfnamefont {S.~M.}\ \bibnamefont {{Rayner}}}, \bibinfo {author} {\bibfnamefont {I.}~\bibnamefont {{Redondo}}}, \bibinfo {author} {\bibfnamefont {A.}~\bibnamefont {{Reimer}}}, \bibinfo {author} {\bibfnamefont {O.}~\bibnamefont {{Reimer}}}, \bibinfo {author} {\bibfnamefont {J.}~\bibnamefont {{Ripken}}}, \bibinfo {author}
  {\bibfnamefont {L.}~\bibnamefont {{Rob}}}, \bibinfo {author} {\bibfnamefont {L.}~\bibnamefont {{Rolland}}}, \bibinfo {author} {\bibfnamefont {G.}~\bibnamefont {{Rowell}}}, \bibinfo {author} {\bibfnamefont {V.}~\bibnamefont {{Sahakian}}}, \bibinfo {author} {\bibfnamefont {L.}~\bibnamefont {{Saug{\'e}}}}, \bibinfo {author} {\bibfnamefont {S.}~\bibnamefont {{Schlenker}}}, \bibinfo {author} {\bibfnamefont {R.}~\bibnamefont {{Schlickeiser}}}, \bibinfo {author} {\bibfnamefont {C.}~\bibnamefont {{Schuster}}}, \bibinfo {author} {\bibfnamefont {U.}~\bibnamefont {{Schwanke}}}, \bibinfo {author} {\bibfnamefont {M.}~\bibnamefont {{Siewert}}}, \bibinfo {author} {\bibfnamefont {H.}~\bibnamefont {{Sol}}}, \bibinfo {author} {\bibfnamefont {R.}~\bibnamefont {{Steenkamp}}}, \bibinfo {author} {\bibfnamefont {C.}~\bibnamefont {{Stegmann}}}, \bibinfo {author} {\bibfnamefont {J.-P.}\ \bibnamefont {{Tavernet}}}, \bibinfo {author} {\bibfnamefont {R.}~\bibnamefont {{Terrier}}}, \bibinfo {author} {\bibfnamefont {C.~G.}\ \bibnamefont
  {{Th{\'e}oret}}}, \bibinfo {author} {\bibfnamefont {M.}~\bibnamefont {{Tluczykont}}}, \bibinfo {author} {\bibfnamefont {D.~J.}\ \bibnamefont {{van der Walt}}}, \bibinfo {author} {\bibfnamefont {G.}~\bibnamefont {{Vasileiadis}}}, \bibinfo {author} {\bibfnamefont {C.}~\bibnamefont {{Venter}}}, \bibinfo {author} {\bibfnamefont {P.}~\bibnamefont {{Vincent}}}, \bibinfo {author} {\bibfnamefont {B.}~\bibnamefont {{Visser}}}, \bibinfo {author} {\bibfnamefont {H.~J.}\ \bibnamefont {{V{\"o}lk}}},\ and\ \bibinfo {author} {\bibfnamefont {S.~J.}\ \bibnamefont {{Wagner}}},\ }\bibfield  {title} {\bibinfo {title} {A new population of very high energy gamma-ray sources in the milky way},\ }\href {https://doi.org/10.1126/science.1108643} {\bibfield  {journal} {\bibinfo  {journal} {Science}\ }\textbf {\bibinfo {volume} {307}},\ \bibinfo {pages} {1938} (\bibinfo {year} {2005})}\BibitemShut {NoStop}%
\bibitem [{\citenamefont {Lorimer}(2008)}]{Lorimer_2008}%
  \BibitemOpen
  \bibfield  {author} {\bibinfo {author} {\bibfnamefont {D.~R.}\ \bibnamefont {Lorimer}},\ }\bibfield  {title} {\bibinfo {title} {{Binary and Millisecond Pulsars}},\ }\href {https://doi.org/10.12942/lrr-2008-8} {\bibfield  {journal} {\bibinfo  {journal} {Living Rev. Rel.}\ }\textbf {\bibinfo {volume} {11}},\ \bibinfo {pages} {8} (\bibinfo {year} {2008})},\ \Eprint {https://arxiv.org/abs/0811.0762} {arXiv:0811.0762 [astro-ph]} \BibitemShut {NoStop}%
\bibitem [{\citenamefont {Wang}\ \emph {et~al.}(2006)\citenamefont {Wang}, \citenamefont {Chakrabarty},\ and\ \citenamefont {Kaplan}}]{Wang_2006}%
  \BibitemOpen
  \bibfield  {author} {\bibinfo {author} {\bibfnamefont {Z.}~\bibnamefont {Wang}}, \bibinfo {author} {\bibfnamefont {D.}~\bibnamefont {Chakrabarty}},\ and\ \bibinfo {author} {\bibfnamefont {D.~L.}\ \bibnamefont {Kaplan}},\ }\bibfield  {title} {\bibinfo {title} {A debris disk around an isolated young neutron star},\ }\href {https://doi.org/10.1038/nature04669} {\bibfield  {journal} {\bibinfo  {journal} {Nature}\ }\textbf {\bibinfo {volume} {440}},\ \bibinfo {pages} {772–775} (\bibinfo {year} {2006})}\BibitemShut {NoStop}%
\bibitem [{\citenamefont {Bailes}\ \emph {et~al.}(2011)\citenamefont {Bailes}, \citenamefont {Bates}, \citenamefont {Bhalerao}, \citenamefont {Bhat}, \citenamefont {Burgay}, \citenamefont {Burke-Spolaor}, \citenamefont {D’Amico}, \citenamefont {Johnston}, \citenamefont {Keith}, \citenamefont {Kramer}, \citenamefont {Kulkarni}, \citenamefont {Levin}, \citenamefont {Lyne}, \citenamefont {Milia}, \citenamefont {Possenti}, \citenamefont {Spitler}, \citenamefont {Stappers},\ and\ \citenamefont {van Straten}}]{Bailes_2011}%
  \BibitemOpen
  \bibfield  {author} {\bibinfo {author} {\bibfnamefont {M.}~\bibnamefont {Bailes}}, \bibinfo {author} {\bibfnamefont {S.~D.}\ \bibnamefont {Bates}}, \bibinfo {author} {\bibfnamefont {V.}~\bibnamefont {Bhalerao}}, \bibinfo {author} {\bibfnamefont {N.~D.~R.}\ \bibnamefont {Bhat}}, \bibinfo {author} {\bibfnamefont {M.}~\bibnamefont {Burgay}}, \bibinfo {author} {\bibfnamefont {S.}~\bibnamefont {Burke-Spolaor}}, \bibinfo {author} {\bibfnamefont {N.}~\bibnamefont {D’Amico}}, \bibinfo {author} {\bibfnamefont {S.}~\bibnamefont {Johnston}}, \bibinfo {author} {\bibfnamefont {M.~J.}\ \bibnamefont {Keith}}, \bibinfo {author} {\bibfnamefont {M.}~\bibnamefont {Kramer}}, \bibinfo {author} {\bibfnamefont {S.~R.}\ \bibnamefont {Kulkarni}}, \bibinfo {author} {\bibfnamefont {L.}~\bibnamefont {Levin}}, \bibinfo {author} {\bibfnamefont {A.~G.}\ \bibnamefont {Lyne}}, \bibinfo {author} {\bibfnamefont {S.}~\bibnamefont {Milia}}, \bibinfo {author} {\bibfnamefont {A.}~\bibnamefont {Possenti}}, \bibinfo {author} {\bibfnamefont
  {L.}~\bibnamefont {Spitler}}, \bibinfo {author} {\bibfnamefont {B.}~\bibnamefont {Stappers}},\ and\ \bibinfo {author} {\bibfnamefont {W.}~\bibnamefont {van Straten}},\ }\bibfield  {title} {\bibinfo {title} {Transformation of a star into a planet in a millisecond pulsar binary},\ }\href {https://doi.org/10.1126/science.1208890} {\bibfield  {journal} {\bibinfo  {journal} {Science}\ }\textbf {\bibinfo {volume} {333}},\ \bibinfo {pages} {1717–1720} (\bibinfo {year} {2011})}\BibitemShut {NoStop}%
\bibitem [{\citenamefont {{Wolszczan}}\ and\ \citenamefont {{Frail}}(1992)}]{Wolszczan1992}%
  \BibitemOpen
  \bibfield  {author} {\bibinfo {author} {\bibfnamefont {A.}~\bibnamefont {{Wolszczan}}}\ and\ \bibinfo {author} {\bibfnamefont {D.~A.}\ \bibnamefont {{Frail}}},\ }\bibfield  {title} {\bibinfo {title} {{A planetary system around the millisecond pulsar PSR1257 + 12}},\ }\href {https://doi.org/10.1038/355145a0} {\bibfield  {journal} {\bibinfo  {journal} {\nat}\ }\textbf {\bibinfo {volume} {355}},\ \bibinfo {pages} {145} (\bibinfo {year} {1992})}\BibitemShut {NoStop}%
\bibitem [{\citenamefont {{Spiewak}}\ \emph {et~al.}(2018)\citenamefont {{Spiewak}}, \citenamefont {{Bailes}}, \citenamefont {{Barr}}, \citenamefont {{Bhat}}, \citenamefont {{Burgay}}, \citenamefont {{Cameron}}, \citenamefont {{Champion}}, \citenamefont {{Flynn}}, \citenamefont {{Jameson}}, \citenamefont {{Johnston}}, \citenamefont {{Keith}}, \citenamefont {{Kramer}}, \citenamefont {{Kulkarni}}, \citenamefont {{Levin}}, \citenamefont {{Lyne}}, \citenamefont {{Morello}}, \citenamefont {{Ng}}, \citenamefont {{Possenti}}, \citenamefont {{Ravi}}, \citenamefont {{Stappers}}, \citenamefont {{van Straten}},\ and\ \citenamefont {{Tiburzi}}}]{Spiewak2018}%
  \BibitemOpen
  \bibfield  {author} {\bibinfo {author} {\bibfnamefont {R.}~\bibnamefont {{Spiewak}}}, \bibinfo {author} {\bibfnamefont {M.}~\bibnamefont {{Bailes}}}, \bibinfo {author} {\bibfnamefont {E.~D.}\ \bibnamefont {{Barr}}}, \bibinfo {author} {\bibfnamefont {N.~D.~R.}\ \bibnamefont {{Bhat}}}, \bibinfo {author} {\bibfnamefont {M.}~\bibnamefont {{Burgay}}}, \bibinfo {author} {\bibfnamefont {A.~D.}\ \bibnamefont {{Cameron}}}, \bibinfo {author} {\bibfnamefont {D.~J.}\ \bibnamefont {{Champion}}}, \bibinfo {author} {\bibfnamefont {C.~M.~L.}\ \bibnamefont {{Flynn}}}, \bibinfo {author} {\bibfnamefont {A.}~\bibnamefont {{Jameson}}}, \bibinfo {author} {\bibfnamefont {S.}~\bibnamefont {{Johnston}}}, \bibinfo {author} {\bibfnamefont {M.~J.}\ \bibnamefont {{Keith}}}, \bibinfo {author} {\bibfnamefont {M.}~\bibnamefont {{Kramer}}}, \bibinfo {author} {\bibfnamefont {S.~R.}\ \bibnamefont {{Kulkarni}}}, \bibinfo {author} {\bibfnamefont {L.}~\bibnamefont {{Levin}}}, \bibinfo {author} {\bibfnamefont {A.~G.}\ \bibnamefont {{Lyne}}},
  \bibinfo {author} {\bibfnamefont {V.}~\bibnamefont {{Morello}}}, \bibinfo {author} {\bibfnamefont {C.}~\bibnamefont {{Ng}}}, \bibinfo {author} {\bibfnamefont {A.}~\bibnamefont {{Possenti}}}, \bibinfo {author} {\bibfnamefont {V.}~\bibnamefont {{Ravi}}}, \bibinfo {author} {\bibfnamefont {B.~W.}\ \bibnamefont {{Stappers}}}, \bibinfo {author} {\bibfnamefont {W.}~\bibnamefont {{van Straten}}},\ and\ \bibinfo {author} {\bibfnamefont {C.}~\bibnamefont {{Tiburzi}}},\ }\bibfield  {title} {\bibinfo {title} {{PSR J2322-2650 - a low-luminosity millisecond pulsar with a planetary-mass companion}},\ }\href {https://doi.org/10.1093/mnras/stx3157} {\bibfield  {journal} {\bibinfo  {journal} {\mnras}\ }\textbf {\bibinfo {volume} {475}},\ \bibinfo {pages} {469} (\bibinfo {year} {2018})},\ \Eprint {https://arxiv.org/abs/1712.04445} {arXiv:1712.04445 [astro-ph.HE]} \BibitemShut {NoStop}%
\bibitem [{\citenamefont {Behrens}\ \emph {et~al.}(2020)\citenamefont {Behrens}, \citenamefont {Ransom}, \citenamefont {Madison}, \citenamefont {Arzoumanian}, \citenamefont {Crowter}, \citenamefont {DeCesar}, \citenamefont {Demorest}, \citenamefont {Dolch}, \citenamefont {Ellis}, \citenamefont {Ferdman}, \citenamefont {Ferrara}, \citenamefont {Fonseca}, \citenamefont {Gentile}, \citenamefont {Jones}, \citenamefont {Jones}, \citenamefont {Lam}, \citenamefont {Levin}, \citenamefont {Lorimer}, \citenamefont {Lynch}, \citenamefont {McLaughlin}, \citenamefont {Ng}, \citenamefont {Nice}, \citenamefont {Pennucci}, \citenamefont {Perera}, \citenamefont {Ray}, \citenamefont {Spiewak}, \citenamefont {Stairs}, \citenamefont {Stovall}, \citenamefont {Swiggum},\ and\ \citenamefont {Zhu}}]{Behrens2020}%
  \BibitemOpen
  \bibfield  {author} {\bibinfo {author} {\bibfnamefont {E.~A.}\ \bibnamefont {Behrens}}, \bibinfo {author} {\bibfnamefont {S.~M.}\ \bibnamefont {Ransom}}, \bibinfo {author} {\bibfnamefont {D.~R.}\ \bibnamefont {Madison}}, \bibinfo {author} {\bibfnamefont {Z.}~\bibnamefont {Arzoumanian}}, \bibinfo {author} {\bibfnamefont {K.}~\bibnamefont {Crowter}}, \bibinfo {author} {\bibfnamefont {M.~E.}\ \bibnamefont {DeCesar}}, \bibinfo {author} {\bibfnamefont {P.~B.}\ \bibnamefont {Demorest}}, \bibinfo {author} {\bibfnamefont {T.}~\bibnamefont {Dolch}}, \bibinfo {author} {\bibfnamefont {J.~A.}\ \bibnamefont {Ellis}}, \bibinfo {author} {\bibfnamefont {R.~D.}\ \bibnamefont {Ferdman}}, \bibinfo {author} {\bibfnamefont {E.~C.}\ \bibnamefont {Ferrara}}, \bibinfo {author} {\bibfnamefont {E.}~\bibnamefont {Fonseca}}, \bibinfo {author} {\bibfnamefont {P.~A.}\ \bibnamefont {Gentile}}, \bibinfo {author} {\bibfnamefont {G.}~\bibnamefont {Jones}}, \bibinfo {author} {\bibfnamefont {M.~L.}\ \bibnamefont {Jones}}, \bibinfo {author}
  {\bibfnamefont {M.~T.}\ \bibnamefont {Lam}}, \bibinfo {author} {\bibfnamefont {L.}~\bibnamefont {Levin}}, \bibinfo {author} {\bibfnamefont {D.~R.}\ \bibnamefont {Lorimer}}, \bibinfo {author} {\bibfnamefont {R.~S.}\ \bibnamefont {Lynch}}, \bibinfo {author} {\bibfnamefont {M.~A.}\ \bibnamefont {McLaughlin}}, \bibinfo {author} {\bibfnamefont {C.}~\bibnamefont {Ng}}, \bibinfo {author} {\bibfnamefont {D.~J.}\ \bibnamefont {Nice}}, \bibinfo {author} {\bibfnamefont {T.~T.}\ \bibnamefont {Pennucci}}, \bibinfo {author} {\bibfnamefont {B.~B.~P.}\ \bibnamefont {Perera}}, \bibinfo {author} {\bibfnamefont {P.~S.}\ \bibnamefont {Ray}}, \bibinfo {author} {\bibfnamefont {R.}~\bibnamefont {Spiewak}}, \bibinfo {author} {\bibfnamefont {I.~H.}\ \bibnamefont {Stairs}}, \bibinfo {author} {\bibfnamefont {K.}~\bibnamefont {Stovall}}, \bibinfo {author} {\bibfnamefont {J.~K.}\ \bibnamefont {Swiggum}},\ and\ \bibinfo {author} {\bibfnamefont {W.~W.}\ \bibnamefont {Zhu}},\ }\bibfield  {title} {\bibinfo {title} {The nanograv 11 yr data
  set: Constraints on planetary masses around 45 millisecond pulsars},\ }\href {https://doi.org/10.3847/2041-8213/ab8121} {\bibfield  {journal} {\bibinfo  {journal} {The Astrophysical Journal Letters}\ }\textbf {\bibinfo {volume} {893}},\ \bibinfo {pages} {L8} (\bibinfo {year} {2020})}\BibitemShut {NoStop}%
\bibitem [{\citenamefont {Niţu}\ \emph {et~al.}(2022)\citenamefont {Niţu}, \citenamefont {Keith}, \citenamefont {Stappers}, \citenamefont {Lyne},\ and\ \citenamefont {Mickaliger}}]{Ni_u_2022}%
  \BibitemOpen
  \bibfield  {author} {\bibinfo {author} {\bibfnamefont {I.~C.}\ \bibnamefont {Niţu}}, \bibinfo {author} {\bibfnamefont {M.~J.}\ \bibnamefont {Keith}}, \bibinfo {author} {\bibfnamefont {B.~W.}\ \bibnamefont {Stappers}}, \bibinfo {author} {\bibfnamefont {A.~G.}\ \bibnamefont {Lyne}},\ and\ \bibinfo {author} {\bibfnamefont {M.~B.}\ \bibnamefont {Mickaliger}},\ }\bibfield  {title} {\bibinfo {title} {A search for planetary companions around 800 pulsars from the jodrell bank pulsar timing programme},\ }\href {https://doi.org/10.1093/mnras/stac593} {\bibfield  {journal} {\bibinfo  {journal} {Monthly Notices of the Royal Astronomical Society}\ }\textbf {\bibinfo {volume} {512}},\ \bibinfo {pages} {2446–2459} (\bibinfo {year} {2022})}\BibitemShut {NoStop}%
\bibitem [{\citenamefont {Dunn}\ \emph {et~al.}(2025)\citenamefont {Dunn}, \citenamefont {Melatos}, \citenamefont {Clearwater}, \citenamefont {Suvorova}, \citenamefont {Sun}, \citenamefont {Moran},\ and\ \citenamefont {Evans}}]{Dunn2025}%
  \BibitemOpen
  \bibfield  {author} {\bibinfo {author} {\bibfnamefont {L.}~\bibnamefont {Dunn}}, \bibinfo {author} {\bibfnamefont {A.}~\bibnamefont {Melatos}}, \bibinfo {author} {\bibfnamefont {P.}~\bibnamefont {Clearwater}}, \bibinfo {author} {\bibfnamefont {S.}~\bibnamefont {Suvorova}}, \bibinfo {author} {\bibfnamefont {L.}~\bibnamefont {Sun}}, \bibinfo {author} {\bibfnamefont {W.}~\bibnamefont {Moran}},\ and\ \bibinfo {author} {\bibfnamefont {R.~J.}\ \bibnamefont {Evans}},\ }\bibfield  {title} {\bibinfo {title} {Search for continuous gravitational waves from neutron stars in five globular clusters with a phase-tracking hidden markov model in the third ligo observing run},\ }\href {https://doi.org/10.1103/PhysRevD.111.083012} {\bibfield  {journal} {\bibinfo  {journal} {Phys. Rev. D}\ }\textbf {\bibinfo {volume} {111}},\ \bibinfo {pages} {083012} (\bibinfo {year} {2025})}\BibitemShut {NoStop}%
\bibitem [{\citenamefont {Aasi}\ \emph {et~al.}(2015{\natexlab{a}})\citenamefont {Aasi} \emph {et~al.}}]{LIGO}%
  \BibitemOpen
  \bibfield  {author} {\bibinfo {author} {\bibfnamefont {J.}~\bibnamefont {Aasi}} \emph {et~al.},\ }\bibfield  {title} {\bibinfo {title} {Advanced ligo},\ }\href@noop {} {\bibfield  {journal} {\bibinfo  {journal} {Class. Quantum Grav.}\ }\textbf {\bibinfo {volume} {32}},\ \bibinfo {pages} {7} (\bibinfo {year} {2015}{\natexlab{a}})}\BibitemShut {NoStop}%
\bibitem [{\citenamefont {Abbott}\ \emph {et~al.}(2016)\citenamefont {Abbott} \emph {et~al.}}]{DetectorPaper}%
  \BibitemOpen
  \bibfield  {author} {\bibinfo {author} {\bibfnamefont {B.}~\bibnamefont {Abbott}} \emph {et~al.},\ }\bibfield  {title} {\bibinfo {title} {Gw150914: The advanced ligo detectors in the era of first discoveries},\ }\href@noop {} {\bibfield  {journal} {\bibinfo  {journal} {Phys. Rev. Lett.}\ }\textbf {\bibinfo {volume} {116}},\ \bibinfo {pages} {131103} (\bibinfo {year} {2016})}\BibitemShut {NoStop}%
\bibitem [{\citenamefont {Abac}\ \emph {et~al.}(2025)\citenamefont {Abac}, \citenamefont {Abouelfettouh}, \citenamefont {Acernese}, \citenamefont {Ackley}, \citenamefont {Adamcewicz}, \citenamefont {Adhicary}, \citenamefont {Adhikari}, \citenamefont {Adhikari}, \citenamefont {Adhikari}, \citenamefont {Adkins}, \citenamefont {Afroz}, \citenamefont {Agapito}, \citenamefont {Agarwal},\ and\ \citenamefont {et~al.}}]{o4a_data}%
  \BibitemOpen
  \bibfield  {author} {\bibinfo {author} {\bibfnamefont {A.~G.}\ \bibnamefont {Abac}}, \bibinfo {author} {\bibfnamefont {I.}~\bibnamefont {Abouelfettouh}}, \bibinfo {author} {\bibfnamefont {F.}~\bibnamefont {Acernese}}, \bibinfo {author} {\bibfnamefont {K.}~\bibnamefont {Ackley}}, \bibinfo {author} {\bibfnamefont {C.}~\bibnamefont {Adamcewicz}}, \bibinfo {author} {\bibfnamefont {S.}~\bibnamefont {Adhicary}}, \bibinfo {author} {\bibfnamefont {D.}~\bibnamefont {Adhikari}}, \bibinfo {author} {\bibfnamefont {N.}~\bibnamefont {Adhikari}}, \bibinfo {author} {\bibfnamefont {R.~X.}\ \bibnamefont {Adhikari}}, \bibinfo {author} {\bibfnamefont {V.~K.}\ \bibnamefont {Adkins}}, \bibinfo {author} {\bibfnamefont {S.}~\bibnamefont {Afroz}}, \bibinfo {author} {\bibfnamefont {A.}~\bibnamefont {Agapito}}, \bibinfo {author} {\bibfnamefont {D.}~\bibnamefont {Agarwal}},\ and\ \bibinfo {author} {\bibnamefont {et~al.}},\ }\href {https://arxiv.org/abs/2508.18079} {\bibinfo {title} {Open data from ligo, virgo, and kagra through the
  first part of the fourth observing run}} (\bibinfo {year} {2025}),\ \Eprint {https://arxiv.org/abs/2508.18079} {arXiv:2508.18079 [gr-qc]} \BibitemShut {NoStop}%
\bibitem [{\citenamefont {Capote}\ \emph {et~al.}(2025)\citenamefont {Capote}, \citenamefont {Jia}, \citenamefont {Aritomi}, \citenamefont {Nakano}, \citenamefont {Xu}, \citenamefont {Abbott},\ and\ \citenamefont {et~al.}}]{Capote_2025}%
  \BibitemOpen
  \bibfield  {author} {\bibinfo {author} {\bibfnamefont {E.}~\bibnamefont {Capote}}, \bibinfo {author} {\bibfnamefont {W.}~\bibnamefont {Jia}}, \bibinfo {author} {\bibfnamefont {N.}~\bibnamefont {Aritomi}}, \bibinfo {author} {\bibfnamefont {M.}~\bibnamefont {Nakano}}, \bibinfo {author} {\bibfnamefont {V.}~\bibnamefont {Xu}}, \bibinfo {author} {\bibfnamefont {R.}~\bibnamefont {Abbott}},\ and\ \bibinfo {author} {\bibnamefont {et~al.}},\ }\bibfield  {title} {\bibinfo {title} {Advanced ligo detector performance in the fourth observing run},\ }\bibfield  {journal} {\bibinfo  {journal} {Physical Review D}\ }\textbf {\bibinfo {volume} {111}},\ \href {https://doi.org/10.1103/physrevd.111.062002} {10.1103/physrevd.111.062002} (\bibinfo {year} {2025})\BibitemShut {NoStop}%
\bibitem [{\citenamefont {Jaranowski}\ \emph {et~al.}(1998{\natexlab{a}})\citenamefont {Jaranowski}, \citenamefont {Krolak},\ and\ \citenamefont {Schutz}}]{JKS}%
  \BibitemOpen
  \bibfield  {author} {\bibinfo {author} {\bibfnamefont {P.}~\bibnamefont {Jaranowski}}, \bibinfo {author} {\bibfnamefont {A.}~\bibnamefont {Krolak}},\ and\ \bibinfo {author} {\bibfnamefont {B.}~\bibnamefont {Schutz}},\ }\bibfield  {title} {\bibinfo {title} {Data analysis of gravitational-wave signals from spinning neutron stars. 1. the signal and its detection},\ }\href@noop {} {\bibfield  {journal} {\bibinfo  {journal} {Phys. Rev. D}\ }\textbf {\bibinfo {volume} {58}},\ \bibinfo {pages} {063001} (\bibinfo {year} {1998}{\natexlab{a}})}\BibitemShut {NoStop}%
\bibitem [{\citenamefont {Abbott}\ \emph {et~al.}(2017)\citenamefont {Abbott} \emph {et~al.}}]{LVC2017GC}%
  \BibitemOpen
  \bibfield  {author} {\bibinfo {author} {\bibfnamefont {B.~P.}\ \bibnamefont {Abbott}} \emph {et~al.} (\bibinfo {collaboration} {LIGO Scientific Collaboration and Virgo Collaboration}),\ }\bibfield  {title} {\bibinfo {title} {Search for continuous gravitational waves from neutron stars in globular cluster ngc 6544},\ }\href {https://doi.org/10.1103/PhysRevD.95.082005} {\bibfield  {journal} {\bibinfo  {journal} {Phys. Rev. D}\ }\textbf {\bibinfo {volume} {95}},\ \bibinfo {pages} {082005} (\bibinfo {year} {2017})}\BibitemShut {NoStop}%
\bibitem [{\citenamefont {Aasi}\ \emph {et~al.}(2015{\natexlab{b}})\citenamefont {Aasi} \emph {et~al.}}]{LIGO2015}%
  \BibitemOpen
  \bibfield  {author} {\bibinfo {author} {\bibfnamefont {J.}~\bibnamefont {Aasi}} \emph {et~al.} (\bibinfo {collaboration} {LSC}),\ }\bibfield  {title} {\bibinfo {title} {{Advanced LIGO}},\ }\href {https://doi.org/10.1088/0264-9381/32/7/074001} {\bibfield  {journal} {\bibinfo  {journal} {Classical and Quantum Gravity}\ }\textbf {\bibinfo {volume} {32}},\ \bibinfo {pages} {074001} (\bibinfo {year} {2015}{\natexlab{b}})}\BibitemShut {NoStop}%
\bibitem [{\citenamefont {Goetz}\ and\ \citenamefont {Riles}(2025)}]{science_mode_O4a}%
  \BibitemOpen
  \bibfield  {author} {\bibinfo {author} {\bibfnamefont {E.}~\bibnamefont {Goetz}}\ and\ \bibinfo {author} {\bibfnamefont {K.}~\bibnamefont {Riles}},\ }\href {https://dcc.ligo.org/T2400058-v2/public} {\bibinfo {title} {{Segments used for creating standard SFTs in O4 data}}} (\bibinfo {year} {2025})\BibitemShut {NoStop}%
\bibitem [{\citenamefont {Acernese}\ \emph {et~al.}(2015)\citenamefont {Acernese} \emph {et~al.}}]{Virgo2014}%
  \BibitemOpen
  \bibfield  {author} {\bibinfo {author} {\bibfnamefont {F.}~\bibnamefont {Acernese}} \emph {et~al.} (\bibinfo {collaboration} {Virgo}),\ }\bibfield  {title} {\bibinfo {title} {{Advanced Virgo: a second-generation interferometric gravitational wave detector}},\ }\href {https://doi.org/10.1088/0264-9381/32/2/024001} {\bibfield  {journal} {\bibinfo  {journal} {Classical and Quantum Gravity}\ }\textbf {\bibinfo {volume} {32}},\ \bibinfo {pages} {024001} (\bibinfo {year} {2015})}\BibitemShut {NoStop}%
\bibitem [{\citenamefont {Akutsu}\ \emph {et~al.}(2020)\citenamefont {Akutsu}, \citenamefont {Ando}, \citenamefont {Arai}, \citenamefont {Arai}, \citenamefont {Araki}, \citenamefont {Araya}, \citenamefont {Aritomi}, \citenamefont {Aso}, \citenamefont {Bae}, \citenamefont {Bae}, \citenamefont {Baiotti}, \citenamefont {Bajpai}, \citenamefont {Barton}, \citenamefont {Cannon}, \citenamefont {Capocasa}, \citenamefont {Chan}, \citenamefont {Chen}, \citenamefont {Chen}, \citenamefont {Chen}, \citenamefont {Chu}, \citenamefont {Chu}, \citenamefont {Eguchi}, \citenamefont {Enomoto}, \citenamefont {Flaminio}, \citenamefont {Fujii}, \citenamefont {Fukunaga}, \citenamefont {Fukushima}, \citenamefont {Ge}, \citenamefont {Hagiwara}, \citenamefont {Haino}, \citenamefont {Hasegawa}, \citenamefont {Hayakawa}, \citenamefont {Hayama}, \citenamefont {Himemoto}, \citenamefont {Hiranuma}, \citenamefont {Hirata}, \citenamefont {Hirose}, \citenamefont {Hong}, \citenamefont {Hsieh}, \citenamefont {Huang}, \citenamefont {Huang},
  \citenamefont {Huang}, \citenamefont {Ikenoue}, \citenamefont {Imam}, \citenamefont {Inayoshi}, \citenamefont {Inoue}, \citenamefont {Ioka}, \citenamefont {Itoh}, \citenamefont {Izumi}, \citenamefont {Jung}, \citenamefont {Jung}, \citenamefont {Kajita}, \citenamefont {Kamiizumi}, \citenamefont {Kanda}, \citenamefont {Kang}, \citenamefont {Kawaguchi}, \citenamefont {Kawai}, \citenamefont {Kawasaki}, \citenamefont {Kim}, \citenamefont {Kim}, \citenamefont {Kim}, \citenamefont {Kim}, \citenamefont {Kimura}, \citenamefont {Kita}, \citenamefont {Kitazawa}, \citenamefont {Kojima}, \citenamefont {Kokeyama}, \citenamefont {Komori}, \citenamefont {Kong}, \citenamefont {Kotake}, \citenamefont {Kozakai}, \citenamefont {Kozu}, \citenamefont {Kumar}, \citenamefont {Kume}, \citenamefont {Kuo}, \citenamefont {Kuo}, \citenamefont {Kuroyanagi}, \citenamefont {Kusayanagi}, \citenamefont {Kwak}, \citenamefont {Lee}, \citenamefont {Lee}, \citenamefont {Lee}, \citenamefont {Leonardi}, \citenamefont {Lin}, \citenamefont {Lin},
  \citenamefont {Lin}, \citenamefont {Liu}, \citenamefont {Luo}, \citenamefont {Marchio}, \citenamefont {Michimura}, \citenamefont {Mio}, \citenamefont {Miyakawa}, \citenamefont {Miyamoto}, \citenamefont {Miyazaki}, \citenamefont {Miyo}, \citenamefont {Miyoki}, \citenamefont {Morisaki}, \citenamefont {Moriwaki}, \citenamefont {Nagano}, \citenamefont {Nagano}, \citenamefont {Nakamura}, \citenamefont {Nakano}, \citenamefont {Nakano}, \citenamefont {Nakashima}, \citenamefont {Narikawa}, \citenamefont {Negishi}, \citenamefont {Ni}, \citenamefont {Nishizawa}, \citenamefont {Obuchi}, \citenamefont {Ogaki}, \citenamefont {Oh}, \citenamefont {Oh}, \citenamefont {Ohashi}, \citenamefont {Ohishi}, \citenamefont {Ohkawa}, \citenamefont {Okutomi}, \citenamefont {Oohara}, \citenamefont {Ooi}, \citenamefont {Oshino}, \citenamefont {Pan}, \citenamefont {Pang}, \citenamefont {Park}, \citenamefont {Arellano}, \citenamefont {Pinto}, \citenamefont {Sago}, \citenamefont {Saito}, \citenamefont {Saito}, \citenamefont {Sakai},
  \citenamefont {Sakai}, \citenamefont {Sakuno}, \citenamefont {Sato}, \citenamefont {Sato}, \citenamefont {Sawada}, \citenamefont {Sekiguchi}, \citenamefont {Sekiguchi}, \citenamefont {Shibagaki}, \citenamefont {Shimizu}, \citenamefont {Shimoda}, \citenamefont {Shimode}, \citenamefont {Shinkai}, \citenamefont {Shishido}, \citenamefont {Shoda}, \citenamefont {Somiya}, \citenamefont {Son}, \citenamefont {Sotani}, \citenamefont {Sugimoto}, \citenamefont {Suzuki}, \citenamefont {Suzuki}, \citenamefont {Tagoshi}, \citenamefont {Takahashi}, \citenamefont {Takahashi}, \citenamefont {Takamori}, \citenamefont {Takano}, \citenamefont {Takeda}, \citenamefont {Takeda}, \citenamefont {Tanaka}, \citenamefont {Tanaka}, \citenamefont {Tanaka}, \citenamefont {Tanaka}, \citenamefont {Tanaka}, \citenamefont {Tanioka}, \citenamefont {Tapia San~Martin}, \citenamefont {Telada}, \citenamefont {Tomaru}, \citenamefont {Tomigami}, \citenamefont {Tomura}, \citenamefont {Travasso}, \citenamefont {Trozzo}, \citenamefont {Tsang},
  \citenamefont {Tsubono}, \citenamefont {Tsuchida}, \citenamefont {Tsuzuki}, \citenamefont {Tuyenbayev}, \citenamefont {Uchikata}, \citenamefont {Uchiyama}, \citenamefont {Ueda}, \citenamefont {Uehara}, \citenamefont {Ueno}, \citenamefont {Ueshima}, \citenamefont {Uraguchi}, \citenamefont {Ushiba}, \citenamefont {van Putten}, \citenamefont {Vocca}, \citenamefont {Wang}, \citenamefont {Wu}, \citenamefont {Wu}, \citenamefont {Wu}, \citenamefont {Xu}, \citenamefont {Yamada}, \citenamefont {Yamamoto}, \citenamefont {Yamamoto}, \citenamefont {Yamamoto}, \citenamefont {Yokogawa}, \citenamefont {Yokoyama}, \citenamefont {Yokozawa}, \citenamefont {Yoshioka}, \citenamefont {Yuzurihara}, \citenamefont {Zeidler}, \citenamefont {Zhao},\ and\ \citenamefont {Zhu}}]{kagra_2021}%
  \BibitemOpen
  \bibfield  {author} {\bibinfo {author} {\bibfnamefont {T.}~\bibnamefont {Akutsu}}, \bibinfo {author} {\bibfnamefont {M.}~\bibnamefont {Ando}}, \bibinfo {author} {\bibfnamefont {K.}~\bibnamefont {Arai}}, \bibinfo {author} {\bibfnamefont {Y.}~\bibnamefont {Arai}}, \bibinfo {author} {\bibfnamefont {S.}~\bibnamefont {Araki}}, \bibinfo {author} {\bibfnamefont {A.}~\bibnamefont {Araya}}, \bibinfo {author} {\bibfnamefont {N.}~\bibnamefont {Aritomi}}, \bibinfo {author} {\bibfnamefont {Y.}~\bibnamefont {Aso}}, \bibinfo {author} {\bibfnamefont {S.}~\bibnamefont {Bae}}, \bibinfo {author} {\bibfnamefont {Y.}~\bibnamefont {Bae}}, \bibinfo {author} {\bibfnamefont {L.}~\bibnamefont {Baiotti}}, \bibinfo {author} {\bibfnamefont {R.}~\bibnamefont {Bajpai}}, \bibinfo {author} {\bibfnamefont {M.~A.}\ \bibnamefont {Barton}}, \bibinfo {author} {\bibfnamefont {K.}~\bibnamefont {Cannon}}, \bibinfo {author} {\bibfnamefont {E.}~\bibnamefont {Capocasa}}, \bibinfo {author} {\bibfnamefont {M.}~\bibnamefont {Chan}}, \bibinfo {author}
  {\bibfnamefont {C.}~\bibnamefont {Chen}}, \bibinfo {author} {\bibfnamefont {K.}~\bibnamefont {Chen}}, \bibinfo {author} {\bibfnamefont {Y.}~\bibnamefont {Chen}}, \bibinfo {author} {\bibfnamefont {H.}~\bibnamefont {Chu}}, \bibinfo {author} {\bibfnamefont {Y.~K.}\ \bibnamefont {Chu}}, \bibinfo {author} {\bibfnamefont {S.}~\bibnamefont {Eguchi}}, \bibinfo {author} {\bibfnamefont {Y.}~\bibnamefont {Enomoto}}, \bibinfo {author} {\bibfnamefont {R.}~\bibnamefont {Flaminio}}, \bibinfo {author} {\bibfnamefont {Y.}~\bibnamefont {Fujii}}, \bibinfo {author} {\bibfnamefont {M.}~\bibnamefont {Fukunaga}}, \bibinfo {author} {\bibfnamefont {M.}~\bibnamefont {Fukushima}}, \bibinfo {author} {\bibfnamefont {G.}~\bibnamefont {Ge}}, \bibinfo {author} {\bibfnamefont {A.}~\bibnamefont {Hagiwara}}, \bibinfo {author} {\bibfnamefont {S.}~\bibnamefont {Haino}}, \bibinfo {author} {\bibfnamefont {K.}~\bibnamefont {Hasegawa}}, \bibinfo {author} {\bibfnamefont {H.}~\bibnamefont {Hayakawa}}, \bibinfo {author} {\bibfnamefont
  {K.}~\bibnamefont {Hayama}}, \bibinfo {author} {\bibfnamefont {Y.}~\bibnamefont {Himemoto}}, \bibinfo {author} {\bibfnamefont {Y.}~\bibnamefont {Hiranuma}}, \bibinfo {author} {\bibfnamefont {N.}~\bibnamefont {Hirata}}, \bibinfo {author} {\bibfnamefont {E.}~\bibnamefont {Hirose}}, \bibinfo {author} {\bibfnamefont {Z.}~\bibnamefont {Hong}}, \bibinfo {author} {\bibfnamefont {B.~H.}\ \bibnamefont {Hsieh}}, \bibinfo {author} {\bibfnamefont {C.~Z.}\ \bibnamefont {Huang}}, \bibinfo {author} {\bibfnamefont {P.}~\bibnamefont {Huang}}, \bibinfo {author} {\bibfnamefont {Y.}~\bibnamefont {Huang}}, \bibinfo {author} {\bibfnamefont {B.}~\bibnamefont {Ikenoue}}, \bibinfo {author} {\bibfnamefont {S.}~\bibnamefont {Imam}}, \bibinfo {author} {\bibfnamefont {K.}~\bibnamefont {Inayoshi}}, \bibinfo {author} {\bibfnamefont {Y.}~\bibnamefont {Inoue}}, \bibinfo {author} {\bibfnamefont {K.}~\bibnamefont {Ioka}}, \bibinfo {author} {\bibfnamefont {Y.}~\bibnamefont {Itoh}}, \bibinfo {author} {\bibfnamefont {K.}~\bibnamefont {Izumi}},
  \bibinfo {author} {\bibfnamefont {K.}~\bibnamefont {Jung}}, \bibinfo {author} {\bibfnamefont {P.}~\bibnamefont {Jung}}, \bibinfo {author} {\bibfnamefont {T.}~\bibnamefont {Kajita}}, \bibinfo {author} {\bibfnamefont {M.}~\bibnamefont {Kamiizumi}}, \bibinfo {author} {\bibfnamefont {N.}~\bibnamefont {Kanda}}, \bibinfo {author} {\bibfnamefont {G.}~\bibnamefont {Kang}}, \bibinfo {author} {\bibfnamefont {K.}~\bibnamefont {Kawaguchi}}, \bibinfo {author} {\bibfnamefont {N.}~\bibnamefont {Kawai}}, \bibinfo {author} {\bibfnamefont {T.}~\bibnamefont {Kawasaki}}, \bibinfo {author} {\bibfnamefont {C.}~\bibnamefont {Kim}}, \bibinfo {author} {\bibfnamefont {J.~C.}\ \bibnamefont {Kim}}, \bibinfo {author} {\bibfnamefont {W.~S.}\ \bibnamefont {Kim}}, \bibinfo {author} {\bibfnamefont {Y.~M.}\ \bibnamefont {Kim}}, \bibinfo {author} {\bibfnamefont {N.}~\bibnamefont {Kimura}}, \bibinfo {author} {\bibfnamefont {N.}~\bibnamefont {Kita}}, \bibinfo {author} {\bibfnamefont {H.}~\bibnamefont {Kitazawa}}, \bibinfo {author}
  {\bibfnamefont {Y.}~\bibnamefont {Kojima}}, \bibinfo {author} {\bibfnamefont {K.}~\bibnamefont {Kokeyama}}, \bibinfo {author} {\bibfnamefont {K.}~\bibnamefont {Komori}}, \bibinfo {author} {\bibfnamefont {A.~K.~H.}\ \bibnamefont {Kong}}, \bibinfo {author} {\bibfnamefont {K.}~\bibnamefont {Kotake}}, \bibinfo {author} {\bibfnamefont {C.}~\bibnamefont {Kozakai}}, \bibinfo {author} {\bibfnamefont {R.}~\bibnamefont {Kozu}}, \bibinfo {author} {\bibfnamefont {R.}~\bibnamefont {Kumar}}, \bibinfo {author} {\bibfnamefont {J.}~\bibnamefont {Kume}}, \bibinfo {author} {\bibfnamefont {C.}~\bibnamefont {Kuo}}, \bibinfo {author} {\bibfnamefont {H.~S.}\ \bibnamefont {Kuo}}, \bibinfo {author} {\bibfnamefont {S.}~\bibnamefont {Kuroyanagi}}, \bibinfo {author} {\bibfnamefont {K.}~\bibnamefont {Kusayanagi}}, \bibinfo {author} {\bibfnamefont {K.}~\bibnamefont {Kwak}}, \bibinfo {author} {\bibfnamefont {H.~K.}\ \bibnamefont {Lee}}, \bibinfo {author} {\bibfnamefont {H.~W.}\ \bibnamefont {Lee}}, \bibinfo {author} {\bibfnamefont
  {R.}~\bibnamefont {Lee}}, \bibinfo {author} {\bibfnamefont {M.}~\bibnamefont {Leonardi}}, \bibinfo {author} {\bibfnamefont {L.~C.~C.}\ \bibnamefont {Lin}}, \bibinfo {author} {\bibfnamefont {C.~Y.}\ \bibnamefont {Lin}}, \bibinfo {author} {\bibfnamefont {F.~L.}\ \bibnamefont {Lin}}, \bibinfo {author} {\bibfnamefont {G.~C.}\ \bibnamefont {Liu}}, \bibinfo {author} {\bibfnamefont {L.~W.}\ \bibnamefont {Luo}}, \bibinfo {author} {\bibfnamefont {M.}~\bibnamefont {Marchio}}, \bibinfo {author} {\bibfnamefont {Y.}~\bibnamefont {Michimura}}, \bibinfo {author} {\bibfnamefont {N.}~\bibnamefont {Mio}}, \bibinfo {author} {\bibfnamefont {O.}~\bibnamefont {Miyakawa}}, \bibinfo {author} {\bibfnamefont {A.}~\bibnamefont {Miyamoto}}, \bibinfo {author} {\bibfnamefont {Y.}~\bibnamefont {Miyazaki}}, \bibinfo {author} {\bibfnamefont {K.}~\bibnamefont {Miyo}}, \bibinfo {author} {\bibfnamefont {S.}~\bibnamefont {Miyoki}}, \bibinfo {author} {\bibfnamefont {S.}~\bibnamefont {Morisaki}}, \bibinfo {author} {\bibfnamefont
  {Y.}~\bibnamefont {Moriwaki}}, \bibinfo {author} {\bibfnamefont {K.}~\bibnamefont {Nagano}}, \bibinfo {author} {\bibfnamefont {S.}~\bibnamefont {Nagano}}, \bibinfo {author} {\bibfnamefont {K.}~\bibnamefont {Nakamura}}, \bibinfo {author} {\bibfnamefont {H.}~\bibnamefont {Nakano}}, \bibinfo {author} {\bibfnamefont {M.}~\bibnamefont {Nakano}}, \bibinfo {author} {\bibfnamefont {R.}~\bibnamefont {Nakashima}}, \bibinfo {author} {\bibfnamefont {T.}~\bibnamefont {Narikawa}}, \bibinfo {author} {\bibfnamefont {R.}~\bibnamefont {Negishi}}, \bibinfo {author} {\bibfnamefont {W.~T.}\ \bibnamefont {Ni}}, \bibinfo {author} {\bibfnamefont {A.}~\bibnamefont {Nishizawa}}, \bibinfo {author} {\bibfnamefont {Y.}~\bibnamefont {Obuchi}}, \bibinfo {author} {\bibfnamefont {W.}~\bibnamefont {Ogaki}}, \bibinfo {author} {\bibfnamefont {J.~J.}\ \bibnamefont {Oh}}, \bibinfo {author} {\bibfnamefont {S.~H.}\ \bibnamefont {Oh}}, \bibinfo {author} {\bibfnamefont {M.}~\bibnamefont {Ohashi}}, \bibinfo {author} {\bibfnamefont {N.}~\bibnamefont
  {Ohishi}}, \bibinfo {author} {\bibfnamefont {M.}~\bibnamefont {Ohkawa}}, \bibinfo {author} {\bibfnamefont {K.}~\bibnamefont {Okutomi}}, \bibinfo {author} {\bibfnamefont {K.}~\bibnamefont {Oohara}}, \bibinfo {author} {\bibfnamefont {C.~P.}\ \bibnamefont {Ooi}}, \bibinfo {author} {\bibfnamefont {S.}~\bibnamefont {Oshino}}, \bibinfo {author} {\bibfnamefont {K.}~\bibnamefont {Pan}}, \bibinfo {author} {\bibfnamefont {H.}~\bibnamefont {Pang}}, \bibinfo {author} {\bibfnamefont {J.}~\bibnamefont {Park}}, \bibinfo {author} {\bibfnamefont {F.~E.~P.}\ \bibnamefont {Arellano}}, \bibinfo {author} {\bibfnamefont {I.}~\bibnamefont {Pinto}}, \bibinfo {author} {\bibfnamefont {N.}~\bibnamefont {Sago}}, \bibinfo {author} {\bibfnamefont {S.}~\bibnamefont {Saito}}, \bibinfo {author} {\bibfnamefont {Y.}~\bibnamefont {Saito}}, \bibinfo {author} {\bibfnamefont {K.}~\bibnamefont {Sakai}}, \bibinfo {author} {\bibfnamefont {Y.}~\bibnamefont {Sakai}}, \bibinfo {author} {\bibfnamefont {Y.}~\bibnamefont {Sakuno}}, \bibinfo {author}
  {\bibfnamefont {S.}~\bibnamefont {Sato}}, \bibinfo {author} {\bibfnamefont {T.}~\bibnamefont {Sato}}, \bibinfo {author} {\bibfnamefont {T.}~\bibnamefont {Sawada}}, \bibinfo {author} {\bibfnamefont {T.}~\bibnamefont {Sekiguchi}}, \bibinfo {author} {\bibfnamefont {Y.}~\bibnamefont {Sekiguchi}}, \bibinfo {author} {\bibfnamefont {S.}~\bibnamefont {Shibagaki}}, \bibinfo {author} {\bibfnamefont {R.}~\bibnamefont {Shimizu}}, \bibinfo {author} {\bibfnamefont {T.}~\bibnamefont {Shimoda}}, \bibinfo {author} {\bibfnamefont {K.}~\bibnamefont {Shimode}}, \bibinfo {author} {\bibfnamefont {H.}~\bibnamefont {Shinkai}}, \bibinfo {author} {\bibfnamefont {T.}~\bibnamefont {Shishido}}, \bibinfo {author} {\bibfnamefont {A.}~\bibnamefont {Shoda}}, \bibinfo {author} {\bibfnamefont {K.}~\bibnamefont {Somiya}}, \bibinfo {author} {\bibfnamefont {E.~J.}\ \bibnamefont {Son}}, \bibinfo {author} {\bibfnamefont {H.}~\bibnamefont {Sotani}}, \bibinfo {author} {\bibfnamefont {R.}~\bibnamefont {Sugimoto}}, \bibinfo {author} {\bibfnamefont
  {T.}~\bibnamefont {Suzuki}}, \bibinfo {author} {\bibfnamefont {T.}~\bibnamefont {Suzuki}}, \bibinfo {author} {\bibfnamefont {H.}~\bibnamefont {Tagoshi}}, \bibinfo {author} {\bibfnamefont {H.}~\bibnamefont {Takahashi}}, \bibinfo {author} {\bibfnamefont {R.}~\bibnamefont {Takahashi}}, \bibinfo {author} {\bibfnamefont {A.}~\bibnamefont {Takamori}}, \bibinfo {author} {\bibfnamefont {S.}~\bibnamefont {Takano}}, \bibinfo {author} {\bibfnamefont {H.}~\bibnamefont {Takeda}}, \bibinfo {author} {\bibfnamefont {M.}~\bibnamefont {Takeda}}, \bibinfo {author} {\bibfnamefont {H.}~\bibnamefont {Tanaka}}, \bibinfo {author} {\bibfnamefont {K.}~\bibnamefont {Tanaka}}, \bibinfo {author} {\bibfnamefont {K.}~\bibnamefont {Tanaka}}, \bibinfo {author} {\bibfnamefont {T.}~\bibnamefont {Tanaka}}, \bibinfo {author} {\bibfnamefont {T.}~\bibnamefont {Tanaka}}, \bibinfo {author} {\bibfnamefont {S.}~\bibnamefont {Tanioka}}, \bibinfo {author} {\bibfnamefont {E.~N.}\ \bibnamefont {Tapia San~Martin}}, \bibinfo {author} {\bibfnamefont
  {S.}~\bibnamefont {Telada}}, \bibinfo {author} {\bibfnamefont {T.}~\bibnamefont {Tomaru}}, \bibinfo {author} {\bibfnamefont {Y.}~\bibnamefont {Tomigami}}, \bibinfo {author} {\bibfnamefont {T.}~\bibnamefont {Tomura}}, \bibinfo {author} {\bibfnamefont {F.}~\bibnamefont {Travasso}}, \bibinfo {author} {\bibfnamefont {L.}~\bibnamefont {Trozzo}}, \bibinfo {author} {\bibfnamefont {T.}~\bibnamefont {Tsang}}, \bibinfo {author} {\bibfnamefont {K.}~\bibnamefont {Tsubono}}, \bibinfo {author} {\bibfnamefont {S.}~\bibnamefont {Tsuchida}}, \bibinfo {author} {\bibfnamefont {T.}~\bibnamefont {Tsuzuki}}, \bibinfo {author} {\bibfnamefont {D.}~\bibnamefont {Tuyenbayev}}, \bibinfo {author} {\bibfnamefont {N.}~\bibnamefont {Uchikata}}, \bibinfo {author} {\bibfnamefont {T.}~\bibnamefont {Uchiyama}}, \bibinfo {author} {\bibfnamefont {A.}~\bibnamefont {Ueda}}, \bibinfo {author} {\bibfnamefont {T.}~\bibnamefont {Uehara}}, \bibinfo {author} {\bibfnamefont {K.}~\bibnamefont {Ueno}}, \bibinfo {author} {\bibfnamefont {G.}~\bibnamefont
  {Ueshima}}, \bibinfo {author} {\bibfnamefont {F.}~\bibnamefont {Uraguchi}}, \bibinfo {author} {\bibfnamefont {T.}~\bibnamefont {Ushiba}}, \bibinfo {author} {\bibfnamefont {M.~H. P.~M.}\ \bibnamefont {van Putten}}, \bibinfo {author} {\bibfnamefont {H.}~\bibnamefont {Vocca}}, \bibinfo {author} {\bibfnamefont {J.}~\bibnamefont {Wang}}, \bibinfo {author} {\bibfnamefont {C.}~\bibnamefont {Wu}}, \bibinfo {author} {\bibfnamefont {H.}~\bibnamefont {Wu}}, \bibinfo {author} {\bibfnamefont {S.}~\bibnamefont {Wu}}, \bibinfo {author} {\bibfnamefont {W.-R.}\ \bibnamefont {Xu}}, \bibinfo {author} {\bibfnamefont {T.}~\bibnamefont {Yamada}}, \bibinfo {author} {\bibfnamefont {K.}~\bibnamefont {Yamamoto}}, \bibinfo {author} {\bibfnamefont {K.}~\bibnamefont {Yamamoto}}, \bibinfo {author} {\bibfnamefont {T.}~\bibnamefont {Yamamoto}}, \bibinfo {author} {\bibfnamefont {K.}~\bibnamefont {Yokogawa}}, \bibinfo {author} {\bibfnamefont {J.}~\bibnamefont {Yokoyama}}, \bibinfo {author} {\bibfnamefont {T.}~\bibnamefont {Yokozawa}},
  \bibinfo {author} {\bibfnamefont {T.}~\bibnamefont {Yoshioka}}, \bibinfo {author} {\bibfnamefont {H.}~\bibnamefont {Yuzurihara}}, \bibinfo {author} {\bibfnamefont {S.}~\bibnamefont {Zeidler}}, \bibinfo {author} {\bibfnamefont {Y.}~\bibnamefont {Zhao}},\ and\ \bibinfo {author} {\bibfnamefont {Z.~H.}\ \bibnamefont {Zhu}},\ }\bibfield  {title} {\bibinfo {title} {Overview of kagra: Detector design and construction history},\ }\href {https://doi.org/10.1093/ptep/ptaa125} {\bibfield  {journal} {\bibinfo  {journal} {Progress of Theoretical and Experimental Physics}\ }\textbf {\bibinfo {volume} {2021}},\ \bibinfo {pages} {05A101} (\bibinfo {year} {2020})},\ \Eprint {https://arxiv.org/abs/https://academic.oup.com/ptep/article-pdf/2021/5/05A101/37974994/ptaa125.pdf} {https://academic.oup.com/ptep/article-pdf/2021/5/05A101/37974994/ptaa125.pdf} \BibitemShut {NoStop}%
\bibitem [{\citenamefont {Abac}\ \emph {et~al.}(2024)\citenamefont {Abac}, \citenamefont {Abbott}, \citenamefont {Abouelfettouh}, \citenamefont {Acernese}, \citenamefont {Ackley}, \citenamefont {Adhicary}, \citenamefont {Adhikari}, \citenamefont {Adhikari},\ and\ \citenamefont {et~al.}}]{Abac_2024}%
  \BibitemOpen
  \bibfield  {author} {\bibinfo {author} {\bibfnamefont {A.~G.}\ \bibnamefont {Abac}}, \bibinfo {author} {\bibfnamefont {R.}~\bibnamefont {Abbott}}, \bibinfo {author} {\bibfnamefont {I.}~\bibnamefont {Abouelfettouh}}, \bibinfo {author} {\bibfnamefont {F.}~\bibnamefont {Acernese}}, \bibinfo {author} {\bibfnamefont {K.}~\bibnamefont {Ackley}}, \bibinfo {author} {\bibfnamefont {S.}~\bibnamefont {Adhicary}}, \bibinfo {author} {\bibfnamefont {N.}~\bibnamefont {Adhikari}}, \bibinfo {author} {\bibfnamefont {R.~X.}\ \bibnamefont {Adhikari}},\ and\ \bibinfo {author} {\bibnamefont {et~al.}},\ }\bibfield  {title} {\bibinfo {title} {Observation of gravitational waves from the coalescence of a 2.5--4.5 $m_{\odot}$ compact object and a neutron star},\ }\href {https://doi.org/10.3847/2041-8213/ad5beb} {\bibfield  {journal} {\bibinfo  {journal} {The Astrophysical Journal Letters}\ }\textbf {\bibinfo {volume} {970}},\ \bibinfo {pages} {L34} (\bibinfo {year} {2024})}\BibitemShut {NoStop}%
\bibitem [{\citenamefont {Jia}\ \emph {et~al.}(2024)\citenamefont {Jia} \emph {et~al.}}]{Jia_2024}%
  \BibitemOpen
  \bibfield  {author} {\bibinfo {author} {\bibfnamefont {W.}~\bibnamefont {Jia}} \emph {et~al.},\ }\bibfield  {title} {\bibinfo {title} {{Squeezing the quantum noise of a gravitational-wave detector below the standard quantum limit}},\ }\href {https://doi.org/10.1126/science.ado8069} {\bibfield  {journal} {\bibinfo  {journal} {Science}\ }\textbf {\bibinfo {volume} {385}},\ \bibinfo {pages} {1318} (\bibinfo {year} {2024})},\ \Eprint {https://arxiv.org/abs/2404.14569} {arXiv:2404.14569 [gr-qc]} \BibitemShut {NoStop}%
\bibitem [{\citenamefont {Ganapathy}\ \emph {et~al.}(2023)\citenamefont {Ganapathy}, \citenamefont {Jia}, \citenamefont {Nakano}, \citenamefont {Xu}, \citenamefont {Aritomi}, \citenamefont {Cullen}, \citenamefont {Kijbunchoo}, \citenamefont {Dwyer}, \citenamefont {Mullavey}, \citenamefont {McCuller}, \citenamefont {Abbott}, \citenamefont {Abouelfettouh}, \citenamefont {Adhikari}, \citenamefont {Ananyeva}, \citenamefont {Appert}, \citenamefont {Arai}, \citenamefont {Aston}, \citenamefont {Ball}, \citenamefont {Ballmer},\ and\ \citenamefont {et~al.}}]{Ganapathy_2023}%
  \BibitemOpen
  \bibfield  {author} {\bibinfo {author} {\bibfnamefont {D.}~\bibnamefont {Ganapathy}}, \bibinfo {author} {\bibfnamefont {W.}~\bibnamefont {Jia}}, \bibinfo {author} {\bibfnamefont {M.}~\bibnamefont {Nakano}}, \bibinfo {author} {\bibfnamefont {V.}~\bibnamefont {Xu}}, \bibinfo {author} {\bibfnamefont {N.}~\bibnamefont {Aritomi}}, \bibinfo {author} {\bibfnamefont {T.}~\bibnamefont {Cullen}}, \bibinfo {author} {\bibfnamefont {N.}~\bibnamefont {Kijbunchoo}}, \bibinfo {author} {\bibfnamefont {S.~E.}\ \bibnamefont {Dwyer}}, \bibinfo {author} {\bibfnamefont {A.}~\bibnamefont {Mullavey}}, \bibinfo {author} {\bibfnamefont {L.}~\bibnamefont {McCuller}}, \bibinfo {author} {\bibfnamefont {R.}~\bibnamefont {Abbott}}, \bibinfo {author} {\bibfnamefont {I.}~\bibnamefont {Abouelfettouh}}, \bibinfo {author} {\bibfnamefont {R.~X.}\ \bibnamefont {Adhikari}}, \bibinfo {author} {\bibfnamefont {A.}~\bibnamefont {Ananyeva}}, \bibinfo {author} {\bibfnamefont {S.}~\bibnamefont {Appert}}, \bibinfo {author} {\bibfnamefont
  {K.}~\bibnamefont {Arai}}, \bibinfo {author} {\bibfnamefont {S.~M.}\ \bibnamefont {Aston}}, \bibinfo {author} {\bibfnamefont {M.}~\bibnamefont {Ball}}, \bibinfo {author} {\bibfnamefont {S.~W.}\ \bibnamefont {Ballmer}},\ and\ \bibinfo {author} {\bibnamefont {et~al.}} (\bibinfo {collaboration} {LIGO O4 Detector Collaboration}),\ }\bibfield  {title} {\bibinfo {title} {Broadband quantum enhancement of the ligo detectors with frequency-dependent squeezing},\ }\href {https://doi.org/10.1103/PhysRevX.13.041021} {\bibfield  {journal} {\bibinfo  {journal} {Phys. Rev. X}\ }\textbf {\bibinfo {volume} {13}},\ \bibinfo {pages} {041021} (\bibinfo {year} {2023})}\BibitemShut {NoStop}%
\bibitem [{\citenamefont {Wade}\ \emph {et~al.}(2025)\citenamefont {Wade}, \citenamefont {Betzwieser}, \citenamefont {Bhattacharjee}, \citenamefont {Dartez}, \citenamefont {Goetz}, \citenamefont {Kissel}, \citenamefont {Sun}, \citenamefont {Viets}, \citenamefont {Carney}, \citenamefont {Makelele},\ and\ \citenamefont {Wade}}]{Wade_2025}%
  \BibitemOpen
  \bibfield  {author} {\bibinfo {author} {\bibfnamefont {M.}~\bibnamefont {Wade}}, \bibinfo {author} {\bibfnamefont {J.}~\bibnamefont {Betzwieser}}, \bibinfo {author} {\bibfnamefont {D.}~\bibnamefont {Bhattacharjee}}, \bibinfo {author} {\bibfnamefont {L.}~\bibnamefont {Dartez}}, \bibinfo {author} {\bibfnamefont {E.}~\bibnamefont {Goetz}}, \bibinfo {author} {\bibfnamefont {J.}~\bibnamefont {Kissel}}, \bibinfo {author} {\bibfnamefont {L.}~\bibnamefont {Sun}}, \bibinfo {author} {\bibfnamefont {A.}~\bibnamefont {Viets}}, \bibinfo {author} {\bibfnamefont {M.}~\bibnamefont {Carney}}, \bibinfo {author} {\bibfnamefont {E.}~\bibnamefont {Makelele}},\ and\ \bibinfo {author} {\bibfnamefont {L.}~\bibnamefont {Wade}},\ }\bibfield  {title} {\bibinfo {title} {Toward low-latency, high-fidelity calibration of the ligo detectors with enhanced monitoring tools},\ }\href {https://doi.org/10.1088/1361-6382/ae1095} {\bibfield  {journal} {\bibinfo  {journal} {Classical and Quantum Gravity}\ }\textbf {\bibinfo {volume} {42}},\
  \bibinfo {pages} {215016} (\bibinfo {year} {2025})}\BibitemShut {NoStop}%
\bibitem [{\citenamefont {Goetz}\ \emph {et~al.}(2009)\citenamefont {Goetz}, \citenamefont {Kalmus}, \citenamefont {Erickson}, \citenamefont {Savage}, \citenamefont {Gonzalez}, \citenamefont {Kawabe}, \citenamefont {Landry}, \citenamefont {Marka}, \citenamefont {O’Reilly}, \citenamefont {Riles}, \citenamefont {Sigg},\ and\ \citenamefont {Willems}}]{photoncalibration}%
  \BibitemOpen
  \bibfield  {author} {\bibinfo {author} {\bibfnamefont {E.}~\bibnamefont {Goetz}}, \bibinfo {author} {\bibfnamefont {P.}~\bibnamefont {Kalmus}}, \bibinfo {author} {\bibfnamefont {S.}~\bibnamefont {Erickson}}, \bibinfo {author} {\bibfnamefont {R.~L.}\ \bibnamefont {Savage}}, \bibinfo {author} {\bibfnamefont {G.}~\bibnamefont {Gonzalez}}, \bibinfo {author} {\bibfnamefont {K.}~\bibnamefont {Kawabe}}, \bibinfo {author} {\bibfnamefont {M.}~\bibnamefont {Landry}}, \bibinfo {author} {\bibfnamefont {S.}~\bibnamefont {Marka}}, \bibinfo {author} {\bibfnamefont {B.}~\bibnamefont {O’Reilly}}, \bibinfo {author} {\bibfnamefont {K.}~\bibnamefont {Riles}}, \bibinfo {author} {\bibfnamefont {D.}~\bibnamefont {Sigg}},\ and\ \bibinfo {author} {\bibfnamefont {P.}~\bibnamefont {Willems}},\ }\bibfield  {title} {\bibinfo {title} {Precise calibration of ligo test mass actuators using photon radiation pressure},\ }\href {https://doi.org/10.1088/0264-9381/26/24/245011} {\bibfield  {journal} {\bibinfo  {journal} {Classical and
  Quantum Gravity}\ }\textbf {\bibinfo {volume} {26}},\ \bibinfo {pages} {245011} (\bibinfo {year} {2009})}\BibitemShut {NoStop}%
\bibitem [{\citenamefont {Karki}\ \emph {et~al.}(2016)\citenamefont {Karki}, \citenamefont {Tuyenbayev}, \citenamefont {Kandhasamy}, \citenamefont {Abbott}, \citenamefont {Abbott}, \citenamefont {Anders}, \citenamefont {Berliner}, \citenamefont {Betzwieser}, \citenamefont {Cahillane}, \citenamefont {Canete}, \citenamefont {Conley}, \citenamefont {Daveloza}, \citenamefont {De~Lillo}, \citenamefont {Gleason}, \citenamefont {Goetz}, \citenamefont {Izumi}, \citenamefont {Kissel}, \citenamefont {Mendell}, \citenamefont {Quetschke}, \citenamefont {Rodruck}, \citenamefont {Sachdev}, \citenamefont {Sadecki}, \citenamefont {Schwinberg}, \citenamefont {Sottile}, \citenamefont {Wade}, \citenamefont {Weinstein}, \citenamefont {West},\ and\ \citenamefont {Savage}}]{Karki_2016}%
  \BibitemOpen
  \bibfield  {author} {\bibinfo {author} {\bibfnamefont {S.}~\bibnamefont {Karki}}, \bibinfo {author} {\bibfnamefont {D.}~\bibnamefont {Tuyenbayev}}, \bibinfo {author} {\bibfnamefont {S.}~\bibnamefont {Kandhasamy}}, \bibinfo {author} {\bibfnamefont {B.~P.}\ \bibnamefont {Abbott}}, \bibinfo {author} {\bibfnamefont {T.~D.}\ \bibnamefont {Abbott}}, \bibinfo {author} {\bibfnamefont {E.~H.}\ \bibnamefont {Anders}}, \bibinfo {author} {\bibfnamefont {J.}~\bibnamefont {Berliner}}, \bibinfo {author} {\bibfnamefont {J.}~\bibnamefont {Betzwieser}}, \bibinfo {author} {\bibfnamefont {C.}~\bibnamefont {Cahillane}}, \bibinfo {author} {\bibfnamefont {L.}~\bibnamefont {Canete}}, \bibinfo {author} {\bibfnamefont {C.}~\bibnamefont {Conley}}, \bibinfo {author} {\bibfnamefont {H.~P.}\ \bibnamefont {Daveloza}}, \bibinfo {author} {\bibfnamefont {N.}~\bibnamefont {De~Lillo}}, \bibinfo {author} {\bibfnamefont {J.~R.}\ \bibnamefont {Gleason}}, \bibinfo {author} {\bibfnamefont {E.}~\bibnamefont {Goetz}}, \bibinfo {author}
  {\bibfnamefont {K.}~\bibnamefont {Izumi}}, \bibinfo {author} {\bibfnamefont {J.~S.}\ \bibnamefont {Kissel}}, \bibinfo {author} {\bibfnamefont {G.}~\bibnamefont {Mendell}}, \bibinfo {author} {\bibfnamefont {V.}~\bibnamefont {Quetschke}}, \bibinfo {author} {\bibfnamefont {M.}~\bibnamefont {Rodruck}}, \bibinfo {author} {\bibfnamefont {S.}~\bibnamefont {Sachdev}}, \bibinfo {author} {\bibfnamefont {T.}~\bibnamefont {Sadecki}}, \bibinfo {author} {\bibfnamefont {P.~B.}\ \bibnamefont {Schwinberg}}, \bibinfo {author} {\bibfnamefont {A.}~\bibnamefont {Sottile}}, \bibinfo {author} {\bibfnamefont {M.}~\bibnamefont {Wade}}, \bibinfo {author} {\bibfnamefont {A.~J.}\ \bibnamefont {Weinstein}}, \bibinfo {author} {\bibfnamefont {M.}~\bibnamefont {West}},\ and\ \bibinfo {author} {\bibfnamefont {R.~L.}\ \bibnamefont {Savage}},\ }\bibfield  {title} {\bibinfo {title} {The advanced ligo photon calibrators},\ }\bibfield  {journal} {\bibinfo  {journal} {Review of Scientific Instruments}\ }\textbf {\bibinfo {volume} {87}},\ \href
  {https://doi.org/10.1063/1.4967303} {10.1063/1.4967303} (\bibinfo {year} {2016})\BibitemShut {NoStop}%
\bibitem [{\citenamefont {Viets}\ \emph {et~al.}(2018)\citenamefont {Viets}, \citenamefont {Wade}, \citenamefont {Urban}, \citenamefont {Kandhasamy}, \citenamefont {Betzwieser}, \citenamefont {Brown}, \citenamefont {Burguet-Castell}, \citenamefont {Cahillane}, \citenamefont {Goetz}, \citenamefont {Izumi}, \citenamefont {Karki}, \citenamefont {Kissel}, \citenamefont {Mendell}, \citenamefont {Savage}, \citenamefont {Siemens}, \citenamefont {Tuyenbayev},\ and\ \citenamefont {Weinstein}}]{Viets_2018}%
  \BibitemOpen
  \bibfield  {author} {\bibinfo {author} {\bibfnamefont {A.~D.}\ \bibnamefont {Viets}}, \bibinfo {author} {\bibfnamefont {M.}~\bibnamefont {Wade}}, \bibinfo {author} {\bibfnamefont {A.~L.}\ \bibnamefont {Urban}}, \bibinfo {author} {\bibfnamefont {S.}~\bibnamefont {Kandhasamy}}, \bibinfo {author} {\bibfnamefont {J.}~\bibnamefont {Betzwieser}}, \bibinfo {author} {\bibfnamefont {D.~A.}\ \bibnamefont {Brown}}, \bibinfo {author} {\bibfnamefont {J.}~\bibnamefont {Burguet-Castell}}, \bibinfo {author} {\bibfnamefont {C.}~\bibnamefont {Cahillane}}, \bibinfo {author} {\bibfnamefont {E.}~\bibnamefont {Goetz}}, \bibinfo {author} {\bibfnamefont {K.}~\bibnamefont {Izumi}}, \bibinfo {author} {\bibfnamefont {S.}~\bibnamefont {Karki}}, \bibinfo {author} {\bibfnamefont {J.~S.}\ \bibnamefont {Kissel}}, \bibinfo {author} {\bibfnamefont {G.}~\bibnamefont {Mendell}}, \bibinfo {author} {\bibfnamefont {R.~L.}\ \bibnamefont {Savage}}, \bibinfo {author} {\bibfnamefont {X.}~\bibnamefont {Siemens}}, \bibinfo {author} {\bibfnamefont
  {D.}~\bibnamefont {Tuyenbayev}},\ and\ \bibinfo {author} {\bibfnamefont {A.~J.}\ \bibnamefont {Weinstein}},\ }\bibfield  {title} {\bibinfo {title} {Reconstructing the calibrated strain signal in the advanced ligo detectors},\ }\href {https://doi.org/10.1088/1361-6382/aab658} {\bibfield  {journal} {\bibinfo  {journal} {Classical and Quantum Gravity}\ }\textbf {\bibinfo {volume} {35}},\ \bibinfo {pages} {095015} (\bibinfo {year} {2018})}\BibitemShut {NoStop}%
\bibitem [{\citenamefont {Dartez}\ \emph {et~al.}(2025)\citenamefont {Dartez} \emph {et~al.}}]{O4cal}%
  \BibitemOpen
  \bibfield  {author} {\bibinfo {author} {\bibfnamefont {L.}~\bibnamefont {Dartez}} \emph {et~al.},\ }\href {https://dcc.ligo.org/P2400219} {\bibinfo {title} {Characterization of systematic error in {Advanced LIGO} calibration in the fourth observing run}} (\bibinfo {year} {2025})\BibitemShut {NoStop}%
\bibitem [{\citenamefont {Soni}\ \emph {et~al.}(2025)\citenamefont {Soni}, \citenamefont {Berger}, \citenamefont {Davis}, \citenamefont {Di~Renzo}, \citenamefont {Effler}, \citenamefont {Ferreira},\ and\ \citenamefont {et~al.}}]{soni_2025}%
  \BibitemOpen
  \bibfield  {author} {\bibinfo {author} {\bibfnamefont {S.}~\bibnamefont {Soni}}, \bibinfo {author} {\bibfnamefont {B.~K.}\ \bibnamefont {Berger}}, \bibinfo {author} {\bibfnamefont {D.}~\bibnamefont {Davis}}, \bibinfo {author} {\bibfnamefont {F.}~\bibnamefont {Di~Renzo}}, \bibinfo {author} {\bibfnamefont {A.}~\bibnamefont {Effler}}, \bibinfo {author} {\bibfnamefont {T.~A.}\ \bibnamefont {Ferreira}},\ and\ \bibinfo {author} {\bibnamefont {et~al.}},\ }\bibfield  {title} {\bibinfo {title} {Ligo detector characterization in the first half of the fourth observing run},\ }\href {https://doi.org/10.1088/1361-6382/adc4b6} {\bibfield  {journal} {\bibinfo  {journal} {Classical and Quantum Gravity}\ }\textbf {\bibinfo {volume} {42}},\ \bibinfo {pages} {085016} (\bibinfo {year} {2025})}\BibitemShut {NoStop}%
\bibitem [{\citenamefont {Covas}\ \emph {et~al.}(2018)\citenamefont {Covas} \emph {et~al.}}]{linespaper}%
  \BibitemOpen
  \bibfield  {author} {\bibinfo {author} {\bibfnamefont {P.}~\bibnamefont {Covas}} \emph {et~al.},\ }\bibfield  {title} {\bibinfo {title} {Identification and mitigation of narrow spectral artifacts that degrade searches for persistent gravitational waves in the first two observing runs of advanced ligo},\ }\href@noop {} {\bibfield  {journal} {\bibinfo  {journal} {Phys. Rev. D}\ }\textbf {\bibinfo {volume} {97}},\ \bibinfo {pages} {082002} (\bibinfo {year} {2018})}\BibitemShut {NoStop}%
\bibitem [{\citenamefont {Goetz}\ \emph {et~al.}(2024)\citenamefont {Goetz} \emph {et~al.}}]{known_lines_O4a}%
  \BibitemOpen
  \bibfield  {author} {\bibinfo {author} {\bibfnamefont {E.}~\bibnamefont {Goetz}} \emph {et~al.},\ }\href {https://dcc.ligo.org/T2400204} {\bibinfo {title} {{O4a lines and combs in found in self-gated C00 cleaned data}}} (\bibinfo {year} {2024})\BibitemShut {NoStop}%
\bibitem [{\citenamefont {Abbott}\ \emph {et~al.}(2021{\natexlab{a}})\citenamefont {Abbott} \emph {et~al.}}]{O3aAllSky}%
  \BibitemOpen
  \bibfield  {author} {\bibinfo {author} {\bibfnamefont {R.}~\bibnamefont {Abbott}} \emph {et~al.} (\bibinfo {collaboration} {LIGO Scientific Collaboration, Virgo Collaboration, and KAGRA Collaboration}),\ }\bibfield  {title} {\bibinfo {title} {All-sky search for continuous gravitational waves from isolated neutron stars in the early o3 ligo data},\ }\href {https://doi.org/10.1103/PhysRevD.104.082004} {\bibfield  {journal} {\bibinfo  {journal} {Phys. Rev. D}\ }\textbf {\bibinfo {volume} {104}},\ \bibinfo {pages} {082004} (\bibinfo {year} {2021}{\natexlab{a}})}\BibitemShut {NoStop}%
\bibitem [{\citenamefont {Davis}\ \emph {et~al.}(2024)\citenamefont {Davis}, \citenamefont {Neunzert}, \citenamefont {Goetz}, \citenamefont {Riles}, \citenamefont {Wette},\ and\ \citenamefont {Lalleman}}]{o4gating}%
  \BibitemOpen
  \bibfield  {author} {\bibinfo {author} {\bibfnamefont {D.}~\bibnamefont {Davis}}, \bibinfo {author} {\bibfnamefont {A.}~\bibnamefont {Neunzert}}, \bibinfo {author} {\bibfnamefont {E.}~\bibnamefont {Goetz}}, \bibinfo {author} {\bibfnamefont {K.}~\bibnamefont {Riles}}, \bibinfo {author} {\bibfnamefont {K.}~\bibnamefont {Wette}},\ and\ \bibinfo {author} {\bibfnamefont {M.}~\bibnamefont {Lalleman}},\ }\href {https://dcc.ligo.org/T2400003} {\bibinfo {title} {{Self-gating of O4a h(t) for use in continuous-wave searches}}} (\bibinfo {year} {2024})\BibitemShut {NoStop}%
\bibitem [{\citenamefont {Allen}\ \emph {et~al.}(2025)\citenamefont {Allen}, \citenamefont {Goetz}, \citenamefont {Keitel}, \citenamefont {Landry}, \citenamefont {Mendell}, \citenamefont {Prix}, \citenamefont {Riles},\ and\ \citenamefont {Wette}}]{sft_spec_2025}%
  \BibitemOpen
  \bibfield  {author} {\bibinfo {author} {\bibfnamefont {B.}~\bibnamefont {Allen}}, \bibinfo {author} {\bibfnamefont {E.}~\bibnamefont {Goetz}}, \bibinfo {author} {\bibfnamefont {D.}~\bibnamefont {Keitel}}, \bibinfo {author} {\bibfnamefont {M.}~\bibnamefont {Landry}}, \bibinfo {author} {\bibfnamefont {G.}~\bibnamefont {Mendell}}, \bibinfo {author} {\bibfnamefont {R.}~\bibnamefont {Prix}}, \bibinfo {author} {\bibfnamefont {K.}~\bibnamefont {Riles}},\ and\ \bibinfo {author} {\bibfnamefont {K.}~\bibnamefont {Wette}} (\bibinfo {collaboration} {{The LIGO Scientific Collaboration}}),\ }\href {https://dcc.ligo.org/LIGO-T040164/public} {\bibinfo {title} {{SFT} {Data} {Format} {Version} 2--3 {Specification}}} (\bibinfo {year} {2025})\BibitemShut {NoStop}%
\bibitem [{\citenamefont {{Katz}}(1975)}]{Katz_1975}%
  \BibitemOpen
  \bibfield  {author} {\bibinfo {author} {\bibfnamefont {J.~I.}\ \bibnamefont {{Katz}}},\ }\bibfield  {title} {\bibinfo {title} {{Two kinds of stellar collapse}},\ }\href {https://doi.org/10.1038/253698a0} {\bibfield  {journal} {\bibinfo  {journal} {\nat}\ }\textbf {\bibinfo {volume} {253}},\ \bibinfo {pages} {698} (\bibinfo {year} {1975})}\BibitemShut {NoStop}%
\bibitem [{\citenamefont {{Clark}}(1975)}]{Clark_1975}%
  \BibitemOpen
  \bibfield  {author} {\bibinfo {author} {\bibfnamefont {G.~W.}\ \bibnamefont {{Clark}}},\ }\bibfield  {title} {\bibinfo {title} {{X-ray binaries in globular clusters.}},\ }\href {https://doi.org/10.1086/181869} {\bibfield  {journal} {\bibinfo  {journal} {\apjl}\ }\textbf {\bibinfo {volume} {199}},\ \bibinfo {pages} {L143} (\bibinfo {year} {1975})}\BibitemShut {NoStop}%
\bibitem [{\citenamefont {{Pooley}}\ \emph {et~al.}(2003)\citenamefont {{Pooley}}, \citenamefont {{Lewin}}, \citenamefont {{Anderson}}, \citenamefont {{Baumgardt}}, \citenamefont {{Filippenko}}, \citenamefont {{Gaensler}}, \citenamefont {{Homer}}, \citenamefont {{Hut}}, \citenamefont {{Kaspi}}, \citenamefont {{Makino}}, \citenamefont {{Margon}}, \citenamefont {{McMillan}}, \citenamefont {{Portegies Zwart}}, \citenamefont {{van der Klis}},\ and\ \citenamefont {{Verbunt}}}]{Pooley_2003}%
  \BibitemOpen
  \bibfield  {author} {\bibinfo {author} {\bibfnamefont {D.}~\bibnamefont {{Pooley}}}, \bibinfo {author} {\bibfnamefont {W.~H.~G.}\ \bibnamefont {{Lewin}}}, \bibinfo {author} {\bibfnamefont {S.~F.}\ \bibnamefont {{Anderson}}}, \bibinfo {author} {\bibfnamefont {H.}~\bibnamefont {{Baumgardt}}}, \bibinfo {author} {\bibfnamefont {A.~V.}\ \bibnamefont {{Filippenko}}}, \bibinfo {author} {\bibfnamefont {B.~M.}\ \bibnamefont {{Gaensler}}}, \bibinfo {author} {\bibfnamefont {L.}~\bibnamefont {{Homer}}}, \bibinfo {author} {\bibfnamefont {P.}~\bibnamefont {{Hut}}}, \bibinfo {author} {\bibfnamefont {V.~M.}\ \bibnamefont {{Kaspi}}}, \bibinfo {author} {\bibfnamefont {J.}~\bibnamefont {{Makino}}}, \bibinfo {author} {\bibfnamefont {B.}~\bibnamefont {{Margon}}}, \bibinfo {author} {\bibfnamefont {S.}~\bibnamefont {{McMillan}}}, \bibinfo {author} {\bibfnamefont {S.}~\bibnamefont {{Portegies Zwart}}}, \bibinfo {author} {\bibfnamefont {M.}~\bibnamefont {{van der Klis}}},\ and\ \bibinfo {author} {\bibfnamefont {F.}~\bibnamefont
  {{Verbunt}}},\ }\bibfield  {title} {\bibinfo {title} {{Dynamical Formation of Close Binary Systems in Globular Clusters}},\ }\href {https://doi.org/10.1086/377074} {\bibfield  {journal} {\bibinfo  {journal} {\apjl}\ }\textbf {\bibinfo {volume} {591}},\ \bibinfo {pages} {L131} (\bibinfo {year} {2003})},\ \Eprint {https://arxiv.org/abs/astro-ph/0305003} {arXiv:astro-ph/0305003 [astro-ph]} \BibitemShut {NoStop}%
\bibitem [{\citenamefont {{Alpar}}\ \emph {et~al.}(1982)\citenamefont {{Alpar}}, \citenamefont {{Cheng}}, \citenamefont {{Ruderman}},\ and\ \citenamefont {{Shaham}}}]{msp_1982}%
  \BibitemOpen
  \bibfield  {author} {\bibinfo {author} {\bibfnamefont {M.~A.}\ \bibnamefont {{Alpar}}}, \bibinfo {author} {\bibfnamefont {A.~F.}\ \bibnamefont {{Cheng}}}, \bibinfo {author} {\bibfnamefont {M.~A.}\ \bibnamefont {{Ruderman}}},\ and\ \bibinfo {author} {\bibfnamefont {J.}~\bibnamefont {{Shaham}}},\ }\bibfield  {title} {\bibinfo {title} {{A new class of radio pulsars}},\ }\href {https://doi.org/10.1038/300728a0} {\bibfield  {journal} {\bibinfo  {journal} {\nat}\ }\textbf {\bibinfo {volume} {300}},\ \bibinfo {pages} {728} (\bibinfo {year} {1982})}\BibitemShut {NoStop}%
\bibitem [{\citenamefont {Wette}\ \emph {et~al.}(2008)\citenamefont {Wette} \emph {et~al.}}]{cwcasamethod}%
  \BibitemOpen
  \bibfield  {author} {\bibinfo {author} {\bibfnamefont {K.}~\bibnamefont {Wette}} \emph {et~al.},\ }\bibfield  {title} {\bibinfo {title} {Searching for gravitational waves from cassiopeia a with ligo},\ }\href@noop {} {\bibfield  {journal} {\bibinfo  {journal} {Class. Quant. Grav.}\ }\textbf {\bibinfo {volume} {25}},\ \bibinfo {pages} {235011} (\bibinfo {year} {2008})}\BibitemShut {NoStop}%
\bibitem [{\citenamefont {{Harris}}(1996)}]{Harris1996}%
  \BibitemOpen
  \bibfield  {author} {\bibinfo {author} {\bibfnamefont {W.~E.}\ \bibnamefont {{Harris}}},\ }\bibfield  {title} {\bibinfo {title} {{A Catalog of Parameters for Globular Clusters in the Milky Way}},\ }\href {https://doi.org/10.1086/118116} {\bibfield  {journal} {\bibinfo  {journal} {\aj}\ }\textbf {\bibinfo {volume} {112}},\ \bibinfo {pages} {1487} (\bibinfo {year} {1996})}\BibitemShut {NoStop}%
\bibitem [{\citenamefont {Verbunt}\ and\ \citenamefont {Freire}(2014)}]{Verbunt_2013}%
  \BibitemOpen
  \bibfield  {author} {\bibinfo {author} {\bibfnamefont {F.}~\bibnamefont {Verbunt}}\ and\ \bibinfo {author} {\bibfnamefont {P.~C.~C.}\ \bibnamefont {Freire}},\ }\bibfield  {title} {\bibinfo {title} {{On the disruption of pulsar and X-ray binaries in globular clusters}},\ }\href {https://doi.org/10.1051/0004-6361/201321177} {\bibfield  {journal} {\bibinfo  {journal} {Astron. Astrophys.}\ }\textbf {\bibinfo {volume} {561}},\ \bibinfo {pages} {A11} (\bibinfo {year} {2014})},\ \Eprint {https://arxiv.org/abs/1310.4669} {arXiv:1310.4669 [astro-ph.SR]} \BibitemShut {NoStop}%
\bibitem [{\citenamefont {Zhang}\ \emph {et~al.}(2022)\citenamefont {Zhang}, \citenamefont {Ridolfi}, \citenamefont {Blumer}, \citenamefont {Freire}, \citenamefont {Manchester}, \citenamefont {McLaughlin}, \citenamefont {Kremer}, \citenamefont {Cameron}, \citenamefont {Zhang}, \citenamefont {Behrend}, \citenamefont {Burgay}, \citenamefont {Buchner}, \citenamefont {Champion}, \citenamefont {Chen}, \citenamefont {Dai}, \citenamefont {Feng}, \citenamefont {Fu}, \citenamefont {Guo}, \citenamefont {Hobbs}, \citenamefont {Keane}, \citenamefont {Kramer}, \citenamefont {Levin}, \citenamefont {Li}, \citenamefont {Ni}, \citenamefont {Pan}, \citenamefont {Padmanabh}, \citenamefont {Possenti}, \citenamefont {Ransom}, \citenamefont {Tsai}, \citenamefont {Venkatraman~Krishnan}, \citenamefont {Wang}, \citenamefont {Zhang}, \citenamefont {Zhi}, \citenamefont {Zhang},\ and\ \citenamefont {Li}}]{Zhang_2022}%
  \BibitemOpen
  \bibfield  {author} {\bibinfo {author} {\bibfnamefont {L.}~\bibnamefont {Zhang}}, \bibinfo {author} {\bibfnamefont {A.}~\bibnamefont {Ridolfi}}, \bibinfo {author} {\bibfnamefont {H.}~\bibnamefont {Blumer}}, \bibinfo {author} {\bibfnamefont {P.~C.~C.}\ \bibnamefont {Freire}}, \bibinfo {author} {\bibfnamefont {R.~N.}\ \bibnamefont {Manchester}}, \bibinfo {author} {\bibfnamefont {M.}~\bibnamefont {McLaughlin}}, \bibinfo {author} {\bibfnamefont {K.}~\bibnamefont {Kremer}}, \bibinfo {author} {\bibfnamefont {A.~D.}\ \bibnamefont {Cameron}}, \bibinfo {author} {\bibfnamefont {Z.}~\bibnamefont {Zhang}}, \bibinfo {author} {\bibfnamefont {J.}~\bibnamefont {Behrend}}, \bibinfo {author} {\bibfnamefont {M.}~\bibnamefont {Burgay}}, \bibinfo {author} {\bibfnamefont {S.}~\bibnamefont {Buchner}}, \bibinfo {author} {\bibfnamefont {D.~J.}\ \bibnamefont {Champion}}, \bibinfo {author} {\bibfnamefont {W.}~\bibnamefont {Chen}}, \bibinfo {author} {\bibfnamefont {S.}~\bibnamefont {Dai}}, \bibinfo {author} {\bibfnamefont
  {Y.}~\bibnamefont {Feng}}, \bibinfo {author} {\bibfnamefont {X.}~\bibnamefont {Fu}}, \bibinfo {author} {\bibfnamefont {M.}~\bibnamefont {Guo}}, \bibinfo {author} {\bibfnamefont {G.}~\bibnamefont {Hobbs}}, \bibinfo {author} {\bibfnamefont {E.~F.}\ \bibnamefont {Keane}}, \bibinfo {author} {\bibfnamefont {M.}~\bibnamefont {Kramer}}, \bibinfo {author} {\bibfnamefont {L.}~\bibnamefont {Levin}}, \bibinfo {author} {\bibfnamefont {X.}~\bibnamefont {Li}}, \bibinfo {author} {\bibfnamefont {M.}~\bibnamefont {Ni}}, \bibinfo {author} {\bibfnamefont {J.}~\bibnamefont {Pan}}, \bibinfo {author} {\bibfnamefont {P.~V.}\ \bibnamefont {Padmanabh}}, \bibinfo {author} {\bibfnamefont {A.}~\bibnamefont {Possenti}}, \bibinfo {author} {\bibfnamefont {S.~M.}\ \bibnamefont {Ransom}}, \bibinfo {author} {\bibfnamefont {C.-W.}\ \bibnamefont {Tsai}}, \bibinfo {author} {\bibfnamefont {V.}~\bibnamefont {Venkatraman~Krishnan}}, \bibinfo {author} {\bibfnamefont {P.}~\bibnamefont {Wang}}, \bibinfo {author} {\bibfnamefont {J.}~\bibnamefont
  {Zhang}}, \bibinfo {author} {\bibfnamefont {Q.}~\bibnamefont {Zhi}}, \bibinfo {author} {\bibfnamefont {Y.}~\bibnamefont {Zhang}},\ and\ \bibinfo {author} {\bibfnamefont {D.}~\bibnamefont {Li}},\ }\bibfield  {title} {\bibinfo {title} {Radio detection of an elusive millisecond pulsar in the globular cluster ngc 6397},\ }\href {https://doi.org/10.3847/2041-8213/ac81c3} {\bibfield  {journal} {\bibinfo  {journal} {The Astrophysical Journal Letters}\ }\textbf {\bibinfo {volume} {934}},\ \bibinfo {pages} {L21} (\bibinfo {year} {2022})}\BibitemShut {NoStop}%
\bibitem [{\citenamefont {Abbott}\ \emph {et~al.}(2008)\citenamefont {Abbott} \emph {et~al.}}]{S4allsky}%
  \BibitemOpen
  \bibfield  {author} {\bibinfo {author} {\bibfnamefont {B.}~\bibnamefont {Abbott}} \emph {et~al.},\ }\bibfield  {title} {\bibinfo {title} {All-sky search for periodic gravitational waves in ligo s4 data},\ }\href@noop {} {\bibfield  {journal} {\bibinfo  {journal} {Phys. Rev. D}\ }\textbf {\bibinfo {volume} {77}},\ \bibinfo {pages} {022001} (\bibinfo {year} {2008})}\BibitemShut {NoStop}%
\bibitem [{\citenamefont {Ushomirsky}\ \emph {et~al.}(2000)\citenamefont {Ushomirsky}, \citenamefont {Cutler},\ and\ \citenamefont {Bildsten}}]{ushomirsky2000deformations}%
  \BibitemOpen
  \bibfield  {author} {\bibinfo {author} {\bibfnamefont {G.}~\bibnamefont {Ushomirsky}}, \bibinfo {author} {\bibfnamefont {C.}~\bibnamefont {Cutler}},\ and\ \bibinfo {author} {\bibfnamefont {L.}~\bibnamefont {Bildsten}},\ }\bibfield  {title} {\bibinfo {title} {Deformations of accreting neutron star crusts and gravitational wave emission},\ }\href {https://doi.org/10.1046/j.1365-8711.2000.03938.x} {\bibfield  {journal} {\bibinfo  {journal} {Monthly Notices of the Royal Astronomical Society}\ }\textbf {\bibinfo {volume} {319}},\ \bibinfo {pages} {902} (\bibinfo {year} {2000})}\BibitemShut {NoStop}%
\bibitem [{\citenamefont {Cutler}(2002)}]{Cutler}%
  \BibitemOpen
  \bibfield  {author} {\bibinfo {author} {\bibfnamefont {C.}~\bibnamefont {Cutler}},\ }\bibfield  {title} {\bibinfo {title} {Gravitational waves from neutron stars with large toroidal b fields},\ }\href@noop {} {\bibfield  {journal} {\bibinfo  {journal} {Physical Review D}\ }\textbf {\bibinfo {volume} {66}},\ \bibinfo {pages} {084025} (\bibinfo {year} {2002})}\BibitemShut {NoStop}%
\bibitem [{\citenamefont {{Lasky}}\ and\ \citenamefont {{Melatos}}(2013)}]{Lasky2013}%
  \BibitemOpen
  \bibfield  {author} {\bibinfo {author} {\bibfnamefont {P.~D.}\ \bibnamefont {{Lasky}}}\ and\ \bibinfo {author} {\bibfnamefont {A.}~\bibnamefont {{Melatos}}},\ }\bibfield  {title} {\bibinfo {title} {{Tilted torus magnetic fields in neutron stars and their gravitational wave signatures}},\ }\href {https://doi.org/10.1103/PhysRevD.88.103005} {\bibfield  {journal} {\bibinfo  {journal} {Physical Review D}\ }\textbf {\bibinfo {volume} {88}},\ \bibinfo {eid} {103005} (\bibinfo {year} {2013})}\BibitemShut {NoStop}%
\bibitem [{\citenamefont {Andersson}(1998)}]{Andersson}%
  \BibitemOpen
  \bibfield  {author} {\bibinfo {author} {\bibfnamefont {N.}~\bibnamefont {Andersson}},\ }\bibfield  {title} {\bibinfo {title} {A new class of unstable modes of rotating relativistic stars},\ }\href@noop {} {\bibfield  {journal} {\bibinfo  {journal} {Astrophysical Journal}\ }\textbf {\bibinfo {volume} {502}},\ \bibinfo {pages} {708} (\bibinfo {year} {1998})}\BibitemShut {NoStop}%
\bibitem [{\citenamefont {Owen}\ \emph {et~al.}(1998)\citenamefont {Owen} \emph {et~al.}}]{OwenEtal}%
  \BibitemOpen
  \bibfield  {author} {\bibinfo {author} {\bibfnamefont {B.}~\bibnamefont {Owen}} \emph {et~al.},\ }\bibfield  {title} {\bibinfo {title} {Gravitational waves from hot young rapidly rotating neutron stars},\ }\href@noop {} {\bibfield  {journal} {\bibinfo  {journal} {Phys. Rev. D}\ }\textbf {\bibinfo {volume} {58}},\ \bibinfo {pages} {084020} (\bibinfo {year} {1998})}\BibitemShut {NoStop}%
\bibitem [{\citenamefont {Idrisy}\ \emph {et~al.}(2015)\citenamefont {Idrisy}, \citenamefont {Owen},\ and\ \citenamefont {Jones}}]{Idrisy_2015}%
  \BibitemOpen
  \bibfield  {author} {\bibinfo {author} {\bibfnamefont {A.}~\bibnamefont {Idrisy}}, \bibinfo {author} {\bibfnamefont {B.~J.}\ \bibnamefont {Owen}},\ and\ \bibinfo {author} {\bibfnamefont {D.~I.}\ \bibnamefont {Jones}},\ }\bibfield  {title} {\bibinfo {title} {$r$-mode frequencies of slowly rotating relativistic neutron stars with realistic equations of state},\ }\href {https://doi.org/10.1103/PhysRevD.91.024001} {\bibfield  {journal} {\bibinfo  {journal} {Phys. Rev. D}\ }\textbf {\bibinfo {volume} {91}},\ \bibinfo {pages} {024001} (\bibinfo {year} {2015})}\BibitemShut {NoStop}%
\bibitem [{\citenamefont {Caride}\ \emph {et~al.}(2019)\citenamefont {Caride}, \citenamefont {Inta}, \citenamefont {Owen},\ and\ \citenamefont {Rajbhandari}}]{caride2019search}%
  \BibitemOpen
  \bibfield  {author} {\bibinfo {author} {\bibfnamefont {S.}~\bibnamefont {Caride}}, \bibinfo {author} {\bibfnamefont {R.}~\bibnamefont {Inta}}, \bibinfo {author} {\bibfnamefont {B.~J.}\ \bibnamefont {Owen}},\ and\ \bibinfo {author} {\bibfnamefont {B.}~\bibnamefont {Rajbhandari}},\ }\bibfield  {title} {\bibinfo {title} {How to search for gravitational waves from $r$-modes of known pulsars},\ }\href {https://doi.org/10.1103/PhysRevD.100.064013} {\bibfield  {journal} {\bibinfo  {journal} {Phys. Rev. D}\ }\textbf {\bibinfo {volume} {100}},\ \bibinfo {pages} {064013} (\bibinfo {year} {2019})}\BibitemShut {NoStop}%
\bibitem [{\citenamefont {Gittins}\ and\ \citenamefont {Andersson}(2023)}]{Gittins_2023}%
  \BibitemOpen
  \bibfield  {author} {\bibinfo {author} {\bibfnamefont {F.}~\bibnamefont {Gittins}}\ and\ \bibinfo {author} {\bibfnamefont {N.}~\bibnamefont {Andersson}},\ }\bibfield  {title} {\bibinfo {title} {The r-modes of slowly rotating, stratified neutron stars},\ }\href {https://doi.org/10.1093/mnras/stad672} {\bibfield  {journal} {\bibinfo  {journal} {Monthly Notices of the Royal Astronomical Society}\ }\textbf {\bibinfo {volume} {521}},\ \bibinfo {pages} {3043} (\bibinfo {year} {2023})},\ \Eprint {https://arxiv.org/abs/https://academic.oup.com/mnras/article-pdf/521/2/3043/56454111/stad672.pdf} {https://academic.oup.com/mnras/article-pdf/521/2/3043/56454111/stad672.pdf} \BibitemShut {NoStop}%
\bibitem [{\citenamefont {Abadie}\ \emph {et~al.}(2010)\citenamefont {Abadie}, \citenamefont {Abbott}, \citenamefont {Abbott}, \citenamefont {Abernathy}, \citenamefont {Adams}, \citenamefont {Adhikari}, \citenamefont {Ajith}, \citenamefont {Allen}, \citenamefont {Allen}, \citenamefont {Amador~Ceron},\ and\ \citenamefont {et~al.}}]{abadie2010first}%
  \BibitemOpen
  \bibfield  {author} {\bibinfo {author} {\bibfnamefont {J.}~\bibnamefont {Abadie}}, \bibinfo {author} {\bibfnamefont {B.~P.}\ \bibnamefont {Abbott}}, \bibinfo {author} {\bibfnamefont {R.}~\bibnamefont {Abbott}}, \bibinfo {author} {\bibfnamefont {M.}~\bibnamefont {Abernathy}}, \bibinfo {author} {\bibfnamefont {C.}~\bibnamefont {Adams}}, \bibinfo {author} {\bibfnamefont {R.}~\bibnamefont {Adhikari}}, \bibinfo {author} {\bibfnamefont {P.}~\bibnamefont {Ajith}}, \bibinfo {author} {\bibfnamefont {B.}~\bibnamefont {Allen}}, \bibinfo {author} {\bibfnamefont {G.}~\bibnamefont {Allen}}, \bibinfo {author} {\bibfnamefont {E.}~\bibnamefont {Amador~Ceron}},\ and\ \bibinfo {author} {\bibnamefont {et~al.}},\ }\bibfield  {title} {\bibinfo {title} {First search for gravitational waves from the youngest known neutron star},\ }\href {https://doi.org/10.1088/0004-637x/722/2/1504} {\bibfield  {journal} {\bibinfo  {journal} {The Astrophysical Journal}\ }\textbf {\bibinfo {volume} {722}},\ \bibinfo {pages} {1504–1513} (\bibinfo
  {year} {2010})}\BibitemShut {NoStop}%
\bibitem [{\citenamefont {Aasi}\ \emph {et~al.}(2015{\natexlab{c}})\citenamefont {Aasi}, \citenamefont {Abbott}, \citenamefont {Abbott}, \citenamefont {Abbott}, \citenamefont {Abernathy}, \citenamefont {Acernese}, \citenamefont {Ackley}, \citenamefont {Adams}, \citenamefont {Adams}, \citenamefont {Addesso} \emph {et~al.}}]{aasi2015searches}%
  \BibitemOpen
  \bibfield  {author} {\bibinfo {author} {\bibfnamefont {J.}~\bibnamefont {Aasi}}, \bibinfo {author} {\bibfnamefont {B.}~\bibnamefont {Abbott}}, \bibinfo {author} {\bibfnamefont {R.}~\bibnamefont {Abbott}}, \bibinfo {author} {\bibfnamefont {T.}~\bibnamefont {Abbott}}, \bibinfo {author} {\bibfnamefont {M.}~\bibnamefont {Abernathy}}, \bibinfo {author} {\bibfnamefont {F.}~\bibnamefont {Acernese}}, \bibinfo {author} {\bibfnamefont {K.}~\bibnamefont {Ackley}}, \bibinfo {author} {\bibfnamefont {C.}~\bibnamefont {Adams}}, \bibinfo {author} {\bibfnamefont {T.}~\bibnamefont {Adams}}, \bibinfo {author} {\bibfnamefont {P.}~\bibnamefont {Addesso}}, \emph {et~al.},\ }\bibfield  {title} {\bibinfo {title} {Searches for continuous gravitational waves from nine young supernova remnants},\ }\href {https://doi.org/10.1088/0004-637X/813/1/39} {\bibfield  {journal} {\bibinfo  {journal} {The Astrophysical Journal}\ }\textbf {\bibinfo {volume} {813}},\ \bibinfo {pages} {39} (\bibinfo {year} {2015}{\natexlab{c}})}\BibitemShut
  {NoStop}%
\bibitem [{\citenamefont {Abbott}\ \emph {et~al.}(2019)\citenamefont {Abbott} \emph {et~al.}}]{Abbott_2019}%
  \BibitemOpen
  \bibfield  {author} {\bibinfo {author} {\bibfnamefont {B.~P.}\ \bibnamefont {Abbott}} \emph {et~al.},\ }\bibfield  {title} {\bibinfo {title} {Searches for continuous gravitational waves from 15 supernova remnants and fomalhaut b with advanced {LIGO}},\ }\href {https://doi.org/10.3847/1538-4357/ab113b} {\bibfield  {journal} {\bibinfo  {journal} {The Astrophysical Journal}\ }\textbf {\bibinfo {volume} {875}},\ \bibinfo {pages} {122} (\bibinfo {year} {2019})}\BibitemShut {NoStop}%
\bibitem [{\citenamefont {Abbott}\ \emph {et~al.}(2021{\natexlab{b}})\citenamefont {Abbott} \emph {et~al.}}]{O3aSNR}%
  \BibitemOpen
  \bibfield  {author} {\bibinfo {author} {\bibfnamefont {R.}~\bibnamefont {Abbott}} \emph {et~al.},\ }\bibfield  {title} {\bibinfo {title} {Searches for continuous gravitational waves from young supernova remnants in the early third observing run of advanced ligo and virgo},\ }\href@noop {} {\bibfield  {journal} {\bibinfo  {journal} {Astrophysical Journal}\ }\textbf {\bibinfo {volume} {921}},\ \bibinfo {pages} {80} (\bibinfo {year} {2021}{\natexlab{b}})}\BibitemShut {NoStop}%
\bibitem [{\citenamefont {Abbott}\ \emph {et~al.}(2022)\citenamefont {Abbott}, \citenamefont {Abbott}, \citenamefont {Acernese}, \citenamefont {Ackley}, \citenamefont {Adams}, \citenamefont {Adhikari}, \citenamefont {Adhikari},\ and\ \citenamefont {et~al.}}]{O3aSNRweave}%
  \BibitemOpen
  \bibfield  {author} {\bibinfo {author} {\bibfnamefont {R.}~\bibnamefont {Abbott}}, \bibinfo {author} {\bibfnamefont {T.~D.}\ \bibnamefont {Abbott}}, \bibinfo {author} {\bibfnamefont {F.}~\bibnamefont {Acernese}}, \bibinfo {author} {\bibfnamefont {K.}~\bibnamefont {Ackley}}, \bibinfo {author} {\bibfnamefont {C.}~\bibnamefont {Adams}}, \bibinfo {author} {\bibfnamefont {N.}~\bibnamefont {Adhikari}}, \bibinfo {author} {\bibfnamefont {R.~X.}\ \bibnamefont {Adhikari}},\ and\ \bibinfo {author} {\bibnamefont {et~al.}} (\bibinfo {collaboration} {LIGO Scientific Collaboration and Virgo Collaboration}),\ }\bibfield  {title} {\bibinfo {title} {Search of the early o3 ligo data for continuous gravitational waves from the cassiopeia a and vela jr. supernova remnants},\ }\href {https://doi.org/10.1103/PhysRevD.105.082005} {\bibfield  {journal} {\bibinfo  {journal} {Phys. Rev. D}\ }\textbf {\bibinfo {volume} {105}},\ \bibinfo {pages} {082005} (\bibinfo {year} {2022})}\BibitemShut {NoStop}%
\bibitem [{\citenamefont {Wang}\ and\ \citenamefont {Riles}(2024)}]{fullo3_cass}%
  \BibitemOpen
  \bibfield  {author} {\bibinfo {author} {\bibfnamefont {J.}~\bibnamefont {Wang}}\ and\ \bibinfo {author} {\bibfnamefont {K.}~\bibnamefont {Riles}},\ }\bibfield  {title} {\bibinfo {title} {Deep search of the full o3 ligo data for continuous gravitational waves from the cassiopeia a central compact object},\ }\href {https://doi.org/10.1103/PhysRevD.110.042006} {\bibfield  {journal} {\bibinfo  {journal} {Phys. Rev. D}\ }\textbf {\bibinfo {volume} {110}},\ \bibinfo {pages} {042006} (\bibinfo {year} {2024})}\BibitemShut {NoStop}%
\bibitem [{\citenamefont {{The LIGO Scientific Collaboration}}\ \emph {et~al.}(2026)\citenamefont {{The LIGO Scientific Collaboration}}, \citenamefont {{the Virgo Collaboration}}, \citenamefont {{the KAGRA Collaboration}}, \citenamefont {Abac}, \citenamefont {Abouelfettouh}, \citenamefont {Acernese}, \citenamefont {Ackley}, \citenamefont {Adamcewicz} \emph {et~al.}}]{O4aSNR}%
  \BibitemOpen
  \bibfield  {author} {\bibinfo {author} {\bibnamefont {{The LIGO Scientific Collaboration}}}, \bibinfo {author} {\bibnamefont {{the Virgo Collaboration}}}, \bibinfo {author} {\bibnamefont {{the KAGRA Collaboration}}}, \bibinfo {author} {\bibfnamefont {A.~G.}\ \bibnamefont {Abac}}, \bibinfo {author} {\bibfnamefont {I.}~\bibnamefont {Abouelfettouh}}, \bibinfo {author} {\bibfnamefont {F.}~\bibnamefont {Acernese}}, \bibinfo {author} {\bibfnamefont {K.}~\bibnamefont {Ackley}}, \bibinfo {author} {\bibfnamefont {C.}~\bibnamefont {Adamcewicz}}, \emph {et~al.},\ }\bibfield  {title} {\bibinfo {title} {Searches for continuous gravitational waves from supernova remnants in the first part of the ligo-virgo-kagra fourth observing run},\ }\href {https://arxiv.org/abs/2603.25808} {\bibfield  {journal} {\bibinfo  {journal} {arXiv}\ } (\bibinfo {year} {2026})},\ \Eprint {https://arxiv.org/abs/2603.25808} {2603.25808 [gr-qc]} \BibitemShut {NoStop}%
\bibitem [{\citenamefont {Palomba}(2005)}]{Palomba_2005}%
  \BibitemOpen
  \bibfield  {author} {\bibinfo {author} {\bibfnamefont {C.}~\bibnamefont {Palomba}},\ }\bibfield  {title} {\bibinfo {title} {Simulation of a population of isolated neutron stars evolving through the emission of gravitational waves},\ }\href {https://doi.org/10.1111/j.1365-2966.2005.08975.x} {\bibfield  {journal} {\bibinfo  {journal} {Monthly Notices of the Royal Astronomical Society}\ }\textbf {\bibinfo {volume} {359}},\ \bibinfo {pages} {1150} (\bibinfo {year} {2005})},\ \Eprint {https://arxiv.org/abs/https://academic.oup.com/mnras/article-pdf/359/3/1150/2944051/359-3-1150.pdf} {https://academic.oup.com/mnras/article-pdf/359/3/1150/2944051/359-3-1150.pdf} \BibitemShut {NoStop}%
\bibitem [{\citenamefont {Cutler}\ and\ \citenamefont {Schutz}(2005)}]{Cutler_2005}%
  \BibitemOpen
  \bibfield  {author} {\bibinfo {author} {\bibfnamefont {C.}~\bibnamefont {Cutler}}\ and\ \bibinfo {author} {\bibfnamefont {B.~F.}\ \bibnamefont {Schutz}},\ }\bibfield  {title} {\bibinfo {title} {{The Generalized F-statistic: Multiple detectors and multiple GW pulsars}},\ }\href {https://doi.org/10.1103/PhysRevD.72.063006} {\bibfield  {journal} {\bibinfo  {journal} {Phys. Rev. D}\ }\textbf {\bibinfo {volume} {72}},\ \bibinfo {pages} {063006} (\bibinfo {year} {2005})},\ \Eprint {https://arxiv.org/abs/gr-qc/0504011} {arXiv:gr-qc/0504011} \BibitemShut {NoStop}%
\bibitem [{\citenamefont {Wette}\ \emph {et~al.}(2018)\citenamefont {Wette} \emph {et~al.}}]{WetteEtal}%
  \BibitemOpen
  \bibfield  {author} {\bibinfo {author} {\bibfnamefont {K.}~\bibnamefont {Wette}} \emph {et~al.},\ }\bibfield  {title} {\bibinfo {title} {Implementing a semicoherent search for continuous gravitational waves using optimally-constructed template banks},\ }\href@noop {} {\bibfield  {journal} {\bibinfo  {journal} {Phys. Rev. D}\ }\textbf {\bibinfo {volume} {97}},\ \bibinfo {pages} {123016} (\bibinfo {year} {2018})}\BibitemShut {NoStop}%
\bibitem [{\citenamefont {Wette}\ and\ \citenamefont {Prix}(2013)}]{WettePrix}%
  \BibitemOpen
  \bibfield  {author} {\bibinfo {author} {\bibfnamefont {K.}~\bibnamefont {Wette}}\ and\ \bibinfo {author} {\bibfnamefont {R.}~\bibnamefont {Prix}},\ }\bibfield  {title} {\bibinfo {title} {Flat parameter-space metric for all-sky searches for gravitational-wave pulsars},\ }\href@noop {} {\bibfield  {journal} {\bibinfo  {journal} {Phys. Rev. D}\ }\textbf {\bibinfo {volume} {88}},\ \bibinfo {pages} {123005} (\bibinfo {year} {2013})}\BibitemShut {NoStop}%
\bibitem [{\citenamefont {Wette}(2015)}]{WetteMetrics}%
  \BibitemOpen
  \bibfield  {author} {\bibinfo {author} {\bibfnamefont {K.}~\bibnamefont {Wette}},\ }\bibfield  {title} {\bibinfo {title} {Parameter-space metric for all-sky semi-coherent searches for gravitational-wave pulsars},\ }\href@noop {} {\bibfield  {journal} {\bibinfo  {journal} {Phys. Rev. D}\ }\textbf {\bibinfo {volume} {92}},\ \bibinfo {pages} {082003} (\bibinfo {year} {2015})}\BibitemShut {NoStop}%
\bibitem [{\citenamefont {Wette}(2014)}]{WetteLattice}%
  \BibitemOpen
  \bibfield  {author} {\bibinfo {author} {\bibfnamefont {K.}~\bibnamefont {Wette}},\ }\bibfield  {title} {\bibinfo {title} {Lattice template placement for coherent all-sky searches for gravitational-wave pulsars},\ }\href@noop {} {\bibfield  {journal} {\bibinfo  {journal} {Phys. Rev. D}\ }\textbf {\bibinfo {volume} {90}},\ \bibinfo {pages} {122010} (\bibinfo {year} {2014})}\BibitemShut {NoStop}%
\bibitem [{\citenamefont {Brady}\ and\ \citenamefont {Creighton}(2000)}]{Brady_1998}%
  \BibitemOpen
  \bibfield  {author} {\bibinfo {author} {\bibfnamefont {P.~R.}\ \bibnamefont {Brady}}\ and\ \bibinfo {author} {\bibfnamefont {T.}~\bibnamefont {Creighton}},\ }\bibfield  {title} {\bibinfo {title} {{Searching for periodic sources with LIGO. 2. Hierarchical searches}},\ }\href {https://doi.org/10.1103/PhysRevD.61.082001} {\bibfield  {journal} {\bibinfo  {journal} {Phys. Rev. D}\ }\textbf {\bibinfo {volume} {61}},\ \bibinfo {pages} {082001} (\bibinfo {year} {2000})},\ \Eprint {https://arxiv.org/abs/gr-qc/9812014} {arXiv:gr-qc/9812014} \BibitemShut {NoStop}%
\bibitem [{\citenamefont {Astone}\ \emph {et~al.}(2002)\citenamefont {Astone}, \citenamefont {Borkowski}, \citenamefont {Jaranowski},\ and\ \citenamefont {Kr\'olak}}]{Astone_2002}%
  \BibitemOpen
  \bibfield  {author} {\bibinfo {author} {\bibfnamefont {P.}~\bibnamefont {Astone}}, \bibinfo {author} {\bibfnamefont {K.~M.}\ \bibnamefont {Borkowski}}, \bibinfo {author} {\bibfnamefont {P.}~\bibnamefont {Jaranowski}},\ and\ \bibinfo {author} {\bibfnamefont {A.}~\bibnamefont {Kr\'olak}},\ }\bibfield  {title} {\bibinfo {title} {Data analysis of gravitational-wave signals from spinning neutron stars. iv. an all-sky search},\ }\href {https://doi.org/10.1103/PhysRevD.65.042003} {\bibfield  {journal} {\bibinfo  {journal} {Phys. Rev. D}\ }\textbf {\bibinfo {volume} {65}},\ \bibinfo {pages} {042003} (\bibinfo {year} {2002})}\BibitemShut {NoStop}%
\bibitem [{\citenamefont {Prix}(2007)}]{Prix_2006}%
  \BibitemOpen
  \bibfield  {author} {\bibinfo {author} {\bibfnamefont {R.}~\bibnamefont {Prix}},\ }\bibfield  {title} {\bibinfo {title} {{Search for continuous gravitational waves: Metric of the multi-detector F-statistic}},\ }\href {https://doi.org/10.1103/PhysRevD.75.023004} {\bibfield  {journal} {\bibinfo  {journal} {Phys. Rev. D}\ }\textbf {\bibinfo {volume} {75}},\ \bibinfo {pages} {023004} (\bibinfo {year} {2007})},\ \bibinfo {note} {[Erratum: Phys.Rev.D 75, 069901 (2007)]},\ \Eprint {https://arxiv.org/abs/gr-qc/0606088} {arXiv:gr-qc/0606088} \BibitemShut {NoStop}%
\bibitem [{\citenamefont {Behnke}\ \emph {et~al.}(2015)\citenamefont {Behnke}, \citenamefont {Papa},\ and\ \citenamefont {Prix}}]{sensitivitydepth}%
  \BibitemOpen
  \bibfield  {author} {\bibinfo {author} {\bibfnamefont {B.}~\bibnamefont {Behnke}}, \bibinfo {author} {\bibfnamefont {M.~A.}\ \bibnamefont {Papa}},\ and\ \bibinfo {author} {\bibfnamefont {R.}~\bibnamefont {Prix}},\ }\bibfield  {title} {\bibinfo {title} {Postprocessing methods used in the search for continuous gravitational-wave signals from the galactic center},\ }\href {https://doi.org/10.1103/PhysRevD.91.064007} {\bibfield  {journal} {\bibinfo  {journal} {Phys. Rev. D}\ }\textbf {\bibinfo {volume} {91}},\ \bibinfo {pages} {064007} (\bibinfo {year} {2015})}\BibitemShut {NoStop}%
\bibitem [{\citenamefont {Jaranowski}\ \emph {et~al.}(1998{\natexlab{b}})\citenamefont {Jaranowski}, \citenamefont {Kr{\'{o}}lak},\ and\ \citenamefont {Schutz}}]{Jaranowski1998}%
  \BibitemOpen
  \bibfield  {author} {\bibinfo {author} {\bibfnamefont {P.}~\bibnamefont {Jaranowski}}, \bibinfo {author} {\bibfnamefont {A.}~\bibnamefont {Kr{\'{o}}lak}},\ and\ \bibinfo {author} {\bibfnamefont {B.~F.}\ \bibnamefont {Schutz}},\ }\bibfield  {title} {\bibinfo {title} {{Data analysis of gravitational-wave signals from spinning neutron stars: The signal and its detection}},\ }\href {https://doi.org/10.1103/PhysRevD.58.063001} {\bibfield  {journal} {\bibinfo  {journal} {Physical Review D}\ }\textbf {\bibinfo {volume} {58}},\ \bibinfo {pages} {063001} (\bibinfo {year} {1998}{\natexlab{b}})}\BibitemShut {NoStop}%
\bibitem [{\citenamefont {Bildsten}(1998)}]{Bildsten}%
  \BibitemOpen
  \bibfield  {author} {\bibinfo {author} {\bibfnamefont {L.}~\bibnamefont {Bildsten}},\ }\bibfield  {title} {\bibinfo {title} {Gravitational radiation and rotation of accreting neutron stars},\ }\href@noop {} {\bibfield  {journal} {\bibinfo  {journal} {Astrophysical Journal Letters}\ }\textbf {\bibinfo {volume} {501}},\ \bibinfo {pages} {L89} (\bibinfo {year} {1998})}\BibitemShut {NoStop}%
\bibitem [{\citenamefont {Friedman}\ and\ \citenamefont {Morsink}(1998)}]{FriedmanMorsink}%
  \BibitemOpen
  \bibfield  {author} {\bibinfo {author} {\bibfnamefont {J.}~\bibnamefont {Friedman}}\ and\ \bibinfo {author} {\bibfnamefont {S.}~\bibnamefont {Morsink}},\ }\bibfield  {title} {\bibinfo {title} {Axial instability of rotating relativistic stars},\ }\href@noop {} {\bibfield  {journal} {\bibinfo  {journal} {Astrophys. J.}\ }\textbf {\bibinfo {volume} {502}},\ \bibinfo {pages} {714} (\bibinfo {year} {1998})}\BibitemShut {NoStop}%
\bibitem [{\citenamefont {Kojima}(1998)}]{Kojima}%
  \BibitemOpen
  \bibfield  {author} {\bibinfo {author} {\bibfnamefont {Y.}~\bibnamefont {Kojima}},\ }\bibfield  {title} {\bibinfo {title} {Quasitoroidal oscillations in rotating relativistic stars},\ }\href@noop {} {\bibfield  {journal} {\bibinfo  {journal} {Mon. Not. Roy. Astron. Soc.}\ }\textbf {\bibinfo {volume} {293}},\ \bibinfo {pages} {49} (\bibinfo {year} {1998})}\BibitemShut {NoStop}%
\bibitem [{\citenamefont {Owen}(2010)}]{Owen2010}%
  \BibitemOpen
  \bibfield  {author} {\bibinfo {author} {\bibfnamefont {B.~J.}\ \bibnamefont {Owen}},\ }\bibfield  {title} {\bibinfo {title} {How to adapt broad-band gravitational-wave searches for $r$-modes},\ }\href {https://doi.org/10.1103/PhysRevD.82.104002} {\bibfield  {journal} {\bibinfo  {journal} {Phys. Rev. D}\ }\textbf {\bibinfo {volume} {82}},\ \bibinfo {pages} {104002} (\bibinfo {year} {2010})}\BibitemShut {NoStop}%
\bibitem [{\citenamefont {Owen}(2005)}]{Owen_2005}%
  \BibitemOpen
  \bibfield  {author} {\bibinfo {author} {\bibfnamefont {B.~J.}\ \bibnamefont {Owen}},\ }\bibfield  {title} {\bibinfo {title} {Maximum elastic deformations of compact stars with exotic equations of state},\ }\bibfield  {journal} {\bibinfo  {journal} {Physical Review Letters}\ }\textbf {\bibinfo {volume} {95}},\ \href {https://doi.org/10.1103/physrevlett.95.211101} {10.1103/physrevlett.95.211101} (\bibinfo {year} {2005})\BibitemShut {NoStop}%
\bibitem [{\citenamefont {Haskell}\ \emph {et~al.}(2007)\citenamefont {Haskell}, \citenamefont {Andersson}, \citenamefont {Jones},\ and\ \citenamefont {Samuelsson}}]{Haskell_2007}%
  \BibitemOpen
  \bibfield  {author} {\bibinfo {author} {\bibfnamefont {B.}~\bibnamefont {Haskell}}, \bibinfo {author} {\bibfnamefont {N.}~\bibnamefont {Andersson}}, \bibinfo {author} {\bibfnamefont {D.~I.}\ \bibnamefont {Jones}},\ and\ \bibinfo {author} {\bibfnamefont {L.}~\bibnamefont {Samuelsson}},\ }\bibfield  {title} {\bibinfo {title} {Are neutron stars with crystalline color-superconducting cores relevant for the ligo experiment?},\ }\bibfield  {journal} {\bibinfo  {journal} {Physical Review Letters}\ }\textbf {\bibinfo {volume} {99}},\ \href {https://doi.org/10.1103/physrevlett.99.231101} {10.1103/physrevlett.99.231101} (\bibinfo {year} {2007})\BibitemShut {NoStop}%
\bibitem [{\citenamefont {Johnson-McDaniel}\ and\ \citenamefont {Owen}(2013)}]{Johnson-McDaniel2013eps}%
  \BibitemOpen
  \bibfield  {author} {\bibinfo {author} {\bibfnamefont {N.~K.}\ \bibnamefont {Johnson-McDaniel}}\ and\ \bibinfo {author} {\bibfnamefont {B.~J.}\ \bibnamefont {Owen}},\ }\bibfield  {title} {\bibinfo {title} {Maximum elastic deformations of relativistic stars},\ }\href {https://doi.org/10.1103/PhysRevD.88.044004} {\bibfield  {journal} {\bibinfo  {journal} {Phys. Rev. D}\ }\textbf {\bibinfo {volume} {88}},\ \bibinfo {pages} {044004} (\bibinfo {year} {2013})}\BibitemShut {NoStop}%
\bibitem [{\citenamefont {Bondarescu}\ \emph {et~al.}(2009)\citenamefont {Bondarescu}, \citenamefont {Teukolsky},\ and\ \citenamefont {Wasserman}}]{Bondarescu2009rmode}%
  \BibitemOpen
  \bibfield  {author} {\bibinfo {author} {\bibfnamefont {R.}~\bibnamefont {Bondarescu}}, \bibinfo {author} {\bibfnamefont {S.~A.}\ \bibnamefont {Teukolsky}},\ and\ \bibinfo {author} {\bibfnamefont {I.}~\bibnamefont {Wasserman}},\ }\bibfield  {title} {\bibinfo {title} {Spinning down newborn neutron stars: Nonlinear development of the $r$-mode instability},\ }\href {https://doi.org/10.1103/PhysRevD.79.104003} {\bibfield  {journal} {\bibinfo  {journal} {Phys. Rev. D}\ }\textbf {\bibinfo {volume} {79}},\ \bibinfo {pages} {104003} (\bibinfo {year} {2009})}\BibitemShut {NoStop}%
\bibitem [{\citenamefont {Bondarescu}\ \emph {et~al.}(2007)\citenamefont {Bondarescu}, \citenamefont {Teukolsky},\ and\ \citenamefont {Wasserman}}]{Bondarescu_2007}%
  \BibitemOpen
  \bibfield  {author} {\bibinfo {author} {\bibfnamefont {R.}~\bibnamefont {Bondarescu}}, \bibinfo {author} {\bibfnamefont {S.~A.}\ \bibnamefont {Teukolsky}},\ and\ \bibinfo {author} {\bibfnamefont {I.}~\bibnamefont {Wasserman}},\ }\bibfield  {title} {\bibinfo {title} {Spin evolution of accreting neutron stars: Nonlinear development of the $r$-mode instability},\ }\href {https://doi.org/10.1103/PhysRevD.76.064019} {\bibfield  {journal} {\bibinfo  {journal} {Phys. Rev. D}\ }\textbf {\bibinfo {volume} {76}},\ \bibinfo {pages} {064019} (\bibinfo {year} {2007})}\BibitemShut {NoStop}%
\bibitem [{\citenamefont {Cheung}(2026)}]{cheung_2026_zenodo}%
  \BibitemOpen
  \bibfield  {author} {\bibinfo {author} {\bibfnamefont {D.}~\bibnamefont {Cheung}},\ }\bibfield  {title} {\bibinfo {title} {Outlier parameters from the study: Search for continuous gravitational waves from neutron stars in five globular clusters in the first part of the fourth ligo–virgo–kagra observing run [data set]},\ }\href {https://doi.org/10.5281/zenodo.18426313} {10.5281/zenodo.18426313} (\bibinfo {year} {2026})\BibitemShut {NoStop}%
\end{thebibliography}%

\end{document}